\documentclass[prd,aps,nofootinbib,showpacs,preprintnumbers,amsmath,amssymb,floatfix]{revtex4}

\usepackage{times}
\usepackage{hyperref}

\usepackage{graphicx}
\usepackage{dcolumn}
\usepackage{bm}
\usepackage{slashed}

\begin{document}

\title{Baryon axial vector current in large-$N_c$ chiral perturbation theory: Complete analysis for $N_c=3$}

\author{
Rub\'en Flores-Mendieta
}
\affiliation{
Instituto de F{\'\i}sica, Universidad Aut\'onoma de San Luis Potos{\'\i}, \'Alvaro Obreg\'on 64, Zona Centro, San Luis Potos{\'\i}, S.L.P.\ 78000, Mexico
}

\author{
Carlos Isaac Garc{\'\i}a
}
\affiliation{
Instituto de F{\'\i}sica, Universidad Aut\'onoma de San Luis Potos{\'\i}, \'Alvaro Obreg\'on 64, Zona Centro, San Luis Potos{\'\i}, S.L.P.\ 78000, Mexico
}

\author{
Johann Hern\'andez
}
\affiliation{
Instituto de F{\'\i}sica, Universidad Aut\'onoma de San Luis Potos{\'\i}, \'Alvaro Obreg\'on 64, Zona Centro, San Luis Potos{\'\i}, S.L.P.\ 78000, Mexico
}

\date{}

\begin{abstract}
The baryon axial vector current is computed in heavy baryon chiral perturbation theory in the large-$N_c$ limit, where $N_c$ is the number of color charges. One-loop nonanalytic corrections of order $m_q \ln m_q$ are comprised in the analysis, with contributions of both intermediate octet and decuplet baryon states, to all orders in the $1/N_c$ expansion of the axial vector current relevant for $N_c=3$. Theoretical expressions are obtained in the limit of vanishing decuplet-octet mass difference only, which allows one to carry out a full comparison with conventional heavy baryon chiral perturbation theory results for three flavors of light quarks and at the physical value $N_c=3$. Both approaches perfectly agree to all orders considered. Furthermore, a numerical analysis via a least-squares fit is performed in order to extract the values of the free parameters of the theory, using the experimental data available. The predictions of formalism are remarkable.
\end{abstract}


\maketitle

\section{Introduction}

The formulation of baryon chiral perturbation theory using an effective Lagrangian for broken $SU(3)_L \times SU(3)_R$ chiral symmetry, in which baryons are assumed to be heavy static fermions, was introduced by Jenkins and Manohar three decades ago \cite{jm255,jm259}. The formulation, usually referred to as heavy baryon chiral perturbation theory (HBChPT), is a nonrelativistic effective theory that has been proven useful in studying meson-baryon interactions at low energies. The success of this effective theory stems from the fact that it possesses a consistent expansion in powers of momentum and light quark masses because the baryon mass does not appear in the effective Lagrangian.

The analysis of the baryon axial vector current was one of the earliest applications of HBChPT \cite{jm255,jm259}, from which two major conclusions were reached: 1) The axial coupling ratios were found to be close to their $SU(6)$ values with $F/D\approx 2/3$, which is the prediction of the nonrelativistic quark model. 2) One-loop corrections became large when only spin-$1/2$ octet baryon intermediate states were kept whereas corrections with the inclusion of both spin-$1/2$ octet and spin-$3/2$ decuplet baryon intermediate states yielded sizable cancellations between them.

The baryon axial vector current has also been tackled in the context of the $1/N_c$ expansion of QCD, where $N_c$ is the number of color charges \cite{dm93,djm94,djm95}. Large-$N_c$ QCD is the generalization of QCD from $N_c=3$ to $N_c \gg 3$. In the large-$N_c$ limit, the baryon sector exhibits an exact $SU(2N_f)$ contracted spin-flavor symmetry, where $N_f$ is the number of light quark flavors. The spin-flavor symmetry becomes a useful tool to classify large-$N_c$ baryon states and matrix elements and to compute static properties of large-$N_c$ baryons in a systematic expansion in 1/$N_c$ \cite{djm94,djm95}.

The joint use of chiral perturbation theory and the $1/N_c$ expansion is another powerful method to study the low-energy consequences of QCD. The resultant formalism, which will be loosely referred to as the combined formalism henceforth, builds on the chiral Lagrangian in the $1/N_c$ expansion constructed in Ref.~\cite{jen96}. Succinctly, the combined formalism describes the interactions between the spin-$1/2$ baryon octet and the spin-$3/2$ baryon decuplet with the pseudoscalar Goldstone boson octet enlarged to include the $\eta^\prime$ to conform a meson nonet. Early analyses of baryon properties based on the chiral Lagrangian of Ref.~\cite{jen96} include mass \cite{jen96} to further continue with axial vector current \cite{rfm00,rfm06,rfm12}, magnetic moment \cite{dai,rfm09,rfm14}, vector current \cite{fmg}, charge radius \cite{fmr}, and more recently $s$-wave amplitude in nonleptonic decay \cite{rfm19}. All these analyses share a common feature: Large cancellations occur between one-loop graphs provided that all baryon states in a complete multiplet of the large-$N_c$ $SU(6)$ spin-flavor symmetry are included in the sum over intermediate states and that the axial coupling ratios predicted by this spin-flavor symmetry are used \cite{djm94,rfm00,rfm06}.

For the specific case of the baryon axial vector current, the structure of large-$N_c$ cancellations was qualitatively discussed in detail in Ref.~\cite{rfm00}. However, it was not until the analyses of Refs.~\cite{rfm06,rfm12} that some calculations were presented, exhibiting explicitly those cancellations. Working to a certain order in the $1/N_c$ expansion, the complicated expressions of commutators and/or anticommutators of the $SU(6)$ spin-flavor operators contained in the one-loop contributions were computed, first under the assumption of degenerate intermediate baryon states in the loops \cite{rfm06}, and next introducing the decuplet-octet mass difference, hereafter represented by $\Delta \equiv M_T - M_B$ \cite{jen96}. In the large-$N_c$ limit, $\Delta \propto 1/N_c$, so the degeneracy limit assumed in Ref.~\cite{rfm06} represents a good first approximation to the problem and its application is legitimate. Although the order in $1/N_c$ considered in these previous works was rather limited, the method paved the way toward a better understanding of baryon properties in terms of a systematic and controlled expansion.

The present work constitutes an attempt to outdo the analyses of Refs.~\cite{rfm06,rfm12} in some aspects. This time, one-loop corrections will be evaluated, for degenerate baryon intermediate states, at {\it all orders} allowed in the expression for the axial vector current relevant for $N_c=3$. Algebraically, this task is extremely difficult due to the considerable amount of group theory involved; nevertheless it is doable. After completing this task, it will be possible to compare the resultant theoretical expressions with the ones obtained within conventional HBChPT (i.e., the effective theory with no $1/N_c$ expansion) to advance in the enterprise outlined through a series of papers a few years ago: To verify whether the combined formalism and conventional HBChPT yield the same results at the physical value $N_c=3$. Up to now, the agreement has been a successful one for mass \cite{jen96}, vector current \cite{fmg}, and $s$-wave amplitude in nonleptonic decays \cite{rfm19}. For other static properties listed above, only partial results have been obtained; it is expected however to complete the comparisons in the near future.

This paper is organized as follows. In Sec.~\ref{sec:overview} some necessary material on large-$N_c$ chiral perturbation theory is reviewed in order to introduce notation and conventions. In Sec.~\ref{sec:renormalization} one-loop graphs that renormalize the baryon axial vector current are reanalyzed under the premises discussed above. Results are classified in terms of the different $SU(3)$ flavor representations. In Sec.~\ref{sec:heavy} the same loop graphs are evaluated within HBChPT; results are arranged in a close parallelism with the ones obtained in the previous section. At this stage, using the linear relations between the operator coefficients of the $1/N_c$ expansion of the axial vector current and the four conventional $SU(3)$ axial vector couplings $D$, $F$, $\mathcal{C}$, and $\mathcal{H}$ \cite{jen96}, theoretical expressions of the two approaches are compared in Sec.~\ref{sec:compara}. There is a remarkable agreement order by order between both approaches. The full expression for the renormalized axial vector current is provided in Sec.~\ref{sec:totalc}; following Ref.~\cite{rfm12}, effects related to $SU(3)$ flavor symmetry breaking are included as follows: On the one hand, at tree level, all relevant operators which explicitly break $SU(3)$ at leading order are accounted for; this introduces four additional free parameters. On the other hand, in the one-loop corrections, $SU(3)$ symmetry breaking is accounted for implicitly, since the loop integrals depend on the meson masses.
The analysis can be extended to attempt a determination of the operator coefficients (or equivalently the axial vector couplings) by performing a comparison with the available experimental data about the semileptonic decays of octet baryons and the strong decays of the decuplet baryons \cite{part} through a least-squares fit. Results are presented in Sec.~\ref{sec:fits}. To close this paper, conclusions are presented in Sec.~\ref{sec:finish}. All the supplementary material necessary to fully construct the main results of the paper are given in various appendices. Concretely, reductions of operators for flavor singlet, flavor $\mathbf{8}$, and flavor $\mathbf{27}$ representations can be found in Appendices \ref{sec:r1}, \ref{sec:r8}, and \ref{sec:r27}, respectively. The operator bases used, along with their matrix elements are listed in Appendix \ref{sec:bases}.

\section{\label{sec:overview}A survey of large-$N_c$ chiral perturbation theory}

This paper builds upon the previous work on large-$N_c$ chiral perturbation theory presented in Ref.~\cite{jen96} to extend the analysis of the renormalization of the baryon axial vector current dealt with in Refs.~\cite{rfm06,rfm12}. In this section the basic elements of the formalism will be summarized; in passing, the notation and conventions used here will be introduced.

The chiral Lagrangian for baryons in the $1/N_c$ expansion, which incorporates nonet symmetry and the contracted spin-flavor symmetry for baryons in the large-$N_c$ limit, can be expressed as \cite{jen96}
\begin{equation}
\mathcal{L}_{\text{baryon}} = i \mathcal{D}^0 - \mathcal{M}_{\text{hyperfine}} + \text{Tr} \left(\mathcal{A}^k \lambda^c \right) A^{kc} + \frac{1}{N_c} \text{Tr} \left(\mathcal{A}^k \frac{2I}{\sqrt 6}\right) A^k + \ldots, \label{eq:ncch}
\end{equation}
where
\begin{equation}
\mathcal{D}^0 = \partial^0 \openone + \text{Tr} \left(\mathcal{V}^0 \lambda^c\right) T^c. \label{eq:kin}
\end{equation}
The ellipses in Eq.~(\ref{eq:ncch}) represents higher partial wave meson couplings which are present at subleading orders in the $1/N_c$ expansions for $N_c > 3$. Due to the fact that these higher partial waves vanish in the large-$N_c$ limit, the meson coupling to baryons is purely $p$ wave. 

Meson fields appear in the Lagrangian (\ref{eq:ncch}) through the relations
\begin{equation}
\mathcal{V}^0 = \frac12 \left(\xi \partial^0 \xi^\dagger + \xi^\dagger \partial^0 \xi\right), \qquad
\mathcal{A}^k = \frac{i}{2} \left(\xi \nabla^k \xi^\dagger - \xi^\dagger \nabla^k \xi\right), \qquad \qquad
\xi(x)=\exp[i\Pi(x)/f],
\end{equation}
where $\Pi(x)$ stands for the nonet of Goldstone boson fields and $f \approx 93$ MeV is the pion decay constant.

Each summand in the Lagrangian (\ref{eq:ncch}) is made up by a baryon operator. Specifically, the baryon kinetic energy term involves the spin-flavor identity, $\mathcal{M}_{\text{hyperfine}}$ represents the hyperfine baryon mass operator which accounts for the spin splittings of the tower of baryon states with spins $1/2,\ldots, N_c/2$ in the flavor representations. On the other hand, $A^k$ and $A^{kc}$ stand for the flavor singlet and flavor octet baryon axial vector currents, respectively.

All these baryon operators are written as polynomials in the $SU(6)$ spin-flavor generators \cite{djm95}
\begin{equation}
J^k = q^\dagger \frac{\sigma^k}{2} q, \qquad T^c = q^\dagger \frac{\lambda^c}{2} q, \qquad G^{kc} = q^\dagger \frac{\sigma^k}{2}\frac{\lambda^c}{2} q, \label{eq:su6gen}
\end{equation}
where $q^\dagger$ and $q$ are $SU(6)$ operators that create and annihilate states in the fundamental representation of $SU(6)$, and
$\sigma^k$ and $\lambda^c$ are the Pauli spin and Gell-Mann flavor matrices, respectively. The spin-flavor generators satisfy the commutation relations displayed in Table \ref{t:surel}.
\begingroup
\begin{table}
\caption{\label{t:surel}${SU}(2N_f)$ commutation relations.}
\bigskip
\label{t:su2fcomm}
\centerline{\vbox{ \tabskip=0pt \offinterlineskip
\halign{
\strut\quad $ # $\quad\hfil&\strut\quad $ # $\quad \hfil\cr
\multispan2\hfil $\left[J^i,T^a\right]=0,$ \hfil \cr
\noalign{\medskip}
\left[J^i,J^j\right]=i\epsilon^{ijk} J^k,
&\left[T^a,T^b\right]=i f^{abc} T^c,\cr
\noalign{\medskip}
\left[J^i,G^{ja}\right]=i\epsilon^{ijk} G^{ka},
&\left[T^a,G^{ib}\right]=i f^{abc} G^{ic},\cr
\noalign{\medskip}
\multispan2\hfil$\displaystyle [G^{ia},G^{jb}] = \frac{i}{4}\delta^{ij}
f^{abc} T^c + \frac{i}{2N_f} \delta^{ab} \epsilon^{ijk} J^k + \frac{i}{2} \epsilon^{ijk} d^{abc} G^{kc}.$ \hfill\cr
}}}
\end{table}
\endgroup

Explicitly, the $1/N_c$ expansion of the baryon mass operator $\mathcal{M}$ is written as \cite{djm95}
\begin{eqnarray}
\mathcal{M} = m_0 N_c \openone + \sum_{n=2,4}^{N_c-1} m_{n} \frac{1}{N_c^{n-1}} J^n, \label{eq:mop}
\end{eqnarray}
where $m_n$ are unknown coefficients. The first term on the right-hand side is the overall spin-independent mass of the baryon multiplet and is removed from the chiral Lagrangian by the heavy baryon field redefinition~\cite{jm255,jm259}. The other terms are spin-dependent and make up $\mathcal{M}_{\text{hyperfine}}$. For $N_c=3$ the hyperfine mass expansion reduces to a single operator
\begin{eqnarray}
\mathcal{M} _{\text{hyperfine}} = \frac{m_2}{N_c} J^2. \label{eq:smop}
\end{eqnarray}

In a similar fashion, $A^k$ is a spin-1 object, odd under time reversal, and a singlet under $SU(3)$. Its $1/N_c$ expansion is given by \cite{djm95}
\begin{equation}
A^k = \sum_{n=1,3}^{N_c} e_n \frac{1}{N_c^{n-1}} \mathcal{D}_n^k, \label{eq:asin}
\end{equation}
where $\mathcal{D}_1^k = J^k$ and $\mathcal{D}_{2m+1}^k = \{J^2,\mathcal{D}_{2m-1}^k\}$ for $m\geq 1$; $e_n$ are unknown coefficients which have expansions in powers of $1/N_c$ and are order unity at leading order in the $1/N_c$ expansion. At the physical value $N_c=3$, $A^k$ reduces to
\begin{equation}
A^k = e_1 J^k + e_3 \frac{1}{N_c^2} \{J^2,J^k\}.
\end{equation}

Similarly, $A^{kc}$ is also a spin-1 object and odd under time reversal, but an octet under $SU(3)$. Its $1/N_c$ expansion reads \cite{djm95}
\begin{equation}
A^{kc} = a_1 G^{kc} + \sum_{n=2,3}^{N_c} b_n \frac{1}{N_c^{n-1}} \mathcal{D}_n^{kc} + \sum_{n=3,5}^{N_c} c_n
\frac{1}{N_c^{n-1}} \mathcal{O}_n^{kc}, \label{eq:akcfull}
\end{equation}
where the unknown coefficients $a_1$, $b_n$, and $c_n$ share the same characteristics in the large-$N_c$ limit with $e_n$ discussed above. Up to 3-body operators, one has
\begin{eqnarray}
\mathcal{D}_2^{kc} & = & J^kT^c, \label{eq:d2kc} \\
\mathcal{D}_3^{kc} & = & \{J^k,\{J^r,G^{rc}\}\}, \label{eq:d3kc} \\
\mathcal{O}_3^{kc} & = & \{J^2,G^{kc}\} - \frac12 \{J^k,\{J^r,G^{rc}\}\}, \label{eq:o3kc}
\end{eqnarray}
while higher order operators are obtained recursively as $\mathcal{D}_n^{kc}=\{J^2,\mathcal{D}_{n-2}^{kc}\}$ and $\mathcal{O}_n^{kc}=\{J^2,\mathcal{O}_{n-2}^{kc}\}$ for $n\geq 4$. $\mathcal{D}_n^{kc}$ [$\mathcal{O}_n^{kc}$] are diagonal [purely off-diagonal] operators with nonzero matrix elements only between states with equal [different] spin. At the physical value $N_c = 3$ the series (\ref{eq:akcfull}) can be truncated as
\begin{equation}
A^{kc} = a_1 G^{kc} + b_2 \frac{1}{N_c} \mathcal{D}_2^{kc} + b_3 \frac{1}{N_c^2} \mathcal{D}_3^{kc} + c_3 \frac{1}{N_c^2} \mathcal{O}_3^{kc}. \label{eq:akc}
\end{equation}
{\it i.e.}, the $1/N_c$ expansion of $A^{kc}$ extends up to 3-body operators.

The matrix elements of the space components of $A^{kc}$ between $SU(6)$ symmetric baryon states yield the values of the axial vector couplings in the $SU(3)$ {\it symmetry limit}. The axial vector couplings are denoted by $g_1$ and $g$ for octet and decuplet baryons, respectively. Going beyond the symmetry limit requires the inclusion of explicit $SU(3)$ flavor symmetry breaking. These effects will be accounted for in the framework of heavy baryon chiral perturbation theory in the large-$N_c$ limit. This is presented in the following sections.

\section{\label{sec:renormalization}Renormalization of the baryon axial vector current}

The $1/N_c$ baryon chiral Lagrangian displayed in Eq.~(\ref{eq:ncch}) has been applied to the calculation of nonanalytic meson-loop corrections to various static properties of baryons: Mass splittings \cite{jen96}, axial vector current \cite{rfm00,rfm06,rfm12}, vector current \cite{fmg}, charge radii \cite{fmr}, $s$-wave decay amplitudes in nonleptonic decays \cite{rfm19}, to name but a few. In all those analyses, theoretical expressions are in very good agreement with both the expectations from the $1/N_c$ expansion and the experimental data.

The aim of this paper is to retrace the issue of the renormalization of the baryon axial vector current within the combined formalism in $1/N_c$ and chiral corrections, partially evaluated in Refs.~\cite{rfm06,rfm12}; therefore, several useful expressions already presented in those works will be freely borrowed here to introduce the guidelines of the calculation.

\begin{figure}[ht]
\scalebox{0.9}{\includegraphics{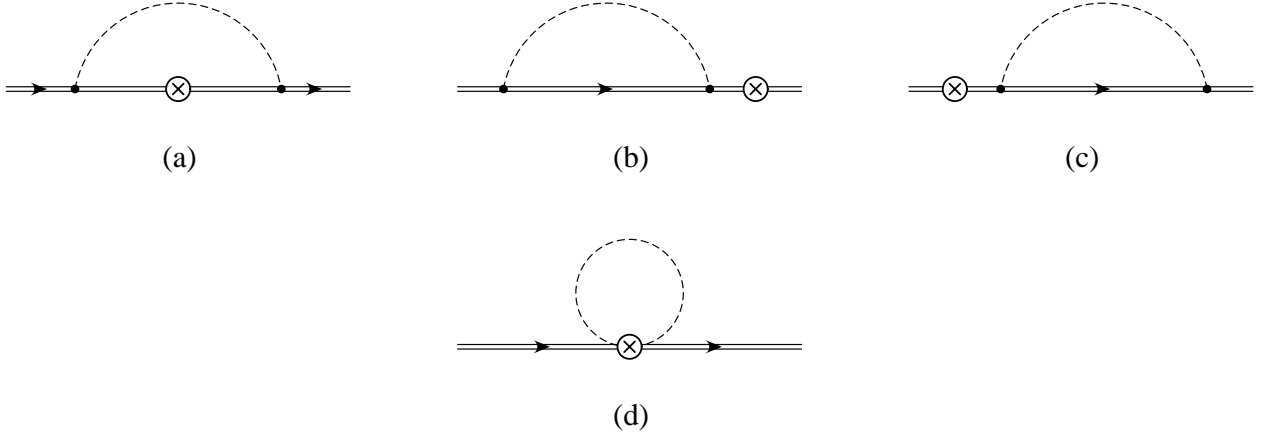}}
\caption{\label{fig:l1}One-loop corrections to the baryon axial vector current. A crossed circle represents an insertion of an axial vector current. Dashed lines and solid lines denote mesons and baryons, respectively.}
\end{figure}

The one-loop diagrams that enter the analysis are depicted in Fig.~\ref{fig:l1}.
These loop graphs have a calculable dependence on the ratio $\Delta/m$, where $m$ denotes the meson mass and $\Delta$ represents the decuplet-octet mass difference. The conditions $m \ll \Lambda_\chi$ and $\Delta \ll \Lambda_\chi$ should be met in order for the theory to be valid, where $\Lambda_\chi \sim 1$ GeV is the scale of chiral symmetry breaking; the ratio $\Delta/m$, on the other hand, can take any value. In the large-$N_c$ limit, $g_1\propto N_c$ and $f\propto\sqrt{N_c}$ \cite{djm95,rfm00}; the meson-baryon vertex is proportional to $g_1/f$, so it is of order $\mathcal{O}(\sqrt{N_c})$. The meson and baryon propagators are independent of $N_c$ and so are the loop integrals because in the $\overline{\mathrm{MS}}$ scheme they are given by the pole structure of the propagators \cite{rfm00}.

The contribution from Fig.~\ref{fig:l1}(a,b,c) to the renormalization of $A^{kc}$ can be written as \cite{rfm00,rfm06,rfm12}
\begin{eqnarray}
\delta A_{1a}^{kc} & = & \frac12 \left[A^{ja},\left[A^{jb},A^{kc}\right]\right] \Pi_{(1)}^{ab} - \frac12 \left\{ A^{ja}, \left[A^{kc},\left[\mathcal{M},A^{jb}\right] \right] \right\} \Pi_{(2)}^{ab} \nonumber \\
&  & \mbox{} + \frac16 \left(\left[A^{ja}, \left[\left[\mathcal{M}, \left[ \mathcal{M},A^{jb}\right]\right],A^{kc}\right] \right] - \frac12 \left[\left[\mathcal{M},A^{ja}\right], \left[\left[\mathcal{M},A^{jb}\right],A^{kc}\right]\right]\right) \Pi_{(3)}^{ab} + \ldots \, . \label{eq:dakc}
\end{eqnarray}
Here $\Pi_{(n)}^{ab}$ is a symmetric tensor which decomposes into flavor singlet, flavor $\mathbf{8}$, and flavor $\mathbf{27}$ representations as \cite{jen96}
\begin{eqnarray}
\Pi_{(n)}^{ab} = F_\mathbf{1}^{(n)} \delta^{ab} + F_\mathbf{8}^{(n)} d^{ab8} + F_\mathbf{27}^{(n)} \left[ \delta^{a8} \delta^{b8} - \frac18 \delta^{ab} - \frac35 d^{ab8} d^{888}\right], \label{eq:pisym}
\end{eqnarray}
where
\begin{subequations}
\begin{eqnarray}
F_\mathbf{1}^{(n)} & = & \frac18 \left[3F^{(n)}(m_\pi,0,\mu) + 4F^{(n)}(m_K,0,\mu) + F^{(n)}(m_\eta,0,\mu) \right], \label{eq:F1} \\
F_\mathbf{8}^{(n)} & = & \frac{2\sqrt 3}{5} \left[\frac32 F^{(n)}(m_\pi,0,\mu) - F^{(n)}(m_K,0,\mu) - \frac12 F^{(n)}(m_\eta,0,\mu) \right], \label{eq:F8} \\
F_\mathbf{27}^{(n)} & = & \frac13 F^{(n)}(m_\pi,0,\mu) - \frac43 F^{(n)}(m_K,0,\mu) + F^{(n)}(m_\eta,0,\mu), \label{eq:F27}
\end{eqnarray}
\end{subequations}
depend linearly on the functions
\begin{equation}
F^{(n)}(m,\Delta,\mu) \equiv \frac{\partial^n F(m,\Delta,\mu)}{\partial \Delta^n},
\end{equation}
which represent loop integrals with the full dependence on the ratio $\Delta/m$ and $\mu$ is the scale parameter of dimensional regularization. The explicit form of the function $F(m,\Delta,\mu)$ is \cite{fmg}\footnote{The function $F(m,\Delta,\mu)$ was first presented in Ref.~\cite{jen96}; however, it contained a wrong sign in one of its terms, which was fixed in Ref.~\cite{fmg}. The correct expression is provided here for the sake of completeness.}
\begin{eqnarray}
24\pi^2 f^2 F(m,\Delta,\mu) & = & -\Delta\left[\Delta^2-\frac32 m^2\right] \lambda_\epsilon + \Delta \left[\Delta^2-\frac32 m^2\right] \ln{\frac{m^2}{\mu^2}} - \frac83 \Delta^3 + \frac72 \Delta m^2 \nonumber \\
&  & \mbox{} + \left\{ \begin{array}{ll}
\displaystyle 2(m^2-\Delta^2)^{3/2} \left[\frac{\pi}{2} - \tan^{-1} \left[\frac{\Delta}{\sqrt{m^2-\Delta ^2}}\right] \right], & |\Delta|< m \\[3mm]
\displaystyle - (\Delta^2-m^2)^{3/2} \left[-2i \pi +\ln \left[\frac{\Delta - \sqrt{\Delta^2-m^2}}{\Delta + \sqrt{\Delta^2-m^2}} \right] \right], & |\Delta| > m. \end{array} \right. \label{eq:ib}
\end{eqnarray}
The $n$th derivatives of $F(m,\Delta,\mu)$ are readily obtained from Eq.~(\ref{eq:ib}).

The renormalization of the axial vector current has already been performed within the combined formalism up to a certain approximation. In Ref.~\cite{rfm06}, corrections up to relative order $\mathcal{O}(1/N_c^3)$ were included, assuming degenerate intermediate baryons in the loops. In Ref.~\cite{rfm12}, the analysis was performed on the same footing as the one before, but this time the degeneracy assumption was lifted.

The aim of this paper is to evaluate the complete contribution $\delta A_{1a}^{kc}$, (\ref{eq:dakc}), to {\it all} relative orders in $1/N_c$ in view of the $1/N_c$ expansion of $A^{kc}$. Due to the considerable amount of group theory involved, only the degeneracy limit $\Delta/m\to 0$ will be considered in an initial stage of the calculation. The reason for doing this is twofold. First, the full analysis will allow one to verify once and for all the agreement between the corrections to the baryon axial vector current computed within the combined formalism and those computed within conventional HBChPT (i.e., the effective theory with no $1/N_c$ expansion) of Refs.~\cite{jm255,jm259}. Similar comparisons have been successfully carried out for baryon vector currents \cite{fmg} and $s$-wave decay amplitudes of baryon nonleptonic decays \cite{rfm19}. Second, with the available experimental information \cite{part}, a detailed numerical analysis can be performed to extract somewhat more reliable values of the free parameters of the theory, namely, the operator coefficients $a_1$, $b_2$, $b_3$, and $c_3$ introduced in Eq.~(\ref{eq:akc}) than those that could have been obtained otherwise \cite{rfm06,rfm12}. These operator coefficients are related to the four conventional $SU(3)$ baryon axial couplings $D$, $F$, $\mathcal{C}$, and $\mathcal{H}$ introduced in HBChPT \cite{jm255,jm259}, at $N_c=3$, by
\begin{subequations}
\label{eq:rel1}
\begin{eqnarray}
&  & D = \frac12 a_1 + \frac16 b_3, \\
&  & F = \frac13 a_1 + \frac16 b_2 + \frac19 b_3, \\
&  & \mathcal{C} = - a_1 - \frac12 c_3, \\
&  & \mathcal{H} = -\frac32 a_1 - \frac32 b_2 - \frac52 b_3.
\end{eqnarray}
\end{subequations}

The only term from $\delta A_{1a}^{kc}$ Eq.~(\ref{eq:dakc}) to be considered here is
\begin{equation}
\frac12 \left[A^{ia},\left[A^{ib},A^{kc}\right]\right] \Pi_{(1)}^{ab}, \label{eq:fsm}
\end{equation}
which has been partially analyzed in Ref.~\cite{rfm06} so it can serve as a reference point. In order to obtain a better idea about the magnitude of the forthcoming calculation, the $1/N_c$ power-counting scheme introduced in previous works \cite{rfm00,rfm06} can come in handy. This counting scheme indicates that, for baryons with spins of order one,
\begin{equation}
T^a \sim N_c, \qquad G^{ia} \sim N_c, \qquad J^i \sim 1, \label{eq:crules}
\end{equation}
which is equivalent to claim that factors of $J^i/N_c$ are $1/N_c$ suppressed relative to factors of $T^a/N_c$ and $G^{ia}/N_c$. This counting scheme is thus suitable for the lowest-lying baryon states, namely, spin-$1/2$ octet and spin-$3/2$ decuplet baryon states, which together constitute the $\mathbf{56}$ dimensional representation of $SU(6)$.

From relation (\ref{eq:crules}) it follows that $A^{kc}$ is order $\mathcal{O}(N_c)$. A simplistic counting suggests that the double commutator in (\ref{eq:fsm}) would be of order $\mathcal{O}(N_c^3)$. Previous work, however, has found that there are large-$N_c$ cancellations between the loop graphs of Figs.~\ref{fig:l1}(a,b,c), provided the sum over all baryon states in a complete multiplet of the large-$N_c$ $SU(6)$ spin-flavor symmetry---i.e.\ over both the octet and decuplet baryon states---is performed and the axial coupling ratios given by the large-$N_c$ spin-flavor symmetry are used \cite{djm94,rfm00,rfm06}. Accordingly, this double commutator should be of order $\mathcal{O}(N_c)$ in such a way that the overall contribution (\ref{eq:fsm}) would be of order $\mathcal{O}(1/N_c)$ times the tree-level value, considering that $f$ is of order $\mathcal{O}(\sqrt{N_c})$. The explicit computation presented in Ref.~\cite{rfm06} shows that this is indeed the case, at least up to the terms retained in that analysis.

The aim is now to include all relative terms in $1/N_c$ contained in expression (\ref{eq:fsm}) up to the highest order terms so
double commutators containing three 3-body operators need be considered; an example of this kind of operators is $[\mathcal{O}_3^{ia},[\mathcal{O}_3^{ib},\mathcal{O}_3^{kc}]]$. Each operator $\mathcal{O}_3^{ia}$ comes along with an explicit suppression factor of $1/N_c^2$; the resultant double commutator should contain up to 7-body operators so that the overall contribution of this structure is again of order $\mathcal{O}(N_c)$, which is consistent with expectations. Explicit calculations up to this order will be presented here.

Following Ref.~\cite{rfm12}, the calculation can be organized by considering separately the three $SU(3)$ flavor representations contained in the tensor structure $\Pi_{(n)}^{ab}$ in Eq.~(\ref{eq:fsm}), namely,
\begin{equation}
\delta A_{1a}^{kc} = \delta A_{\mathbf{1}}^{kc} + \delta A_{\mathbf{8}}^{kc} + \delta A_{\mathbf{27}}^{kc}, \label{eq:dasplit}
\end{equation}
where the subscript $\mathbf{rep}$ attached to each $\delta A_{\mathbf{rep}}^{kc}$ term indicates the flavor representation it comes from.

The mathematical reduction of the operators contained in $\delta A_{\mathbf{rep}}^{kc}$ requires a rather involved, long but otherwise standard calculation. Although the resulting expressions are rather lengthy and not particularly illuminating, they are nevertheless listed in Appendices \ref{sec:r1}, \ref{sec:r8}, and \ref{sec:r27} for all three flavor representations of interest here.

Therefore, the different $\delta A_{\mathbf{rep}}^{kc}$ operators can be cast into
\begin{equation}
\delta A_{\mathbf{1}}^{kc} = \sum_{i=1}^{10} s_i S_i^{kc}, \label{eq:das}
\end{equation}
\begin{equation}
\delta A_{\mathbf{8}}^{kc} = \sum_{i=1}^{51} o_i O_i^{kc}, \label{eq:dao}
\end{equation}
and
\begin{equation}
\delta A_{\mathbf{27}}^{kc} = \sum_{i=1}^{165} t_i T_i^{kc}. \label{eq:dat}
\end{equation}
Each term in $\delta A_{\mathbf{rep}}^{kc}$ is given as the product of a coefficient times an operator in a basis which belongs to the flavor representation $\mathbf{rep}$. For the singlet case, for instance, the $s_i$ coefficients, at the physical values $N_f=N_c=3$, read
\begin{equation}
s_{1} = \left[ \frac{23}{24} a_1^3 - \frac23 a_1^2b_2 - \frac{29}{18} a_1^2b_3 - \frac43 a_1^2c_3 - \frac14 a_1b_2^2 - \frac43 a_1b_2b_3 - \frac{11}{9} a_1b_3^2 - \frac14 a_1c_3^2 \right] F_{\mathbf{1}}^{(1)},
\end{equation}
\begin{equation}
s_{2} = \left[ \frac{101}{72} a_1^2b_2 - a_1^2b_3 - \frac16 a_1^2c_3 + \frac49 a_1b_2^2 + \frac{29}{54} a_1b_2b_3 - \frac{1}{12} a_1b_2c_3 + \frac{1}{36} b_2^3 + \frac29 a_1b_3^2 - \frac89 a_1b_3c_3 - \frac{1}{108} b_2c_3^2 - \frac{8}{81} b_3c_3^2 \right] F_{\mathbf{1}}^{(1)},
\end{equation}
\begin{eqnarray}
s_{3} & = & \left[ \frac{17}{24} a_1^2b_3 + \frac19 a_1^2c_3 + \frac{11}{72} a_1b_2^2 + \frac{17}{27} a_1b_2b_3 - \frac12 a_1b_2c_3 + \frac{31}{54} a_1b_3^2 - \frac{13}{18} a_1b_3c_3 - \frac{1}{72} a_1c_3^2 - \frac{1}{36} b_2^2b_3 - \frac{2}{27} b_2b_3^2 \right. \nonumber \\
&  & \mbox{} \left. - \frac{2}{27} b_2c_3^2 - \frac{11}{243} b_3^3 - \frac{37}{324} b_3c_3^2 \right] F_{\mathbf{1}}^{(1)},
\end{eqnarray}
\begin{eqnarray}
s_{4} & = & \left[ \frac{7}{54} a_1^2b_3 + \frac{167}{216} a_1^2c_3 + \frac{1}{12} a_1b_2^2 + \frac{10}{27} a_1b_2b_3 - \frac{2}{27} a_1b_2c_3 + \frac{125}{486} a_1b_3^2 - \frac{29}{162} a_1b_3c_3 - \frac{25}{72} a_1c_3^2 - \frac{1}{36} b_2^2c_3 - \frac{4}{27} b_2b_3c_3 \right. \nonumber \\
&  & \mbox{} \left. - \frac{11}{81} b_3^2c_3 - \frac{1}{36}c_3^3 \right] F_{\mathbf{1}}^{(1)},
\end{eqnarray}
\begin{eqnarray}
s_{5} & = & \left[ \frac{11}{162} a_1b_2b_3 + \frac{19}{54} a_1b_2c_3 + \frac{5}{108} b_2^3 - \frac19 a_1b_3^2 - \frac{5}{27} a_1b_3c_3 - \frac{1}{36} a_1c_3^2 + \frac{8}{81} b_2^2b_3 + \frac{29}{486} b_2b_3^2 + \frac{29}{648} b_2c_3^2 + \frac{4}{243} b_3^3 \right. \nonumber \\
&  & \mbox{} \left. - \frac{26}{243} b_3c_3^2 \right] F_{\mathbf{1}}^{(1)},
\end{eqnarray}
\begin{equation}
s_{6} = \left[ \frac{11}{324} a_1b_3^2 + \frac{13}{81} a_1b_3c_3 + \frac{1}{108} a_1c_3^2 + \frac{11}{324} b_2^2b_3 + \frac{17}{243} b_2b_3^2 - \frac{11}{324} b_2c_3^2 + \frac{31}{729} b_3^3 - \frac{5}{243} b_3c_3^2 \right] F_{\mathbf{1}}^{(1)},
\end{equation}
\begin{equation}
s_{7} = \left[ \frac{13}{486} a_1b_3^2 + \frac{7}{486} a_1b_3c_3 + \frac{49}{324} a_1c_3^2 + \frac{1}{108} b_2^2c_3 + \frac{10}{243} b_2b_3c_3 + \frac{125}{4374} b_3^2c_3 - \frac{43}{1944}c_3^3 \right] F_{\mathbf{1}}^{(1)},
\end{equation}
\begin{equation}
s_{8} = \left[ \frac{11}{1458} b_2b_3^2 + \frac{23}{972} b_2c_3^2 - \frac{2}{243} b_3^3 - \frac{7}{486} b_3c_3^2 \right] F_{\mathbf{1}}^{(1)},
\end{equation}
\begin{equation}
s_{9} = \left[ \frac{11}{4374} b_3^3 + \frac{8}{729} b_3c_3^2 \right] F_{\mathbf{1}}^{(1)},
\end{equation}
\begin{equation}
s_{10} = \left[ \frac{13}{4374} b_3^2c_3 + \frac{25}{2916}c_3^3 \right] F_{\mathbf{1}}^{(1)},
\end{equation}
where the basis of singlet operators is constituted by
\begin{eqnarray}
\begin{array}{lllll}
S_{1}^{kc} = G^{kc}, & S_{2}^{kc} = \mathcal{D}_2^{kc}, & S_{3}^{kc} = \mathcal{D}_3^{kc}, & S_{4}^{kc} = \mathcal{O}_3^{kc}, & S_{5}^{kc} = \mathcal{D}_4^{kc}, \nonumber \\
S_{6}^{kc} = \mathcal{D}_5^{kc}, & S_{7}^{kc} = \mathcal{O}_5^{kc}, & S_{8}^{kc} = \mathcal{D}_6^{kc}, & S_{9}^{kc} = \mathcal{D}_7^{kc}, & S_{10}^{kc} = \mathcal{O}_7^{kc}.
\end{array}
\end{eqnarray}

The corresponding coefficients $o_j$ and $t_k$ that come along with operators in the $\mathbf{8}$ and $\mathbf{27}$ flavor representations can be obtained in a similar manner and will not be listed here. The remaining operator bases, at any rate, can be found in Appendix \ref{sec:bases} for completeness.

As for the one-loop graph of Fig.~\ref{fig:l1}(d), it has been analyzed in Ref.~\cite{rfm06} in detail and will not be repeated here. Its contribution to the renormalization of the axial vector current will nevertheless be taken into account in the numerical analysis in the next sections.

\subsection{One-loop corrections from Fig.~\ref{fig:l1}(a,b,c)}

The matrix elements of $\delta A_{1a}^{kc}$ between $SU(6)$ symmetric baryon states, $|B_1\rangle$ and $|B_2\rangle$ for definiteness, yield the contribution of loop graphs from Fig.~\ref{fig:l1}(a,b,c) to the renormalization of the axial vector coupling $g_1$, namely,
\begin{equation}
\delta g_1^{B_1B_2} = \langle B_2|\delta A_{1a}^{kc}|B_1\rangle.
\end{equation}

Using the matrix elements listed in Tables \ref{t:mtx1}, \ref{t:mtx8}, and \ref{t:mtx27} allows one to organize $\delta g_1$ as
\begin{equation}
\delta g_1^{B_1B_2} = \alpha_{\mathbf{1}}^{B_1B_2} F_{\mathbf{1}}^{(1)} + \alpha_{\mathbf{8}}^{B_1B_2} F_{\mathbf{8}}^{(1)} + \alpha_{\mathbf{27}}^{B_1B_2} F_{\mathbf{27}}^{(1)}. \label{eq:galn}
\end{equation}

The coefficients $\alpha_{\mathbf{rep}}^{B_1B_2}$ are expressed in terms of the operator coefficients $a_1$, $b_2$, $b_3$, and $c_3$. For some kinematically allowed processes involving octet baryons, they read
\begin{eqnarray}
\alpha_{\mathbf{1}}^{np} & = & \frac{115}{144} a_1^3 + \frac{7}{48} a_1^2b_2 - \frac{31}{432} a_1^2b_3 - \frac{11}{12} a_1^2c_3 + \frac{19}{48} a_1b_2^2 + \frac{169}{216} a_1b_2b_3 - \frac{37}{36} a_1b_2c_3 + \frac{7}{144} b_2^3 + \frac{247}{432} a_1b_3^2 - \frac{193}{108} a_1b_3c_3 \nonumber \\
&  & \mbox{} - \frac{11}{48} a_1c_3^2 + \frac{19}{144} b_2^2b_3 + \frac{169}{1296} b_2b_3^2 - \frac{37}{144} b_2c_3^2 + \frac{247}{3888} b_3^3 - \frac{193}{432} b_3c_3^2,
\end{eqnarray}
\begin{eqnarray}
\alpha_{\mathbf{8}}^{np} & = & \frac{55}{288 \sqrt{3}} a_1^3 - \frac{127}{288 \sqrt{3}} a_1^2b_2 - \frac{419}{864 \sqrt{3}} a_1^2b_3 - \frac{47}{72 \sqrt{3}} a_1^2c_3 - \frac{11}{288 \sqrt{3}} a_1b_2^2 + \frac{109}{432 \sqrt{3}} a_1b_2b_3 - \frac{59}{72 \sqrt{3}} a_1b_2c_3 \nonumber \\
&  & \mbox{} - \frac{5}{288 \sqrt{3}} b_2^3 + \frac{3 \sqrt{3}}{32} a_1b_3^2 - \frac{287}{216 \sqrt{3}} a_1b_3c_3 - \frac{47}{288 \sqrt{3}} a_1c_3^2 - \frac{11}{864 \sqrt{3}} b_2^2b_3 + \frac{109}{2592 \sqrt{3}} b_2b_3^2 - \frac{59}{288 \sqrt{3}} b_2c_3^2 \nonumber \\
&  & \mbox{} + \frac{1}{32 \sqrt{3}} b_3^3 - \frac{287}{864 \sqrt{3}} b_3c_3^2, \nonumber \\
\end{eqnarray}
\begin{eqnarray}
\alpha_{\mathbf{27}}^{np} & = & \frac{1}{128} a_1^3 + \frac{89}{1920} a_1^2b_2 - \frac{107}{5760} a_1^2b_3 - \frac{23}{1440} a_1^2c_3 + \frac{77}{1920} a_1b_2^2 + \frac{157}{2880} a_1b_2b_3 - \frac{17}{480} a_1b_2c_3 + \frac{5}{1152} b_2^3 + \frac{137}{17280} a_1b_3^2 \nonumber \\
&  & \mbox{} - \frac{61}{1440} a_1b_3c_3 - \frac{23}{5760} a_1c_3^2 + \frac{77}{5760} b_2^2b_3 + \frac{157}{17280} b_2b_3^2 - \frac{17}{1920} b_2c_3^2 + \frac{137}{155520} b_3^3 - \frac{61}{5760} b_3c_3^2,
\end{eqnarray}
\begin{eqnarray}
\alpha_{\mathbf{1}}^{\Sigma^\pm\Lambda} & = & \frac{23}{24 \sqrt{6}} a_1^3 - \frac13 \sqrt{\frac23} a_1^2b_2 + \frac{37}{72 \sqrt{6}} a_1^2b_3 - \frac{1}{\sqrt{6}} a_1^2c_3 + \frac{5}{24 \sqrt{6}} a_1b_2^2 + \frac{5}{9 \sqrt{6}} a_1b_2b_3 - \frac12 \sqrt{\frac32} a_1b_2c_3 + \frac{47}{72 \sqrt{6}} a_1b_3^2 \nonumber \\
&  & \mbox{} - \frac{13}{9 \sqrt{6}} a_1b_3c_3 - \frac{1}{4 \sqrt{6}} a_1c_3^2 + \frac{5}{72 \sqrt{6}} b_2^2b_3 + \frac{5}{54 \sqrt{6}} b_2b_3^2 - \frac18 \sqrt{\frac32} b_2c_3^2 + \frac{47}{648 \sqrt{6}} b_3^3 - \frac{13}{36 \sqrt{6}} b_3c_3^2,
\end{eqnarray}
\begin{eqnarray}
\alpha_{\mathbf{8}}^{\Sigma^\pm\Lambda} & = & \frac{11}{144 \sqrt{2}} a_1^3 - \frac{\sqrt{2}}{9} a_1^2b_2 - \frac{71}{432 \sqrt{2}} a_1^2b_3 + \frac{1}{12 \sqrt{2}} a_1^2c_3 - \frac{1}{48 \sqrt{2}} a_1b_2^2 - \frac{1}{18 \sqrt{2}} a_1b_2b_3 - \frac{5}{36 \sqrt{2}} a_1b_2c_3 - \frac{1}{432 \sqrt{2}} a_1b_3^2 \nonumber \\
&  & \mbox{} - \frac{17}{108 \sqrt{2}} a_1b_3c_3 + \frac{1}{48 \sqrt{2}} a_1c_3^2 - \frac{1}{144 \sqrt{2}} b_2^2b_3 - \frac{1}{108 \sqrt{2}} b_2b_3^2 - \frac{5}{144 \sqrt{2}} b_2c_3^2 - \frac{1}{3888 \sqrt{2}} b_3^3 - \frac{17}{432 \sqrt{2}} b_3c_3^2,
\end{eqnarray}
\begin{eqnarray}
\alpha_{\mathbf{27}}^{\Sigma^\pm\Lambda} & = & \frac{1}{320} \sqrt{\frac32} a_1^3 - \frac{1}{20 \sqrt{6}} a_1^2b_2 - \frac{23}{960 \sqrt{6}} a_1^2b_3 + \frac{11}{120 \sqrt{6}} a_1^2c_3 - \frac{37}{960 \sqrt{6}} a_1b_2^2 - \frac{37}{360 \sqrt{6}} a_1b_2b_3 + \frac{5}{48 \sqrt{6}} a_1b_2c_3 \nonumber \\
&  & \mbox{} - \frac{79}{2880 \sqrt{6}} a_1b_3^2 + \frac{7}{120 \sqrt{6}} a_1b_3c_3 + \frac{11}{480 \sqrt{6}} a_1c_3^2 - \frac{37}{2880 \sqrt{6}} b_2^2b_3 - \frac{37}{2160 \sqrt{6}} b_2b_3^2 + \frac{5}{192 \sqrt{6}} b_2c_3^2 - \frac{79}{25920 \sqrt{6}} b_3^3 \nonumber \\
&  & \mbox{} + \frac{7}{480 \sqrt{6}} b_3c_3^2,
\end{eqnarray}
\begin{eqnarray}
\alpha_{\mathbf{1}}^{\Lambda p} & = & - \frac{23}{16 \sqrt{6}} a_1^3 - \frac{53}{48 \sqrt{6}} a_1^2b_2 + \frac{35}{48 \sqrt{6}} a_1^2b_3 + \frac{7}{4 \sqrt{6}} a_1^2c_3 - \frac{47}{48 \sqrt{6}} a_1b_2^2 - \frac{43}{24 \sqrt{6}} a_1b_2b_3 + \frac{19}{12 \sqrt{6}} a_1b_2c_3 - \frac{7}{48 \sqrt{6}} b_2^3 \nonumber \\
&  & \mbox{} - \frac{17}{16 \sqrt{6}} a_1b_3^2 + \frac{47}{12 \sqrt{6}} a_1b_3c_3 + \frac{7}{16 \sqrt{6}} a_1c_3^2 - \frac{47}{144 \sqrt{6}} b_2^2b_3 - \frac{43}{144 \sqrt{6}} b_2b_3^2 + \frac{19}{48 \sqrt{6}} b_2c_3^2 - \frac{17}{144 \sqrt{6}} b_3^3 + \frac{47}{48 \sqrt{6}} b_3c_3^2, \nonumber \\
\end{eqnarray}
\begin{eqnarray}
\alpha_{\mathbf{8}}^{\Lambda p} & = & \frac{11}{192 \sqrt{2}} a_1^3 + \frac{65}{576 \sqrt{2}} a_1^2b_2 + \frac{233}{576 \sqrt{2}} a_1^2b_3 + \frac{5}{16 \sqrt{2}} a_1^2c_3 + \frac{35}{576 \sqrt{2}} a_1b_2^2 - \frac{59}{864 \sqrt{2}} a_1b_2b_3 + \frac{31}{144 \sqrt{2}} a_1b_2c_3  \nonumber \\
&  & \mbox{} + \frac{11}{576 \sqrt{2}} b_2^3 - \frac{49}{576 \sqrt{2}} a_1b_3^2 + \frac{95}{144 \sqrt{2}} a_1b_3c_3 + \frac{5}{64 \sqrt{2}} a_1c_3^2 + \frac{35}{1728 \sqrt{2}} b_2^2b_3 - \frac{59}{5184 \sqrt{2}} b_2b_3^2 + \frac{31}{576 \sqrt{2}} b_2c_3^2 \nonumber \\
&  & \mbox{} - \frac{49}{5184 \sqrt{2}} b_3^3 + \frac{95}{576 \sqrt{2}} b_3c_3^2,
\end{eqnarray}
\begin{eqnarray}
\alpha_{\mathbf{27}}^{\Lambda p} & = & \frac{9}{640} \sqrt{\frac32} a_1^3 + \frac{5}{128} \sqrt{\frac32} a_1^2b_2 - \frac{13}{640 \sqrt{6}} a_1^2b_3 - \frac{1}{32 \sqrt{6}} a_1^2c_3 + \frac{13}{128 \sqrt{6}} a_1b_2^2 + \frac{119}{960 \sqrt{6}} a_1b_2b_3 - \frac{11}{160 \sqrt{6}} a_1b_2c_3 \nonumber \\
&  & \mbox{} + \frac{3}{640} \sqrt{\frac32} b_2^3 + \frac{47}{1920 \sqrt{6}} a_1b_3^2 - \frac{1}{32} \sqrt{\frac32} a_1b_3c_3 - \frac{1}{128 \sqrt{6}} a_1c_3^2 + \frac{13}{384 \sqrt{6}} b_2^2b_3 + \frac{119}{5760 \sqrt{6}} b_2b_3^2 - \frac{11}{640 \sqrt{6}} b_2c_3^2 \nonumber \\
&  & \mbox{} + \frac{47}{17280 \sqrt{6}} b_3^3 - \frac{1}{128} \sqrt{\frac32} b_3c_3^2,
\end{eqnarray}
\begin{eqnarray}
\alpha_{\mathbf{1}}^{\Sigma^- n} & = & \frac{23}{144} a_1^3 - \frac{13}{16} a_1^2b_2 + \frac{253}{432} a_1^2b_3 - \frac{1}{12} a_1^2c_3 - \frac{3}{16} a_1b_2^2 - \frac{49}{216} a_1b_2b_3 - \frac{17}{36} a_1b_2c_3 - \frac{7}{144} b_2^3 + \frac{35}{432} a_1b_3^2 \nonumber \\
&  & \mbox{} + \frac{37}{108} a_1b_3c_3 - \frac{1}{48} a_1c_3^2 - \frac{1}{16} b_2^2b_3 - \frac{49}{1296} b_2b_3^2 - \frac{17}{144} b_2c_3^2 + \frac{35}{3888} b_3^3 + \frac{37}{432} b_3c_3^2,
\end{eqnarray}
\begin{eqnarray}
\alpha_{\mathbf{8}}^{\Sigma^- n} & = & - \frac{11}{576 \sqrt{3}} a_1^3 - \frac{47}{576 \sqrt{3}} a_1^2b_2 + \frac{407}{1728 \sqrt{3}} a_1^2b_3 - \frac{5}{48 \sqrt{3}} a_1^2c_3 - \frac{49}{576 \sqrt{3}} a_1b_2^2 - \frac{43}{864 \sqrt{3}} a_1b_2b_3 - \frac{1}{144 \sqrt{3}} a_1b_2c_3 \nonumber \\
&  & \mbox{} - \frac{5}{576 \sqrt{3}} b_2^3 + \frac{49}{1728 \sqrt{3}} a_1b_3^2 + \frac{65}{432 \sqrt{3}} a_1b_3c_3 - \frac{5}{192 \sqrt{3}} a_1c_3^2 - \frac{49}{1728 \sqrt{3}} b_2^2b_3 - \frac{43}{5184 \sqrt{3}} b_2b_3^2 - \frac{1}{576 \sqrt{3}} b_2c_3^2 \nonumber \\
&  & \mbox{} + \frac{49}{15552 \sqrt{3}} b_3^3 + \frac{65}{1728 \sqrt{3}} b_3c_3^2,
\end{eqnarray}
\begin{eqnarray}
\alpha_{\mathbf{27}}^{\Sigma^- n} & = & - \frac{3}{640} a_1^3 - \frac{53}{1920} a_1^2b_2 + \frac{13}{5760} a_1^2b_3 - \frac{7}{288} a_1^2c_3 - \frac{11}{1920} a_1b_2^2 + \frac{23}{2880} a_1b_2b_3 - \frac{19}{480} a_1b_2c_3 - \frac{1}{1152} b_2^3 \nonumber \\
&  & \mbox{} + \frac{113}{17280} a_1b_3^2 - \frac{5}{288} a_1b_3c_3 - \frac{7}{1152} a_1c_3^2 - \frac{11}{5760} b_2^2b_3 + \frac{23}{17280} b_2b_3^2 - \frac{19}{1920} b_2c_3^2 + \frac{113}{155520} b_3^3 - \frac{5}{1152} b_3c_3^2, \nonumber \\
\end{eqnarray}
\begin{eqnarray}
\alpha_{\mathbf{1}}^{\Xi^- \Lambda} & = & \frac{23}{48 \sqrt{6}} a_1^3 + \frac{85}{48 \sqrt{6}} a_1^2b_2 - \frac{179}{144 \sqrt{6}} a_1^2b_3 - \frac14 \sqrt{\frac32} a_1^2c_3 + \frac{37}{48 \sqrt{6}} a_1b_2^2 + \frac{89}{72 \sqrt{6}} a_1b_2b_3 - \frac{1}{12 \sqrt{6}} a_1b_2c_3 + \frac{7}{48 \sqrt{6}} b_2^3 \nonumber \\
&  & \mbox{} + \frac{59}{144 \sqrt{6}} a_1b_3^2 - \frac{89}{36 \sqrt{6}} a_1b_3c_3 - \frac{1}{16} \sqrt{\frac32} a_1c_3^2 + \frac{37}{144 \sqrt{6}} b_2^2b_3 + \frac{89}{432 \sqrt{6}} b_2b_3^2 - \frac{1}{48 \sqrt{6}} b_2c_3^2 + \frac{59}{1296 \sqrt{6}} b_3^3 \nonumber \\
&  & \mbox{} - \frac{89}{144 \sqrt{6}} b_3c_3^2,
\end{eqnarray}
\begin{eqnarray}
\alpha_{\mathbf{8}}^{\Xi^- \Lambda} & = & - \frac{11}{576 \sqrt{2}} a_1^3 - \frac{241}{576 \sqrt{2}} a_1^2b_2 - \frac{49}{192 \sqrt{2}} a_1^2b_3 + \frac{11}{144 \sqrt{2}} a_1^2c_3 - \frac{97}{576 \sqrt{2}} a_1b_2^2 - \frac{7}{32 \sqrt{2}} a_1b_2b_3 - \frac{13}{144 \sqrt{2}} a_1b_2c_3 \nonumber \\
&  & \mbox{} - \frac{11}{576 \sqrt{2}} b_2^3 - \frac{55}{1728 \sqrt{2}} a_1b_3^2 - \frac{23}{144 \sqrt{2}} a_1b_3c_3 + \frac{11}{576 \sqrt{2}} a_1c_3^2 - \frac{97}{1728 \sqrt{2}} b_2^2b_3 - \frac{7}{192 \sqrt{2}} b_2b_3^2 - \frac{13}{576 \sqrt{2}} b_2c_3^2 \nonumber \\
&  & \mbox{} - \frac{55}{15552 \sqrt{2}} b_3^3 - \frac{23}{576 \sqrt{2}} b_3c_3^2,
\end{eqnarray}
\begin{eqnarray}
\alpha_{\mathbf{27}}^{\Xi^- \Lambda} & = & - \frac{3}{640} \sqrt{\frac32} a_1^3 + \frac{7}{640} \sqrt{\frac32} a_1^2b_2 + \frac{127}{640 \sqrt{6}} a_1^2b_3 - \frac{11}{160 \sqrt{6}} a_1^2c_3 - \frac{43}{640 \sqrt{6}} a_1b_2^2 - \frac{31}{960 \sqrt{6}} a_1b_2b_3 + \frac{13}{160 \sqrt{6}} a_1b_2c_3 \nonumber \\
&  & \mbox{} - \frac{3}{640} \sqrt{\frac32} b_2^3 + \frac{7}{384 \sqrt{6}} a_1b_3^2 + \frac{23}{160 \sqrt{6}} a_1b_3c_3 - \frac{11}{640 \sqrt{6}} a_1c_3^2 - \frac{43}{1920 \sqrt{6}} b_2^2b_3 - \frac{31}{5760 \sqrt{6}} b_2b_3^2 + \frac{13}{640 \sqrt{6}} b_2c_3^2 \nonumber \\
&  & \mbox{} + \frac{7}{3456 \sqrt{6}} b_3^3 + \frac{23}{640 \sqrt{6}} b_3c_3^2,
\end{eqnarray}
\begin{eqnarray}
\alpha_{\mathbf{1}}^{\Xi^- \Sigma^0} & = & \frac{115}{144 \sqrt{2}} a_1^3 + \frac{7}{48 \sqrt{2}} a_1^2b_2 - \frac{31}{432 \sqrt{2}} a_1^2b_3 - \frac{11}{12 \sqrt{2}} a_1^2c_3 + \frac{19}{48 \sqrt{2}} a_1b_2^2 + \frac{169}{216 \sqrt{2}} a_1b_2b_3 - \frac{37}{36 \sqrt{2}} a_1b_2c_3 \nonumber \\
&  & \mbox{} + \frac{7}{144 \sqrt{2}} b_2^3 + \frac{247}{432 \sqrt{2}} a_1b_3^2 - \frac{193}{108 \sqrt{2}} a_1b_3c_3 - \frac{11}{48 \sqrt{2}} a_1c_3^2 + \frac{19}{144 \sqrt{2}} b_2^2b_3 + \frac{169}{1296 \sqrt{2}} b_2b_3^2 - \frac{37}{144 \sqrt{2}} b_2c_3^2 \nonumber \\
&  & \mbox{} + \frac{247}{3888 \sqrt{2}} b_3^3 - \frac{193}{432 \sqrt{2}} b_3c_3^2,
\end{eqnarray}
\begin{eqnarray}
\alpha_{\mathbf{8}}^{\Xi^- \Sigma^0} & = & - \frac{55}{576 \sqrt{6}} a_1^3 + \frac{127}{576 \sqrt{6}} a_1^2b_2 + \frac{419}{1728 \sqrt{6}} a_1^2b_3 + \frac{47}{144 \sqrt{6}} a_1^2c_3 + \frac{11}{576 \sqrt{6}} a_1b_2^2 - \frac{109}{864 \sqrt{6}} a_1b_2b_3 + \frac{59}{144 \sqrt{6}} a_1b_2c_3 \nonumber \\
&  & \mbox{} + \frac{5}{576 \sqrt{6}} b_2^3 - \frac{3}{64} \sqrt{\frac32} a_1b_3^2 + \frac{287}{432 \sqrt{6}} a_1b_3c_3 + \frac{47}{576 \sqrt{6}} a_1c_3^2 + \frac{11}{1728 \sqrt{6}} b_2^2b_3 - \frac{109}{5184 \sqrt{6}} b_2b_3^2 + \frac{59}{576 \sqrt{6}} b_2c_3^2 \nonumber \\
&  & \mbox{} - \frac{1}{64 \sqrt{6}} b_3^3 + \frac{287}{1728 \sqrt{6}} b_3c_3^2,
\end{eqnarray}
\begin{eqnarray}
\alpha_{\mathbf{27}}^{\Xi^- \Sigma^0} & = & - \frac{3}{128 \sqrt{2}} a_1^3 - \frac{107}{1920 \sqrt{2}} a_1^2b_2 - \frac{719}{5760 \sqrt{2}} a_1^2b_3 + \frac{59}{1440 \sqrt{2}} a_1^2c_3 + \frac{3}{640 \sqrt{2}} a_1b_2^2 - \frac{31}{2880 \sqrt{2}} a_1b_2b_3 \nonumber \\
&  & \mbox{} - \frac{19}{480 \sqrt{2}} a_1b_2c_3 + \frac{1}{1152 \sqrt{2}} b_2^3 - \frac{371}{17280 \sqrt{2}} a_1b_3^2 - \frac{29}{480 \sqrt{2}} a_1b_3c_3 + \frac{59}{5760 \sqrt{2}} a_1c_3^2 + \frac{1}{640 \sqrt{2}} b_2^2b_3 \nonumber \\
&  & \mbox{} - \frac{31}{17280 \sqrt{2}} b_2b_3^2 - \frac{19}{1920 \sqrt{2}} b_2c_3^2 - \frac{371}{155520 \sqrt{2}} b_3^3 - \frac{29}{1920 \sqrt{2}} b_3c_3^2,
\end{eqnarray}
\begin{eqnarray}
\alpha_{\mathbf{1}}^{\Xi^0 \Sigma^+} & = & \frac{115}{144} a_1^3 + \frac{7}{48} a_1^2b_2 - \frac{31}{432} a_1^2b_3 - \frac{11}{12} a_1^2c_3 + \frac{19}{48} a_1b_2^2 + \frac{169}{216} a_1b_2b_3 - \frac{37}{36} a_1b_2c_3 + \frac{7}{144} b_2^3 + \frac{247}{432} a_1b_3^2 \nonumber \\
&  & \mbox{} - \frac{193}{108} a_1b_3c_3 - \frac{11}{48} a_1c_3^2 + \frac{19}{144} b_2^2b_3 + \frac{169}{1296} b_2b_3^2 - \frac{37}{144} b_2c_3^2 + \frac{247}{3888} b_3^3 - \frac{193}{432} b_3c_3^2,
\end{eqnarray}
\begin{eqnarray}
\alpha_{\mathbf{8}}^{\Xi^0 \Sigma^+} & = & - \frac{55}{576 \sqrt{3}} a_1^3 + \frac{127}{576 \sqrt{3}} a_1^2b_2 + \frac{419}{1728 \sqrt{3}} a_1^2b_3 + \frac{47}{144 \sqrt{3}} a_1^2c_3 + \frac{11}{576 \sqrt{3}} a_1b_2^2 - \frac{109}{864 \sqrt{3}} a_1b_2b_3 + \frac{59}{144 \sqrt{3}} a_1b_2c_3 \nonumber \\
&  & \mbox{} + \frac{5}{576 \sqrt{3}} b_2^3 - \frac{3 \sqrt{3}}{64} a_1b_3^2 + \frac{287}{432 \sqrt{3}} a_1b_3c_3 + \frac{47}{576 \sqrt{3}} a_1c_3^2 + \frac{11}{1728 \sqrt{3}} b_2^2b_3 - \frac{109}{5184 \sqrt{3}} b_2b_3^2 + \frac{59}{576 \sqrt{3}} b_2c_3^2 \nonumber \\
&  & \mbox{} - \frac{1}{64 \sqrt{3}} b_3^3 + \frac{287}{1728 \sqrt{3}} b_3c_3^2,
\end{eqnarray}
\begin{eqnarray}
\alpha_{\mathbf{27}}^{\Xi^0 \Sigma^+} & = & - \frac{3}{128} a_1^3 - \frac{107}{1920} a_1^2b_2 - \frac{719}{5760} a_1^2b_3 + \frac{59}{1440} a_1^2c_3 + \frac{3}{640} a_1b_2^2 - \frac{31}{2880} a_1b_2b_3 - \frac{19}{480} a_1b_2c_3 + \frac{1}{1152} b_2^3 \nonumber \\
&  & \mbox{} - \frac{371}{17280} a_1b_3^2 - \frac{29}{480} a_1b_3c_3 + \frac{59}{5760} a_1c_3^2 + \frac{1}{640} b_2^2b_3 - \frac{31}{17280} b_2b_3^2 - \frac{19}{1920} b_2c_3^2 - \frac{371}{155520} b_3^3 - \frac{29}{1920} b_3c_3^2. \nonumber \\
\end{eqnarray}

The functions $F_{\mathbf{rep}}^{(1)}$ depend on the loop integral $F^{(1)}(m,0,\mu)$, which is given by \cite{fmg}
\begin{equation}
F^{(1)}(m,0,\mu) = \frac{m^2}{16\pi^2 f^2} \left[ \lambda_\epsilon + 1 - \ln{\frac{m^2}{\mu^2}} \right]. \label{eq:ib1}
\end{equation}

\section{\label{sec:heavy}One-loop corrections within HBChPT}

The very same contribution from loop-graphs of Fig.~\ref{fig:l1}(a,b,c) can also be evaluated within HBChPT. In this case, the expression can be cast into
\begin{equation}
\delta g_1^{B_1B_2} = \gamma_\pi^{B_1B_2} F^{(1)}(m_\pi,0,\mu) + \gamma_K^{B_1B_2} F^{(1)}(m_K,0,\mu) + \gamma_\eta^{B_1B_2} F^{(1)}(m_\eta,0,\mu), \label{eq:gach}
\end{equation}
where the $\gamma_\Pi^{B_1B_2}$ coefficients are given by
\begin{equation}
\gamma_{\pi}^{np} = - 2(D + F)^3 - \frac29 (D + F) \mathcal{C}^2 - \frac{50}{81} \mathcal{C}^2 \mathcal{H},
\end{equation}
\begin{equation}
\gamma_{K}^{np} = -\frac16 (13 D^3 - D^2F + 3 DF^2 + 33 F^3) + \frac{1}{18} (3D - 5F) \mathcal{C}^2 - \frac{10}{81} \mathcal{C}^2 \mathcal{H},
\end{equation}
\begin{equation}
\gamma_{\eta}^{np} = -\frac13 (D^3 - 5 D^2F + 3D F^2 + 9 F^3),
\end{equation}
\begin{equation}
\gamma_{\pi}^{\Sigma^- \Lambda} = - \frac13 \sqrt{\frac23} \left[ D(7D^2 + 3F^2) + \frac{1}{12} (29D - 24F) \mathcal{C}^2 + \frac59 \mathcal{C}^2 \mathcal{H} \right],
\end{equation}
\begin{equation}
\gamma_{K}^{\Sigma^- \Lambda} = -\frac{1}{\sqrt{6}} \left[ D(3 D^2 + 13 F^2) + \frac89(D - 3 F) \mathcal{C}^2 + \frac{5}{27} \mathcal{C}^2 \mathcal{H} \right],
\end{equation}
\begin{equation}
\gamma_{\eta}^{\Sigma^- \Lambda} = - \frac{1}{6\sqrt{6}} D(8 D^2 - \mathcal{C}^2),
\end{equation}
\begin{equation}
\gamma_{\pi}^{\Lambda p} = \frac{1}{\sqrt{6}} \left[ \frac18 (9 D^3 + 81 D^2F + 75 D F^2 + 27 F^3) - \frac{1}{12} (23 D - 51 F) \mathcal{C}^2 + \frac{10}{9} \mathcal{C}^2 \mathcal{H} \right],
\end{equation}
\begin{equation}
\gamma_{K}^{\Lambda p} = \frac{1}{\sqrt{6}} \left[ \frac{1}{12} (31 D^3 + 15 D^2F + 9 D F^2 + 297 F^3) - \frac14 (D - 5 F) \mathcal{C}^2 + \frac59 \mathcal{H} \mathcal{C}^2 \right],
\end{equation}
\begin{equation}
\gamma_{\eta}^{\Lambda p} = \frac{1}{24 \sqrt{6}} (19 D^3 + 27D^2F - 63D F^2 + 81F^3),
\end{equation}
\begin{equation}
\gamma_{\pi}^{\Sigma^- n} = - \frac{1}{24} (35 D^3 + 23D^2F + 33 D F^2 - 123 F^3) - \frac{1}{18} (17 D - 41 F) \mathcal{C}^2 + \frac{10}{81} \mathcal{H} \mathcal{C}^2,
\end{equation}
\begin{equation}
\gamma_{K}^{\Sigma^- n} = - \frac{1}{12} (31 D^3 - 53 D^2F + 57 D F^2 - 51F^3) - \frac{1}{36} (35 D - 59 F) \mathcal{C}^2 + \frac{5}{81} \mathcal{H} \mathcal{C}^2,
\end{equation}
\begin{equation}
\gamma_{\eta}^{\Sigma^- n} = - \frac{1}{24} (11D^3 - 17 D^2F + 33 D F^2 - 27F^3) - \frac{1}{36} (13 D - 21 F) \mathcal{C}^2,
\end{equation}
\begin{equation}
\gamma_{\pi}^{\Xi^- \Lambda} = \frac{1}{\sqrt{6}} \left[ \frac18 (9 D^3 - 81 D^2F + 75 D F^2 - 27 F^3) + \frac43 (D - 3F) \mathcal{C}^2 - \frac59 \mathcal{H} \mathcal{C}^2 \right],
\end{equation}
\begin{equation}
\gamma_{K}^{\Xi^- \Lambda} = \frac{1}{\sqrt{6}} \left[ \frac{1}{12} (31 D^3 - 15 D^2F + 9 D F^2 - 297 F^3) + \frac14 (9 D - 19 F) \mathcal{C}^2 - \frac59 \mathcal{H} \mathcal{C}^2 \right],
\end{equation}
\begin{equation}
\gamma_{\eta}^{\Xi^- \Lambda} = \frac{1}{\sqrt{6}} \left[ \frac{1}{24} (19 D^3 - 27D^2F - 63 D F^2 - 81F^3) + \frac{1}{12} (11 D - 9 F) \mathcal{C}^2 \right],
\end{equation}
\begin{equation}
\gamma_{\pi}^{\Xi^- \Sigma^0} = \frac{1}{\sqrt{2}} \left[ -\frac{1}{24} (35 D^3 - 23 D^2F + 33D F^2 + 123 F^3) + \frac{1}{36} (D - 7 F) \mathcal{C}^2 - \frac{10}{81} \mathcal{H} \mathcal{C}^2 \right],
\end{equation}
\begin{equation}
\gamma_{K}^{\Xi^- \Sigma^0} = \frac{1}{\sqrt{2}} \left[ -\frac{1}{12} (31 D^3 + 53 D^2F + 57D F^2 + 51 F^3) + \frac{1}{36} (3 D - 5 F) \mathcal{C}^2 - \frac{35}{81} \mathcal{H} \mathcal{C}^2 \right],
\end{equation}
\begin{equation}
\gamma_{\eta}^{\Xi^- \Sigma^0} = \frac{1}{\sqrt{2}} \left[ -\frac{1}{24} (11 D^3 + 17 D^2F + 33 D F^2 + 27 F^3) - \frac16 (D + F) \mathcal{C}^2 - \frac{5}{27} \mathcal{H} \mathcal{C}^2 \right],
\end{equation}
\begin{equation}
\gamma_{\pi}^{\Xi^0 \Sigma^+} = -\frac{1}{24} (35 D^3 - 23 D^2F + 33 D F^2 + 123 F^3) + \frac{1}{36} (D - 7F) \mathcal{C}^2 - \frac{10}{81} \mathcal{H} \mathcal{C}^2,
\end{equation}
\begin{equation}
\gamma_{K}^{\Xi^0 \Sigma^+} = -\frac{1}{12} (31 D^3 + 53 D^2F + 57 D F^2 + 51F^3) + \frac{1}{36} (3 D - 5 F) \mathcal{C}^2 - \frac{35}{81} \mathcal{H} \mathcal{C}^2,
\end{equation}
\begin{equation}
\gamma_{\eta}^{\Xi^0 \Sigma^+} = -\frac{1}{24} (11 D^3 + 17 D^2F + 33 D F^2 + 27F^3) - \frac16 (D + F) \mathcal{C}^2 - \frac{5}{27} \mathcal{H} \mathcal{C}^2.
\end{equation}

Expression (\ref{eq:gach}) reduces to the corresponding one presented in Refs.~\cite{jm255,jm259} when the limit $m_u=m_d=0$ is implemented and the Gell-Mann-Okubo formula is used to rewrite $m_\eta$ as $(4/3)m_K^2$, retaining only the chiral logs in the loop integrals.

\section{\label{sec:compara}Comparison with HBChPt in the degeneracy limit $\Delta/m\to 0$}

Expression (\ref{eq:gach}) can also be organized in terms of singlet, $\mathbf{8}$, and $\mathbf{27}$ flavor contributions. The full expression, in a close analogy with Eq.~(\ref{eq:galn}), can be cast into
\begin{equation}
\delta g_1^{B_1B_2} = \gamma_{\mathbf{1}}^{B_1B_2} F_{\mathbf{1}}^{(1)} + \gamma_{\mathbf{8}}^{B_1B_2} F_{\mathbf{8}}^{(1)} + \gamma_{\mathbf{27}}^{B_1B_2} F_{\mathbf{27}}^{(1)},
\end{equation}
where the $\gamma_{\mathbf{rep}}^{B_1B_2}$ coefficients read
\begin{subequations}
\begin{equation}
\gamma_{\mathbf{1}}^{B_1B_2} = -(\gamma_{\pi}^{B_1B_2} + \gamma_{K}^{B_1B_2} + \gamma_{\eta}^{B_1B_2}),
\end{equation}
\begin{equation}
\gamma_{\mathbf{8}}^{B_1B_2} = -\frac{1}{\sqrt{3}} \left[ \gamma_{\pi}^{B_1B_2} - \frac12 \gamma_{K}^{B_1B_2} - \gamma_{\eta}^{B_1B_2} \right],
\end{equation}
\begin{equation}
\gamma_{\mathbf{27}}^{B_1B_2} = -\frac{3}{40} \left[ \gamma_{\pi}^{B_1B_2} - 3\gamma_{K}^{B_1B_2} + 9\gamma_{\eta}^{B_1B_2} \right].
\end{equation}
\end{subequations}

The above coefficients {\it coincide} with the ones obtained in the combined formalism, namely,
\begin{equation}
\alpha_{\mathbf{rep}}^{B_1B_2} = \gamma_{\mathbf{rep}}^{B_1B_2},
\end{equation}
when relations (\ref{eq:rel1}) are used. This agreement, term by term, successfully completes another stage of the computational scheme traced back a few years ago to check whether the combined formalism and conventional HBChPT yield the same results at the physical value $N_c=3$. For the axial vector current it is indeed the case, at least for degenerate intermediate baryon states in the one-loop corrections.

\section{\label{sec:totalc}Total correction to the baryon axial vector current}

Following Ref.~\cite{rfm12}, corrections to the baryon axial vector current $A^{kc}$, Eq.~(\ref{eq:akc}), will be made up by one-loop and perturbative $SU(3)$ symmetry breaking contributions, namely,
\begin{equation}
A^{kc} + \delta A^{kc} = A^{kc} + \delta A_{\mathrm{1L}}^{kc} + \delta A_{\mathrm{SB}}^{kc}. \label{eq:totalakc}
\end{equation}
Here $\delta A_{\mathrm{1L}}^{kc}$ stands for the one-loop correction and is constituted by two terms: $\delta A_{1a}^{kc}$ and $\delta A_{1d}^{kc}$, which arise from Fig.~\ref{fig:l1}(a,b,c) and Fig.~\ref{fig:l1}(d), respectively; the former is given by Eq.~(\ref{eq:dakc}), although only results in the degeneracy limit are contained in it; the later can be identified with Eq.~(42) of Ref.~\cite{rfm06}.

On the other hand, perturbative $SU(3)$ symmetry breaking contributions to the axial vector current to first order in $\epsilon \sim m_s-\hat{m}$ has the $1/N_c$ expansion \cite{rfm12,rfm14}
\begin{equation}
\delta A_{\mathrm{SB}}^{kc} = d_1 d^{c8e} G^{ke} + d_2 \frac{1}{N_c} d^{c8e} \mathcal{D}_2^{ke} + d_3 \frac{1}{N_c} \left( \{G^{kc},T^8\}-\{G^{k8},T^c\} \right) + d_4 \frac{1}{N_c} \left( \{G^{kc},T^8\}+\{G^{k8},T^c\} \right). \label{eq:dakcsb}
\end{equation}
The appearance of this term can be understood as follows. In the conventional chiral momentum counting scheme, tree diagrams involving higher-order vertices will contribute to the axial vector current in addition to one-loop contributions. Some of them are needed as counterterms for the divergent parts of the loop integrals and come along with extra low-energy constants, with the subsequent introduction of more unknowns in the low-energy expansion. The leading $SU(3)$ breaking effects of the axial vector current will thus yield contributions linear in $m_q$. In the combined formalism, terms of order $\mathcal{O}(m_q)$ can formally be accounted for via perturbative flavor $SU(3)$ symmetry breaking, which transforms as a flavor octet. This is where expansion (\ref{eq:dakcsb}) plays a role. However, in order to have predictive power, terms up to relative order $N_c^{-1}$ are retained in the series and yet, there are four operator coefficients $d_i$ to be considered.

Some numerical analyses will be provided in the next section in order to compare theoretical expressions with experimental information \cite{part}.

\section{\label{sec:fits}Preliminary numerical analysis}

Having derived the full set of results for the renormalized axial vector current, it is now timely to study their effects numerically. 

A complete evaluation of the axial vector couplings requires a reliable determination of the free parameters in the formalism, i.e., the operator coefficients $a_1$, $b_2$, $b_3$, and $c_3$ introduced in the definition of the axial vector current Eq.~(\ref{eq:akc}). Information about these parameters can be obtained through a least-squares fit using the available experimental data \cite{part}. For octet baryons, data come from semileptonic decays in terms of decays rates, $g_1/f_1$ ratios ($f_1$ is the vector coupling), angular correlation coefficients, and spin-asymmetry coefficients \cite{part}. Those data have been collected in Table II of Ref.~\cite{rfm12}. For decuplet baryons, physically the transitions $\Delta \to N \pi$, $\Sigma^* \to \Lambda \pi$, $\Sigma^*\to \Sigma \pi$, and $\Xi^* \to \Xi \pi$ are all kinematically allowed and their axial vector couplings $g$ are extracted via Goldberger-Treiman relations from the widths of the strong decays \cite{dai}. Those data are summarized in Table IX of that reference.

In previous works \cite{rfm06,rfm12,fmg}, a number of fits were carried out under various assumptions. The analysis of Ref.~\cite{rfm06}, a rather limited one because of the order of approximation in $1/N_c$ implemented in the axial vector coupling, used the measured decay rates and the $g_1/f_1$ ratios in the fit. Although the predicted observables were in accord with the experimental data, the fit was not entirely satisfactory. The analysis, however, became a useful starting point for a more precise one presented in Ref.~\cite{rfm12}, where the effects of perturbative $SU(3)$ symmetry breaking and a nonzero decuplet-octet mass difference were incorporated. Although the quality of the fit improved slightly with respect to the previous one, still it was not entirely satisfactory. Both numerical analyses limited to determine the axial vector couplings $g_1$ and $g$ only whereas the vector couplings $f_1$ were given at their $SU(3)$ symmetric values. This limitation was removed in Ref.~\cite{fmg}, where the effects of $SU(3)$ symmetry breaking in $f_1$ to second order in $\epsilon$ were evaluated. There was an overall improvement in the output of the fit when all these effects were gathered together.

It is now viable to perform a more refined numerical analysis by using the expressions obtained here, following the lines of Ref.~\cite{fmg}. From the experimental bent, only the data on $g_1/f_1$ and $g$ will be used in the fit. From the theoretical bent, the vector coupling $f_1$ will incorporate effects of $SU(3)$ symmetry breaking using the expressions of Ref.~\cite{fmg}. The axial vector couplings will contain tree, one-loop, and perturbative symmetry breaking effects as indicated in Eq.~(\ref{eq:totalakc}); the latter will introduce four more parameters, namely, $d_1,\ldots,d_4$. One-loop effects will comprise all possible terms allowed in the double commutator structures in $\delta A_{\mathrm{1L}}^{kc}$ for $N_c=3$ {\it in the degeneracy limit}. In this sense, the present analysis will partially surpass the ones of Refs.~\cite{rfm06,rfm12}. For definiteness, the physical masses of the mesons and baryons listed in Ref.~\cite{part} are used; additional inputs are $\Delta = 0.231$ GeV, $f = 93$ MeV, $\mu = 1$ GeV, and the suggested values of the CKM matrix elements $V_{ud}$ and $V_{us}$.

The numerical analysis here consists of two cases of interest. The first one, labeled as fit A, corresponds to the degeneracy limit. The second one, labeled as fit B, makes use of the partial terms with insertions of one and two mass operators available in Ref.~\cite{rfm12}, which have not been computed at the same level as in the degeneracy case. In this sense, fit B is a preliminary one. At any rate, it will be possible to contrast both outputs to draw some conclusions.

Without further ado, the fits yield the best-fit parameters listed in Table \ref{t:summary}. The predicted axial vector couplings, separated in tree and symmetry breaking corrections (explicit and implicit ones), are listed in Tables \ref{t:fita} and \ref{t:fitb} for the best-fit parameters of fit A and B, respectively.

A close inspection to Table \ref{t:summary} reveals that both fits yield parameters consistent with the expectations of the $1/N_c$ expansion: $a_1$, the leading order coefficient, is of order $\mathcal{O}(N_c^0)$; subleading order terms should exhibit a relative suppression in $1/N_c$ with respect to the leading order one. This is more evident in fit A. On the other hand, the quality of the fits can be evaluated based on standard statistical goodness of fit criteria and the value of $\chi^2/\mathrm{dof}$ becomes a useful indicator. Consequently, fit B is by far better than fit A: The overall behavior of fit A is good because the predicted observables displayed in Table \ref{t:fita} are in good agreement with their experimental counterparts. In contrast, the overall behavior of fit B is excellent because its predictions are closer to physics. Notice that symmetry breaking corrections (both implicit and explicit) are roughly suppressed by a factor of $\epsilon\sim 0.3$ with respect to the leading ones (except for the process $\Xi^-\to \Lambda$), which is consistent with first-order symmetry breaking. Notice also the existence of suppression factors in the different flavor contributions that make up loop corrections. The flavor-$\mathbf{27}$ contribution is highly suppressed relative to the flavor $\mathbf{8}$ contributions, which in turn is suppressed relative to flavor $\mathbf{1}$ contribution.

As for $f_1$, the resultant symmetry breaking pattern is consistent with recent results obtained in the context of the combined formalism in $1/N_c$ and chiral corrections in terms of the $\xi$ power counting of Ref.~\cite{fg} and lattice QCD \cite{sha}. The consistency is better appreciated for the output of fit B.

\begingroup
\squeezetable
\begin{table}
\caption{\label{t:summary}Best-fit parameters for the different fits performed. The pertinent values of the equivalent $SU(3)$ couplings $D$, $F$, $\mathcal{C}$, and $\mathcal{H}$ are also listed. The quoted errors come from the fits only.}
\begin{center}
\begin{tabular}{lrr}
\hline\hline
      &         Fit A &         Fit B         \\ \hline
$a_1$ &         $ 1.11(0.05)$ & $ 1.20(0.07)$ \\
$b_2$ &         $-0.53(0.09)$ & $-1.60(0.18)$ \\
$b_3$ &         $-0.62(0.21)$ & $ 1.25(0.07)$ \\
$c_3$ &         $-0.20(0.19)$ & $ 0.46(0.09)$ \\
$d_1$ &         $ 0.33(0.05)$ & $ 0.76(0.12)$ \\
$d_2$ &         $-1.45(0.23)$ & $-0.65(0.25)$ \\
$d_3$ &         $ 0.37(0.03)$ & $ 0.35(0.08)$ \\
$d_4$ &         $ 0.30(0.07)$ & $-0.01(0.07)$ \\

$D$   &         $ 0.45(0.01)$ & $ 0.81(0.04)$ \\
$F$   &         $ 0.21(0.02)$ & $ 0.27(0.01)$ \\
$\mathcal{C}$ & $-1.01(0.05)$ & $-1.43(0.04)$ \\
$\mathcal{H}$ & $ 0.68(0.54)$ & $-2.52(0.06)$ \\
$F/D$  &        $ 0.47(0.03)$ & $ 0.34(0.02)$ \\
$3F-D$ &        $ 0.19(0.05)$ & $ 0.01(0.05)$ \\
$\chi^2/\mathrm{dof}$ & $28.8/4$ & $8.8/4$    \\
\hline\hline
\end{tabular}
\end{center}
\end{table}
\endgroup

\begingroup
\squeezetable
\begin{table}
\caption{\label{t:fita}Predicted axial vector couplings for vanishing $\Delta$. The output of fit A is used in the evaluation. The experimental data about $g_1/f_1$ and $g$ are used in the fit. $SU(3)$ flavor symmetry breaking is taken into account in two ways: explicitly through perturbative symmetry breaking and implicitly through the integrals occurring in the one-loop corrections.}
\begin{center}
\begin{tabular}{lrrrrrrrrr}
\hline\hline
  &  &  &  & \multicolumn{3}{c}{Fig.~\ref{fig:l1}(a,b,c), $\mathcal{O}(\Delta^0)$} & \multicolumn{3}{c}{Fig.~\ref{fig:l1}(d)} \\
Process & Total & Tree & SB & $\mathbf{1}$ & $\mathbf{8}$ & $\mathbf{27}$ & $\mathbf{1}$ & $\mathbf{8}$ & $\mathbf{27}$ \\ \hline
$np$                 & $ 1.270$ & $ 0.664$ & $ 0.336$ & $ 0.269$ & $-0.119$ & $-0.001$ & $ 0.179$ & $-0.060$ & $ 0.001$ \\
$\Sigma^\pm \Lambda$ & $ 0.598$ & $ 0.369$ & $ 0.078$ & $ 0.127$ & $-0.043$ & $ 0.000$ & $ 0.100$ & $-0.033$ & $ 0.001$ \\
$\Lambda p$          & $-0.890$ & $-0.444$ & $-0.150$ & $-0.203$ & $ 0.046$ & $-0.001$ & $-0.120$ & $-0.020$ & $ 0.003$ \\
$\Sigma^- n$         & $ 0.328$ & $ 0.239$ & $-0.042$ & $ 0.042$ & $ 0.016$ & $-0.001$ & $ 0.065$ & $ 0.011$ & $-0.001$ \\
$\Xi^- \Lambda$      & $ 0.187$ & $ 0.076$ & $ 0.048$ & $ 0.076$ & $-0.039$ & $ 0.004$ & $ 0.020$ & $ 0.003$ & $ 0.000$ \\
$\Xi^- \Sigma^0$     & $ 0.726$ & $ 0.470$ & $-0.119$ & $ 0.190$ & $ 0.042$ & $-0.002$ & $ 0.127$ & $ 0.021$ & $-0.003$ \\
$\Xi^0 \Sigma^+$     & $ 1.027$ & $ 0.664$ & $-0.168$ & $ 0.269$ & $ 0.060$ & $-0.003$ & $ 0.179$ & $ 0.030$ & $-0.004$ \\
$\Delta N$           & $-2.030$ & $-1.008$ & $-0.579$ & $-0.408$ & $ 0.151$ & $-0.002$ & $-0.272$ & $ 0.091$ & $-0.002$ \\
$\Sigma^*\Lambda$    & $-1.737$ & $-1.008$ & $-0.191$ & $-0.408$ & $ 0.053$ & $ 0.000$ & $-0.272$ & $ 0.091$ & $-0.002$ \\
$\Sigma^*\Sigma$     & $-1.694$ & $-1.008$ & $-0.104$ & $-0.408$ & $ 0.019$ & $-0.010$ & $-0.272$ & $ 0.091$ & $-0.002$ \\
$\Xi^*\Xi$           & $-1.413$ & $-1.008$ & $ 0.241$ & $-0.408$ & $-0.062$ & $ 0.008$ & $-0.272$ & $ 0.091$ & $-0.002$ \\
\hline\hline
\end{tabular}
\end{center}
\end{table}
\endgroup

\begingroup
\squeezetable
\begin{table}
\caption{\label{t:fitb}Predicted axial vector couplings for non-vanishing $\Delta$. The output of fit B is used in the evaluation. The experimental data about $g_1f_1$ and $g$ are used in the fit. $SU(3)$ flavor symmetry breaking is taken into account in two ways: explicitly through perturbative symmetry breaking and implicitly through the integrals occurring in the one-loop corrections.}
\begin{center}
\begin{tabular}{lrrrrrrrrrrrrrrr}
\hline\hline
  &  &  &  & \multicolumn{3}{c}{Fig.~\ref{fig:l1}(a,b,c), $\mathcal{O}(\Delta^0)$} & \multicolumn{3}{c}{Fig.~\ref{fig:l1}(a,b,c), $\mathcal{O}(\Delta)$} & \multicolumn{3}{c}{Fig.~\ref{fig:l1}(a,b,c), $\mathcal{O}(\Delta^2)$} & \multicolumn{3}{c}{Fig.~\ref{fig:l1}(d)} \\
Process & Total & Tree & SB & $\mathbf{1}$ & $\mathbf{8}$ & $\mathbf{27}$ & $\mathbf{1}$ & $\mathbf{8}$ & $\mathbf{27}$ & $\mathbf{1}$ & $\mathbf{8}$ & $\mathbf{27}$ & $\mathbf{1}$ & $\mathbf{8}$ & $\mathbf{27}$ \\ \hline
$np$                  & $ 1.270$ & $ 1.080$ & $ 0.426$ & $ 0.061$ & $ 0.049$ & $ 0.007$ & $ 0.232$ & $-0.111$ & $-0.002$ & $-0.446$ & $-0.220$ & $-0.003$ & $ 0.292$ & $-0.097$ & $ 0.002$ \\
$\Sigma^\pm \Lambda$  & $ 0.600$ & $ 0.660$ & $ 0.178$ & $ 0.324$ & $-0.072$ & $-0.004$ & $-0.269$ & $-0.028$ & $ 0.000$ & $-0.281$ & $-0.031$ & $ 0.002$ & $ 0.178$ & $-0.059$ & $ 0.001$ \\
$\Lambda p$           & $-0.816$ & $-0.663$ & $ 0.016$ & $ 0.250$ & $-0.186$ & $ 0.004$ & $-0.553$ & $ 0.173$ & $-0.002$ & $ 0.265$ & $ 0.087$ & $-0.002$ & $-0.179$ & $-0.030$ & $ 0.004$ \\
$\Sigma^- n$          & $ 0.326$ & $ 0.535$ & $ 0.001$ & $ 0.733$ & $-0.061$ & $-0.002$ & $-0.891$ & $ 0.094$ & $ 0.000$ & $-0.242$ & $-0.007$ & $-0.001$ & $ 0.145$ & $ 0.024$ & $-0.003$ \\
$\Xi^- \Lambda$       & $ 0.221$ & $ 0.004$ & $ 0.080$ & $-0.574$ & $-0.062$ & $-0.003$ & $ 0.822$ & $-0.045$ & $ 0.004$ & $ 0.016$ & $-0.024$ & $ 0.003$ & $ 0.001$ & $ 0.000$ & $ 0.000$ \\
$\Xi^- \Sigma^0$      & $ 0.840$ & $ 0.764$ & $-0.151$ & $ 0.043$ & $-0.017$ & $ 0.004$ & $ 0.164$ & $ 0.039$ & $-0.002$ & $-0.315$ & $ 0.078$ & $-0.003$ & $ 0.206$ & $ 0.034$ & $-0.005$ \\
$\Xi^0 \Sigma^+$      & $ 1.188$ & $ 1.080$ & $-0.213$ & $ 0.061$ & $-0.025$ & $ 0.006$ & $ 0.232$ & $ 0.056$ & $-0.002$ & $-0.446$ & $ 0.110$ & $-0.004$ & $ 0.292$ & $ 0.049$ & $-0.006$ \\
$\Delta N$            & $-2.037$ & $-1.428$ & $-0.627$ & $-0.647$ & $-0.012$ & $-0.010$ & $ 0.223$ & $-0.157$ & $-0.018$ & $ 0.630$ & $ 0.266$ & $ 0.004$ & $-0.386$ & $ 0.128$ & $-0.003$ \\
$\Sigma^*\Lambda$     & $-1.712$ & $-1.428$ & $-0.436$ & $-0.647$ & $ 0.030$ & $ 0.020$ & $ 0.223$ & $-0.001$ & $ 0.007$ & $ 0.630$ & $ 0.154$ & $-0.003$ & $-0.386$ & $ 0.128$ & $-0.003$ \\
$\Sigma^*\Sigma$      & $-1.738$ & $-1.428$ & $-0.019$ & $-0.647$ & $ 0.185$ & $-0.012$ & $ 0.223$ & $-0.228$ & $-0.013$ & $ 0.630$ & $-0.154$ & $-0.014$ & $-0.386$ & $ 0.128$ & $-0.003$ \\
$\Xi^*\Xi$            & $-1.390$ & $-1.428$ & $-0.037$ & $-0.647$ & $ 0.149$ & $ 0.009$ & $ 0.223$ & $ 0.042$ & $ 0.045$ & $ 0.630$ & $-0.112$ & $-0.005$ & $-0.386$ & $ 0.128$ & $-0.003$ \\
\hline\hline
\end{tabular}
\end{center}
\end{table}
\endgroup

\begingroup
\squeezetable
\begin{table}
\caption{\label{t:f1sb}Symmetry breaking pattern $f_1/f_1^{SU(3)}-1$ for the vector coupling $f_1$.}
\begin{center}
\begin{tabular}{lrrrr}
\hline\hline
                 & Fit A  & Fit B  & Ref.~\cite{fg} & LQCD \cite{sha} \\ \hline
$\Lambda p$      & $-0.028$ & $-0.067$ & $-0.067(15)$   & $-0.05(2)$ \\
$\Sigma^- n$     & $-0.043$ & $-0.028$ & $-0.025(10)$   & $-0.02(3)$ \\
$\Xi^- \Lambda$  & $-0.034$ & $-0.060$ & $-0.053(10)$   & $-0.06(4)$ \\
$\Xi^0 \Sigma^+$ & $-0.024$ & $-0.049$ & $-0.068(17)$   & $-0.05(2)$ \\
$\Xi^0 \Sigma^+$ & $-0.024$ & $-0.049$ &                &            \\
\hline\hline
\end{tabular}
\end{center}
\end{table}
\endgroup

\section{\label{sec:finish}Concluding remarks}

The renormalization to the axial vector current has been dealt with in the combined formalism in $1/N_c$ and chiral corrections; the material presented upgrades the one of previous works \cite{rfm06,rfm12} in various aspects. A meson-baryon vertex in loop graphs can be represented by the insertion of an axial vector operator $A^{ia}$ whose most general form is given in Eq.~(\ref{eq:akcfull}). Unlike previous analyses, here $A^{ia}$ retained up to 3-body operators so its $1/N_c$ expansion is given by Eq.~(\ref{eq:akc}). One specific loop graph is proportional to the double commutator $[A^{ia},[A^{ib},A^{kc}]]$, which is a $7$-body operator according to naive counting rules. The explicit evaluation of this operator structure was rigorously performed for flavor singlet, flavor $\mathbf{8}$,  and $\mathbf{27}$ representations; results are given in full in Appendices \ref{sec:r1}, \ref{sec:r8}, and \ref{sec:r27}, respectively. An additional loop correction has been dealt with in previous works \cite{rfm06,rfm12} and was not repeated here.

The total one-loop correction to the axial vector coupling for degenerate intermediate baryon states (a limitation of the present calculation) was compared to its counterpart obtained within heavy baryon chiral perturbation theory of Refs.~\cite{jm255,jm259}. The comparison was a successful one term by term. This is indeed a remarkable outcome: {\it Results of large-$N_c$ chiral perturbation theory and conventional heavy baryon chiral perturbation theory agree at the physical value $N_c=3$}. The agreement has been partially observed in analyses of Refs.~\cite{rfm06,rfm12} and fully observed for other quantities \cite{fmg,rfm19}. For the first time, this agreement has been totally achieved for the axial vector current for degenerate intermediate baryons.

The renormalized axial vector current was thus constructed by adding together one-loop and perturbative flavor symmetry breaking contributions. The effects of symmetry breaking were accounted for implicitly since the loop integrals depend on the meson masses and explicitly through perturbative first-order symmetry breaking, with the introduction of additional coefficients that serve as counterterms for the divergent parts of the loop integrals. All these contributions were gathered together into a single expression, Eq.~(\ref{eq:totalakc}). It was instructive to perform a numerical analysis via a least-squares fit in order to determine the free parameters of the theory, namely, $a_1$, $b_2$, $b_3$, and $c_3$ for $A^{ia}$, and $d_1,\ldots d_4$ for perturbative symmetry breaking, under two possible scenarios. The first one, labeled as fit A, used the expression for $g_1$ obtained here, along with the expression for the vector coupling $f_1$ computed in Ref.~\cite{fmg}, which included second-order symmetry breaking effects. The second scenario, labeled as fit B, followed the lines of the previous one, and also included effects of a nonzero decuplet-octet mass difference $\Delta$, partially evaluated in Ref.~\cite{rfm12}. 

The idea behind the study of these two possible scenarios sprang from the interest of finding out how important the effects of $\Delta$ are, although partial results are only available. The best-fit parameters are listed in Table \ref{t:summary}. Clearly, fit A produced results consistent with expectations both from the large-$N_c$ and experimental perspectives. Fit B, however, can be considered much better than fit A based on standard statistical goodness of fit criteria and the value of $\chi^2/\mathrm{dof}$, which turned out to be $28.8/4$ and $8.8/4$, respectively. Globally, fit B lead to more consistent values of the equivalent $SU(3)$ couplings $D$, $F$, $\mathcal{C}$, and $\mathcal{H}$ introduced in HBChPT \cite{jm255,jm259}; specifically, the magnitudes of $\mathcal{C}$ and $\mathcal{H}$ were consistent with expectations and the sign of the latter went in the right direction. On the other hand, the $3F-D$ factor, which plays a role in the extraction of the value of the strange quark spin and of the total quark spin from the measured value of the spin-dependent deep inelastic structure functions of the proton and neutron, showed a significant reduction to almost a zero value. Particular attention should be paid to this fact so results should be interpreted with care.

In spite of the relative success of the above mentioned fits, one cannot yet consider the theoretical issues as closed. It is most important that the full evaluation of $\Delta$ effects be performed, and the procedure presented here may provide useful guidance for this enterprise. This calculation, however, requires a non-negligible effort and will be attempted in the near future.

\begin{acknowledgments}
The authors are grateful to Consejo Nacional de Ciencia y Tecnolog{\'\i}a (Mexico) for partial support.
\end{acknowledgments}

\appendix

\section{\label{sec:r1}Reduction of singlet operators}

Here, the complete reduction of the operator structure
\begin{equation*}
[A^{ia},[A^{ia},A^{kc}]]
\end{equation*}
at the physical value $N_c=3$ is provided. Individual structures read
\begin{equation}
[G^{ia},[G^{ia},G^{kc}]] = \frac{3N_f^2-4}{4N_f} G^{kc},
\end{equation}

\begin{equation}
[G^{ia},[G^{ia},\mathcal{D}_2^{kc}]] + [\mathcal{D}_2^{ia},[G^{ia},G^{kc}]] + [G^{ia},[\mathcal{D}_2^{ia},G^{kc}]] = - \frac{2(N_c+N_f)}{N_f} G^{kc} + \frac{9 N_f^2+8N_f-4}{4N_f} \mathcal{D}_2^{kc}, 
\end{equation}

\begin{eqnarray}
&  & [G^{ia},[G^{ia},\mathcal{D}_3^{kc}]] + [\mathcal{D}_3^{ia},[G^{ia},G^{kc}]] + [G^{ia},[\mathcal{D}_3^{ia},G^{kc}]] \nonumber \\
&  &\mbox{\hglue0.2truecm} = - [N_c(N_c+2N_f)-2N_f+8] G^{kc} - 3(N_c+N_f) \mathcal{D}_2^{kc} + \frac{13N_f^2+16N_f-12}{4N_f} \mathcal{D}_3^{kc} + \frac{N_f^2+2N_f-8}{N_f} \mathcal{O}_3^{kc},
\end{eqnarray}

\begin{eqnarray}
&  & [G^{ia},[G^{ia},\mathcal{O}_3^{kc}]] + [\mathcal{O}_3^{ia},[G^{ia},G^{kc}]] + [G^{ia},[\mathcal{O}_3^{ia},G^{kc}]] \nonumber \\
&  &\mbox{\hglue0.2truecm} = - [N_c(N_c+2N_f)-N_f] G^{kc} - \frac12 (N_c+N_f) \mathcal{D}_2^{kc} + \frac12 (N_f+1) \mathcal{D}_3^{kc} + \frac{15 N_f^2+12 N_f-4}{4N_f} \mathcal{O}_3^{kc},
\end{eqnarray}

\begin{eqnarray}
&  & [G^{ia},[\mathcal{D}_2^{ia},\mathcal{D}_2^{kc}]] + [\mathcal{D}_2^{ia},[G^{ia},\mathcal{D}_2^{kc}]] + [\mathcal{D}_2^{ia},[\mathcal{D}_2^{ia},G^{kc}]] \nonumber \\
&  &\mbox{\hglue0.2truecm} = \frac{N_c(N_c+2N_f)(N_f-2)-6 N_f^2}{2N_f} G^{kc} + \frac{2 (N_c+N_f)(N_f-1)}{N_f} \mathcal{D}_2^{kc} + \frac14 (3N_f+2) \mathcal{D}_3^{kc} + \frac12 N_f \mathcal{O}_3^{kc},
\end{eqnarray}

\begin{eqnarray}
&  & [G^{ia},[\mathcal{D}_2^{ia},\mathcal{D}_3^{kc}]] + [\mathcal{D}_2^{ia},[G^{ia},\mathcal{D}_3^{kc}]] + [\mathcal{D}_2^{ia},[\mathcal{D}_3^{ia},G^{kc}]] + [\mathcal{D}_3^{ia},[\mathcal{D}_2^{ia},G^{kc}]] + [G^{ia},[\mathcal{D}_3^{ia},\mathcal{D}_2^{kc}]] + [\mathcal{D}_3^{ia},[G^{ia},\mathcal{D}_2^{kc}]] \nonumber \\
&  &\mbox{\hglue0.2truecm} = -12(N_c+N_f) G^{kc} + [N_c(N_c+2N_f)-2N_f+8] \mathcal{D}_2^{kc} + \frac{(N_c+N_f)(7N_f-4)}{N_f} \mathcal{D}_3^{kc} + \frac{2(N_c+N_f)(3N_f-4)}{N_f} \mathcal{O}_3^{kc} \nonumber \\
&  & \mbox{\hglue0.6truecm} + \frac{3N_f^2-4 N_f-4}{N_f} \mathcal{D}_4^{kc},
\end{eqnarray}

\begin{eqnarray}
&  & [G^{ia},[\mathcal{D}_2^{ia},\mathcal{O}_3^{kc}]] + [\mathcal{D}_2^{ia},[G^{ia},\mathcal{O}_3^{kc}]] + [\mathcal{D}_2^{ia},[\mathcal{O}_3^{ia},G^{kc}]] + [\mathcal{O}_3^{ia},[\mathcal{D}_2^{ia},G^{kc}]] + [G^{ia},[\mathcal{O}_3^{ia},\mathcal{D}_2^{kc}]] + [\mathcal{O}_3^{ia},[G^{ia},\mathcal{D}_2^{kc}]] \nonumber \\
&  &\mbox{\hglue0.2truecm} = - \frac32 [N_c(N_c+2N_f)-8 N_f] \mathcal{D}_2^{kc} -\frac92 (N_c+N_f) \mathcal{D}_3^{kc} - \frac{2(N_c+N_f)}{N_f} \mathcal{O}_3^{kc} + (3N_f+10) \mathcal{D}_4^{kc},
\end{eqnarray}

\begin{equation}
[\mathcal{D}_2^{ia},[\mathcal{D}_2^{ia},\mathcal{D}_2^{kc}]] = \frac{N_c(N_c+2N_f)(N_f-2)-2N_f^2}{2N_f} \mathcal{D}_2^{kc} + \frac12 (N_f+2) \mathcal{D}_4^{kc},
\end{equation}

\begin{eqnarray}
&  & [G^{ia},[\mathcal{D}_3^{ia},\mathcal{D}_3^{kc}]] + [\mathcal{D}_3^{ia},[G^{ia},\mathcal{D}_3^{kc}]] + [\mathcal{D}_3^{ia},[\mathcal{D}_3^{ia},G^{kc}]] \nonumber \\
&  &\mbox{\hglue0.2truecm} = - 6 [N_c(N_c+2N_f)+2N_f] G^{kc} + 6(N_c+N_f) \mathcal{D}_2^{kc} + 3 [N_c(N_c+2N_f)+2N_f-2] \mathcal{D}_3^{kc} \nonumber \\
&  & \mbox{\hglue0.6truecm} + \frac{N_cN_f(N_c+2N_f)+12N_f(N_f-2)+8}{N_f} \mathcal{O}_3^{kc} - 3(N_c+N_f) \mathcal{D}_4^{kc} + \frac{3(N_f^2+2N_f-4)}{2N_f} \mathcal{D}_5^{kc} \nonumber \\
&  & \mbox{\hglue0.6truecm} + \frac{N_f^2+8N_f-20}{N_f} \mathcal{O}_5^{kc},
\end{eqnarray}

\begin{eqnarray}
&  & [G^{ia},[\mathcal{D}_3^{ia},\mathcal{O}_3^{kc}]] + [\mathcal{D}_3^{ia},[G^{ia},\mathcal{O}_3^{kc}]] + [\mathcal{D}_3^{ia},[\mathcal{O}_3^{ia},G^{kc}]] + [\mathcal{O}_3^{ia},[\mathcal{D}_3^{ia},G^{kc}]] + [G^{ia},[\mathcal{O}_3^{ia},\mathcal{D}_3^{kc}]] + [\mathcal{O}_3^{ia},[G^{ia},\mathcal{D}_3^{kc}]] \nonumber \\
&  & \mbox{\hglue0.2truecm} = - 24(N_c+N_f) \mathcal{D}_2^{kc} - 3[2 N_c(N_c+2N_f)-5N_f] \mathcal{D}_3^{kc} - [N_c(N_c+2N_f)-2N_f+8] \mathcal{O}_3^{kc} - 5(N_c+N_f) \mathcal{D}_4^{kc} \nonumber \\
&  & \mbox{\hglue0.6truecm} + (5N_f+11) \mathcal{D}_5^{kc} + \frac{N_f^2+2N_f-8}{N_f} \mathcal{O}_5^{kc},
\end{eqnarray}

\begin{eqnarray}
&  & [G^{ia},[\mathcal{O}_3^{ia},\mathcal{O}_3^{kc}]] + [\mathcal{O}_3^{ia},[G^{ia},\mathcal{O}_3^{kc}]] + [\mathcal{O}_3^{ia},[\mathcal{O}_3^{ia},G^{kc}]] \nonumber \\
&  & \mbox{\hglue0.2truecm} = - \frac32 N_c(N_c+2N_f) G^{kc} - \frac14 [N_c(N_c+2N_f)-6N_f] \mathcal{D}_3^{kc} - \frac14 [13N_c(N_c+2N_f)-38N_f-12] \mathcal{O}_3^{kc} \nonumber \\
&  & \mbox{\hglue0.6truecm} - \frac34 (N_c+N_f) \mathcal{D}_4^{kc} + \frac14 (N_f+3) \mathcal{D}_5^{kc} + \frac12 (11N_f+16) \mathcal{O}_5^{kc},
\end{eqnarray}

\begin{eqnarray}
&  & [\mathcal{D}_2^{ia},[\mathcal{D}_2^{ia},\mathcal{D}_3^{kc}]] + [\mathcal{D}_2^{ia},[\mathcal{D}_3^{ia},\mathcal{D}_2^{kc}]] + [\mathcal{D}_3^{ia},[\mathcal{D}_2^{ia},\mathcal{D}_2^{kc}]] \nonumber \\
&  & \mbox{\hglue0.2truecm} = \frac{N_c(N_c+2N_f)(N_f-2)-6N_f^2}{2N_f} \mathcal{D}_3^{kc} + \frac{4(N_c+N_f)(N_f-1)}{N_f} \mathcal{D}_4^{kc} + \frac12 (3N_f+2) \mathcal{D}_5^{kc},
\end{eqnarray}

\begin{equation}
[\mathcal{D}_2^{ia},[\mathcal{D}_2^{ia},\mathcal{O}_3^{kc}]] + [\mathcal{D}_2^{ia},[\mathcal{O}_3^{ia},\mathcal{D}_2^{kc}]] + [\mathcal{O}_3^{ia},[\mathcal{D}_2^{ia},\mathcal{D}_2^{kc}]] = \frac{N_c(N_c+2N_f)(N_f-2)-6N_f^2}{2N_f} \mathcal{O}_3^{kc} + \frac12 N_f \mathcal{O}_5^{kc} ,
\end{equation}

\begin{eqnarray}
&  & [\mathcal{D}_2^{ia},[\mathcal{D}_3^{ia},\mathcal{D}_3^{kc}]] + [\mathcal{D}_3^{ia},[\mathcal{D}_2^{ia},\mathcal{D}_3^{kc}]] + [\mathcal{D}_3^{ia},[\mathcal{D}_3^{ia},\mathcal{D}_2^{kc}]] \nonumber \\
&  & \mbox{\hglue0.2truecm} = - 6(N_c+N_f) \mathcal{D}_3^{kc} + [N_c(N_c+2N_f)-2N_f+8] \mathcal{D}_4^{kc} + \frac{(N_c+N_f)(7N_f-4)}{N_f} \mathcal{D}_5^{kc} + \frac{3 N_f^2-4N_f-4}{N_f} \mathcal{D}_6^{kc},
\end{eqnarray}

\begin{eqnarray}
&  & [\mathcal{D}_2^{ia},[\mathcal{D}_3^{ia},\mathcal{O}_3^{kc}]] + [\mathcal{D}_3^{ia},[\mathcal{D}_2^{ia},\mathcal{O}_3^{kc}]] + [\mathcal{D}_3^{ia},[\mathcal{O}_3^{ia},\mathcal{D}_2^{kc}]] + [\mathcal{O}_3^{ia},[\mathcal{D}_3^{ia},\mathcal{D}_2^{kc}]] + [\mathcal{D}_2^{ia},[\mathcal{O}_3^{ia},\mathcal{D}_3^{kc}]] + [\mathcal{O}_3^{ia},[\mathcal{D}_2^{ia},\mathcal{D}_3^{kc}]] \nonumber \\
&  & \mbox{\hglue0.2truecm} = - 12(N_c+N_f) \mathcal{O}_3^{kc} + \frac{2(N_c+N_f)(3N_f-4)}{N_f} \mathcal{O}_5^{kc},
\end{eqnarray}

\begin{eqnarray}
&  & [\mathcal{D}_2^{ia},[\mathcal{O}_3^{ia},\mathcal{O}_3^{kc}]] + [\mathcal{O}_3^{ia},[\mathcal{D}_2^{ia},\mathcal{O}_3^{kc}]] + [\mathcal{O}_3^{ia},[\mathcal{O}_3^{ia},\mathcal{D}_2^{kc}]] \nonumber \\
&  & \mbox{\hglue0.2truecm} = - \frac32 [N_c(N_c+2N_f)-8N_f] \mathcal{D}_2^{kc} - 6(N_c+N_f) \mathcal{D}_3^{kc} - \frac14 [5 N_c(N_c+2N_f)-58 N_f-48] \mathcal{D}_4^{kc} - \frac{11}{4} (N_c+N_f) \mathcal{D}_5^{kc} \nonumber \\
&  & \mbox{\hglue0.6truecm} + \frac12(3N_f+14) \mathcal{D}_6^{kc} ,
\end{eqnarray}

\begin{eqnarray}
[\mathcal{D}_3^{ia},[\mathcal{D}_3^{ia},\mathcal{D}_3^{kc}]] & = & - 2[N_c(N_c+2N_f)+2N_f] \mathcal{D}_3^{kc} + 4 (N_c+N_f) \mathcal{D}_4^{kc} + 2[N_c(N_c+2N_f)+2N_f-2] \mathcal{D}_5^{kc} \nonumber \\
&  & \mbox{} - 2 (N_c+N_f) \mathcal{D}_6^{kc} + \frac{N_f^2+2N_f-4}{N_f} \mathcal{D}_7^{kc},
\end{eqnarray}

\begin{eqnarray}
&  & [\mathcal{D}_3^{ia},[\mathcal{D}_3^{ia},\mathcal{O}_3^{kc}]] + [\mathcal{O}_3^{ia},[\mathcal{D}_3^{ia},\mathcal{D}_3^{kc}]] + [\mathcal{D}_3^{ia},[\mathcal{O}_3^{ia},\mathcal{D}_3^{kc}]] \nonumber \\
&  & \mbox{\hglue0.2truecm} = -6[N_c(N_c+2N_f)+2N_f] \mathcal{O}_3^{kc} + \frac{[N_cN_f(N_c+2N_f)+12N_f^2-24N_f+8]}{N_f} \mathcal{O}_5^{kc} 
+ \frac{N_f^2+8 N_f-20}{N_f} \mathcal{O}_7^{kc},
\end{eqnarray}

\begin{eqnarray}
&  & [\mathcal{D}_3^{ia},[\mathcal{O}_3^{ia},\mathcal{O}_3^{kc}]] + [\mathcal{O}_3^{ia},[\mathcal{D}_3^{ia},\mathcal{O}_3^{kc}]] + [\mathcal{O}_3^{ia},[\mathcal{O}_3^{ia},\mathcal{D}_3^{kc}]] \nonumber \\
&  & \mbox{\hglue0.2truecm} = - 24(N_c+N_f) \mathcal{D}_2^{kc} - \frac32 [5 N_c(N_c+2N_f)-8N_f] \mathcal{D}_3^{kc} - 26(N_c+N_f) \mathcal{D}_4^{kc} - 2 [2N_c(N_c+2N_f)-11N_f-6] \mathcal{D}_5^{kc} \nonumber \\
&  & \mbox{\hglue0.6truecm} - \frac72 (N_c+N_f) \mathcal{D}_6^{kc} + \frac12 (5 N_f+17) \mathcal{D}_7^{kc}, 
\end{eqnarray}

\begin{eqnarray}
[\mathcal{O}_3^{ia},[\mathcal{O}_3^{ia},\mathcal{O}_3^{kc}]] = -\frac32 N_c(N_c+2N_f) \mathcal{O}_3^{kc} - \frac14 [9N_c(N_c+2N_f)-34N_f-12] \mathcal{O}_5^{kc} +\frac52 (N_f+2) \mathcal{O}_7^{kc}.
\end{eqnarray}

\section{\label{sec:r8}Reduction of flavor $\mathbf{8}$ operators}

Here, the complete reduction of the operator structure
\begin{equation*}
d^{ab8} [A^{ia},[A^{ib},A^{kc}]]
\end{equation*}
at the physical value $N_c=3$ is provided. Individual structures read
\begin{equation}
d^{ab8} [G^{ia},[G^{ib},G^{kc}]] = \frac{3N_f^2-16}{8 N_f} d^{c8e} G^{ke} + \frac{N_f^2-4}{2N_f^2} \delta^{c8} J^k,
\end{equation}

\begin{eqnarray}
&  & d^{ab8} ([G^{ia},[G^{ib},\mathcal{D}_2^{kc}]] + [\mathcal{D}_2^{ia},[G^{ib},G^{kc}]] + [G^{ia},[\mathcal{D}_2^{ib},G^{kc}]]) \nonumber \\
&  & \mbox{\hglue0.2truecm} = - \frac{2(N_c+N_f)}{N_f} d^{c8e} G^{ke} + \frac{(N_c+N_f)(N_f-2)}{N_f^2} \delta^{c8} J^k + \frac18 (5N_f+8) d^{c8e} \mathcal{D}_2^{ke} - \frac{2}{N_f} \{G^{kc},T^8\} \nonumber \\
&  & \mbox{\hglue0.6truecm} + \frac{N_f^2+2N_f-4}{2N_f} \{G^{k8},T^c\} + \frac{N_f+2}{4} [J^2,[T^8,G^{kc}]],
\end{eqnarray}

\begin{eqnarray}
&  & d^{ab8} ([G^{ia},[G^{ib},\mathcal{D}_3^{kc}]] + [\mathcal{D}_3^{ia},[G^{ib},G^{kc}]] + [G^{ia},[\mathcal{D}_3^{ib},G^{kc}]]) \nonumber \\
&  & \mbox{\hglue0.2truecm} = (N_f-8) d^{c8e} G^{ke} - \frac{3N_c(N_c+2N_f)-8N_f+16}{2N_f} \delta^{c8} J^k - \frac32 (N_c+N_f) d^{c8e} \mathcal{D}_2^{ke} - (N_c+N_f) \{G^{kc},T^8\} \nonumber \\
&  & \mbox{\hglue0.6truecm} + \frac32 (N_c+N_f) [J^2,[T^8,G^{kc}]] + \frac{5N_f^2+12N_f-16}{8N_f} d^{c8e} \mathcal{D}_3^{ke} + \frac{(N_f+6)(N_f-4)}{2N_f} d^{c8e} \mathcal{O}_3^{ke} \nonumber \\
&  & \mbox{\hglue0.6truecm} + \frac{N_f-4}{N_f} \{G^{kc},\{J^r,G^{r8}\}\} + \frac{(N_f+4)(N_f-1)}{N_f} \{G^{k8},\{J^r,G^{rc}\}\} - \frac34 \{J^k,\{T^c,T^8\}\} \nonumber \\
&  & \mbox{\hglue0.6truecm} + (N_f+1) \{J^k,\{G^{r8},G^{rc}\}\} + \frac{(N_f+4)(N_f-1)}{N_f^2} \delta^{c8}\{J^2,J^k\},
\end{eqnarray}

\begin{eqnarray}
&  & d^{ab8} ([G^{ia},[G^{ib},\mathcal{O}_3^{kc}]] + [\mathcal{O}_3^{ia},[G^{ib},G^{kc}]] + [G^{ia},[\mathcal{O}_3^{ib},G^{kc}]]) \nonumber \\
&  & \mbox{\hglue0.2truecm} = \frac{N_f}{2} d^{c8e} G^{ke} - \frac{N_c(N_c+2N_f)}{4N_f} \delta^{c8} J^k - \frac14 (N_c+N_f) d^{c8e} \mathcal{D}_2^{ke} - (N_c+N_f) \{G^{kc},T^8\} - (N_c+N_f) [J^2,[T^8,G^{kc}]] \nonumber \\
&  & \mbox{\hglue0.6truecm} + \frac{N_f^2+N_f-8}{4N_f} d^{c8e} \mathcal{D}_3^{ke} + \frac{7N_f^2+8N_f-16}{8N_f} d^{c8e} \mathcal{O}_3^{ke} + (N_f+2) \{G^{kc},\{J^r,G^{r8}\}\} - \frac12 (N_f+2) \{G^{k8},\{J^r,G^{rc}\}\} \nonumber \\
&  & \mbox{\hglue0.6truecm} - \frac18 \{J^k,\{T^c,T^8\}\} - \frac{N_f^2+N_f-8}{2N_f} \{J^k,\{G^{r8},G^{rc}\}\} + \frac{2N_f^2+N_f-8}{2N_f^2} \delta^{c8}\{J^2,J^k\},
\end{eqnarray}

\begin{eqnarray}
&  & d^{ab8} ([G^{ia},[\mathcal{D}_2^{ib},\mathcal{D}_2^{kc}]] + [\mathcal{D}_2^{ia},[G^{ib},\mathcal{D}_2^{kc}]] + [\mathcal{D}_2^{ia},[\mathcal{D}_2^{ib},G^{kc}]]) \nonumber \\
&  & \mbox{\hglue0.2truecm} = - \frac32 N_f d^{c8e} G^{ke} + \frac{(N_c+N_f)(N_f-4)}{2N_f} \{G^{kc},T^8\} + \frac{(N_c+N_f)(N_f-2)}{N_f} \{G^{k8},T^c\} + \frac14 (N_c+N_f) [J^2,[T^8,G^{kc}]] \nonumber \\
&  & \mbox{\hglue0.6truecm} + \frac38 N_f d^{c8e} \mathcal{D}_3^{ke} + \frac14 (N_f-2) d^{c8e} \mathcal{O}_3^{ke} + \frac12 \{G^{kc},\{J^r,G^{r8}\}\} + \frac12 \{G^{k8},\{J^r,G^{rc}\}\} + \frac{N_f-2}{2N_f} \{J^k,\{T^c,T^8\}\},
\end{eqnarray}

\begin{eqnarray}
&  & d^{ab8} ([G^{ia},[\mathcal{D}_2^{ib},\mathcal{D}_3^{kc}]] + [\mathcal{D}_2^{ia},[G^{ib},\mathcal{D}_3^{kc}]] + [\mathcal{D}_2^{ia},[\mathcal{D}_3^{ib},G^{kc}]] + [\mathcal{D}_3^{ia},[\mathcal{D}_2^{ib},G^{kc}]] + [G^{ia},[\mathcal{D}_3^{ib},\mathcal{D}_2^{kc}]] + [\mathcal{D}_3^{ia},[G^{ib},\mathcal{D}_2^{kc}]]) \nonumber \\
&  &\mbox{\hglue0.2truecm} = - 6(N_c+N_f) d^{c8e} G^{ke} - 3(N_f-2) d^{c8e} \mathcal{D}_2^{ke} -6 \{G^{kc},T^8\} + 2(N_f+1) \{G^{k8},T^c\} + \frac{N_f^2+8}{N_f} [J^2,[T^8,G^{kc}]] \nonumber \\
&  & \mbox{\hglue0.6truecm} + \frac32 (N_c+N_f) d^{c8e} \mathcal{D}_3^{ke} + \frac{(N_c+N_f)(N_f-4)}{N_f} d^{c8e} \mathcal{O}_3^{ke} + \frac{2(N_c+N_f)(N_f-2)}{N_f} \{G^{kc},\{J^r,G^{r8}\}\} \nonumber \\
&  & \mbox{\hglue0.6truecm} + \frac{2(N_c+N_f)(N_f-2)}{N_f} \{G^{k8},\{J^r,G^{rc}\}\} + \frac{N_c+N_f}{2} \{J^k,\{T^c,T^8\}\} + \frac32 (N_f-2) d^{c8e} \mathcal{D}_4^{ke} \nonumber \\
&  & \mbox{\hglue0.6truecm} -(N_f+4) \{\mathcal{D}_2^{kc},\{J^r,G^{r8}\}\} + 4 \{\mathcal{D}_2^{k8},\{J^r,G^{rc}\}\} + \frac{3N_f-8}{N_f} \{J^2,\{G^{kc},T^8\}\} + \frac{N_f^2+3N_f-8}{N_f} \{J^2,\{G^{k8},T^c\}\} \nonumber \\
&  & \mbox{\hglue0.6truecm} + \frac{N_f^2+4N_f-8}{2N_f} \{J^2,[J^2,[T^8,G^{kc}]]\} - \frac18 \{J^2,[G^{kc},\{J^r,G^{r8}\}]\} + \frac18 \{J^2,[G^{k8},\{J^r,G^{rc}\}]\} \nonumber \\
&  & \mbox{\hglue0.6truecm} - \frac18 \{[J^2,G^{kc}],\{J^r,G^{r8}\}\} + \frac18 \{[J^2,G^{k8}],\{J^r,G^{rc}\}\} + \frac18 \{J^k,[\{J^m,G^{mc}\},\{J^r,G^{r8}\}]\},
\end{eqnarray}

\begin{eqnarray}
&  & d^{ab8} [G^{ia},[\mathcal{D}_2^{ib},\mathcal{O}_3^{kc}]] + [\mathcal{D}_2^{ia},[G^{ib},\mathcal{O}_3^{kc}]] + [\mathcal{D}_2^{ia},[\mathcal{O}_3^{ib},G^{kc}]] + [\mathcal{O}_3^{ia},[\mathcal{D}_2^{ib},G^{kc}]] + [G^{ia},[\mathcal{O}_3^{ib},\mathcal{D}_2^{kc}]] + [\mathcal{O}_3^{ia},[G^{ib},\mathcal{D}_2^{kc}]]) \nonumber \\
&  &\mbox{\hglue0.2truecm} = 6N_f d^{c8e} \mathcal{D}_2^{ke} - \frac{(N_c+N_f)(5N_f+8)}{4N_f} d^{c8e} \mathcal{D}_3^{ke} - \frac{2(N_c+N_f)}{N_f} d^{c8e} \mathcal{O}_3^{ke} - \frac34 (N_c+N_f) \{J^k,\{T^c,T^8\}\} \nonumber \\
&  & \mbox{\hglue0.6truecm} - \frac{2(N_c+N_f)(N_f-2)}{N_f} \{J^k,\{G^{r8},G^{rc}\}\} + \frac{2(N_c+N_f)(N_f-2)}{N_f^2} \delta^{c8}\{J^2,J^k\} + \frac12 (N_f+9) d^{c8e} \mathcal{D}_4^{ke} \nonumber \\
&  & \mbox{\hglue0.6truecm} + \frac{N_f^2+9N_f+4}{2N_f} \{\mathcal{D}_2^{kc},\{J^r,G^{r8}\}\} - \frac{9N_f-4}{2N_f} \{\mathcal{D}_2^{k8},\{J^r,G^{rc}\}\} - \frac{2}{N_f} \{J^2,\{G^{kc},T^8\}\} \nonumber \\
&  & \mbox{\hglue0.6truecm} + \frac{N_f^2+2N_f-4}{2N_f} \{J^2,\{G^{k8},T^c\}\} + \frac14 (N_f+2) \{J^2,[J^2,[T^8,G^{kc}]]\} - \frac{15}{32} \{J^2,[G^{kc},\{J^r,G^{r8}\}]\} \nonumber \\
&  & \mbox{\hglue0.6truecm} + \frac{15}{32} \{J^2,[G^{k8},\{J^r,G^{rc}\}]\} - \frac{15}{32} \{[J^2,G^{kc}],\{J^r,G^{r8}\}\} + \frac{15}{32} \{[J^2,G^{k8}],\{J^r,G^{rc}\}\} \nonumber \\
&  & \mbox{\hglue0.6truecm} + \frac{15}{32} \{J^k,[\{J^m,G^{mc}\},\{J^r,G^{r8}\}]\},
\end{eqnarray}

\begin{equation}
d^{ab8} [\mathcal{D}_2^{ia},[\mathcal{D}_2^{ib},\mathcal{D}_2^{kc}]] = - \frac{N_f}{2} d^{c8e} \mathcal{D}_2^{ke} + \frac{(N_c+N_f)(N_f-4)}{4N_f} \{J^k,\{T^c,T^8\}\} + \frac{N_f}{4} d^{c8e} \mathcal{D}_4^{ke} + \{\mathcal{D}_2^{kc},\{J^r,G^{r8}\}\},
\end{equation}

\begin{eqnarray}
&  & d^{ab8} ([G^{ia},[\mathcal{D}_3^{ib},\mathcal{D}_3^{kc}]] + [\mathcal{D}_3^{ia},[G^{ib},\mathcal{D}_3^{kc}]] + [\mathcal{D}_3^{ia},[\mathcal{D}_3^{ib},G^{kc}]]) \nonumber \\
&  &\mbox{\hglue0.2truecm} = - 6N_f d^{c8e} G^{ke} + \frac{3N_c(N_c+2N_f)}{N_f} \delta^{c8} J^k + 3(N_c+N_f) d^{c8e} \mathcal{D}_2^{ke} - 6(N_c+N_f) \{G^{kc},T^8\} \nonumber \\
&  & \mbox{\hglue0.6truecm} + 2(N_c+N_f) [J^2,[T^8,G^{kc}]] - 3 d^{c8e} \mathcal{D}_3^{ke} + (N_f-14) d^{c8e} \mathcal{O}_3^{ke} + \frac{5N_f^2-10N_f+16}{N_f} \{G^{kc},\{J^r,G^{r8}\}\} \nonumber \\
&  & \mbox{\hglue0.6truecm} + \frac{7N_f^2-2N_f-16}{N_f} \{G^{k8},\{J^r,G^{rc}\}\} + \frac32 \{J^k,\{T^c,T^8\}\} - 6(N_f-1) \{J^k,\{G^{r8},G^{rc}\}\} \nonumber \\
&  & \mbox{\hglue0.6truecm} - \frac{3[N_c(N_c+2N_f)+4]}{2N_f} \delta^{c8}\{J^2,J^k\} - \frac32 (N_c+N_f) d^{c8e} \mathcal{D}_4^{ke} + 5(N_c+N_f) \{\mathcal{D}_2^{k8},\{J^r,G^{rc}\}\} \nonumber \\
&  & \mbox{\hglue0.6truecm} + (N_c+N_f) \{J^2,\{G^{kc},T^8\}\} + 2(N_c+N_f) \{J^2,[J^2,[T^8,G^{kc}]]\} + \frac{9N_f-8}{8N_f} \{J^2,[G^{kc},\{J^r,G^{r8}\}]\} \nonumber \\
&  & \mbox{\hglue0.6truecm} - \frac{9N_f-8}{8N_f} \{J^2,[G^{k8},\{J^r,G^{rc}\}]\} + \frac{9N_f-8}{8N_f} \{[J^2,G^{kc}],\{J^r,G^{r8}\}\} - \frac{9N_f-8}{8N_f} \{[J^2,G^{k8}],\{J^r,G^{rc}\}\} \nonumber \\
&  & \mbox{\hglue0.6truecm} - \frac{9N_f-8}{8N_f} \{J^k,[\{J^m,G^{mc}\},\{J^r,G^{r8}\}]\} + \frac34 (N_f+2) d^{c8e} \mathcal{D}_5^{ke} + \frac{(N_f+8)(N_f-4)}{2N_f} d^{c8e} \mathcal{O}_5^{ke} \nonumber \\
&  & \mbox{\hglue0.6truecm} + \frac{6(N_f-4)}{N_f} \{J^2,\{G^{kc},\{J^r,G^{r8}\}\}\} + \frac{N_f^2+6N_f-8}{N_f} \{J^2,\{G^{k8},\{J^r,G^{rc}\}\}\} - \frac34 \{J^2,\{J^k,\{T^c,T^8\}\}\} \nonumber \\
&  & \mbox{\hglue0.6truecm} + 3(N_f-1) \{J^2,\{J^k,\{G^{r8},G^{rc}\}\}\} - \frac{2N_f^2+3N_f-4}{N_f} \{J^k,\{\{J^r,G^{rc}\},\{J^m,G^{m8}\}\}\} + \frac{3}{N_f} \delta^{c8}\{J^2,\{J^2,J^k\}\}, \nonumber \\
\end{eqnarray}

\begin{eqnarray}
&  & d^{ab8} [G^{ia},[\mathcal{D}_3^{ib},\mathcal{O}_3^{kc}]] + [\mathcal{D}_3^{ia},[G^{ib},\mathcal{O}_3^{kc}]] + [\mathcal{D}_3^{ia},[\mathcal{O}_3^{ib},G^{kc}]] + [\mathcal{O}_3^{ia},[\mathcal{D}_3^{ib},G^{kc}]] + [G^{ia},[\mathcal{O}_3^{ib},\mathcal{D}_3^{kc}]] + [\mathcal{O}_3^{ia},[G^{ib},\mathcal{D}_3^{kc}]]) \nonumber \\
&  & \mbox{\hglue0.2truecm} = - \frac{12N_c(N_c+2N_f)}{N_f} \delta^{c8} J^k - 12 (N_c+N_f) d^{c8e} \mathcal{D}_2^{ke} + \frac12 (7N_f-8) d^{c8e} \mathcal{D}_3^{ke} + (N_f-8) d^{c8e} \mathcal{O}_3^{ke} - 6 \{J^k,\{T^c,T^8\}\} \nonumber \\
&  & \mbox{\hglue0.6truecm} + 8 (N_f+1) \{J^k,\{G^{r8},G^{rc}\}\} - \frac{5N_c(N_c+2N_f)-32N_f+16}{2N_f} \delta^{c8}\{J^2,J^k\} - \frac52 (N_c+N_f) d^{c8e} \mathcal{D}_4^{ke} \nonumber \\
&  & \mbox{\hglue0.6truecm} - 11 (N_c+N_f) \{\mathcal{D}_2^{k8},\{J^r,G^{rc}\}\} - (N_c+N_f) \{J^2,\{G^{kc},T^8\}\} + \frac32 (N_c+N_f) \{J^2,[J^2,[T^8,G^{kc}]]\} \nonumber \\
&  & \mbox{\hglue0.6truecm} + \frac{3N_c N_f+18N_f^2-12N_f-116}{32N_f} \{J^2,[G^{kc},\{J^r,G^{r8}\}]\} - \frac{3N_c N_f+18N_f^2-12N_f-116}{32N_f} \{J^2,[G^{k8},\{J^r,G^{rc}\}]\} \nonumber \\
&  & \mbox{\hglue0.6truecm} + \frac{3N_c N_f+18N_f^2-12N_f-116}{32N_f} \{[J^2,G^{kc}],\{J^r,G^{r8}\}\} - \frac{3N_c N_f+18N_f^2-12N_f-116}{32N_f} \{[J^2,G^{k8}],\{J^r,G^{rc}\}\} \nonumber \\
&  & \mbox{\hglue0.6truecm} - \frac{3N_c N_f+18N_f^2-12N_f-116}{32N_f} \{J^k,[\{J^m,G^{mc}\},\{J^r,G^{r8}\}]\} + \frac{N_f^2+5N_f-8}{2N_f} d^{c8e} \mathcal{D}_5^{ke} \nonumber \\
&  & \mbox{\hglue0.6truecm} + \frac{(N_f+6)(N_f-4)}{2N_f} d^{c8e} \mathcal{O}_5^{ke} + \frac{N_f-4}{N_f} \{J^2,\{G^{kc},\{J^r,G^{r8}\}\}\} + \frac{(N_f+4)(N_f-1)}{N_f} \{J^2,\{G^{k8},\{J^r,G^{rc}\}\}\} \nonumber \\
&  & \mbox{\hglue0.6truecm} - \frac54 \{J^2,\{J^k,\{T^c,T^8\}\}\} - \frac{N_f^2-3N_f-8}{N_f} \{J^2,\{J^k,\{G^{r8},G^{rc}\}\}\} \nonumber \\
&  & \mbox{\hglue0.6truecm} + \frac{2N_f^2+5N_f+4}{N_f} \{J^k,\{\{J^r,G^{rc}\},\{J^m,G^{m8}\}\}\} + \frac{2N_f^2+5N_f-8}{N_f^2} \delta^{c8}\{J^2,\{J^2,J^k\}\},
\end{eqnarray}

\begin{eqnarray}
&  & d^{ab8} [G^{ia},[\mathcal{O}_3^{ib},\mathcal{O}_3^{kc}]] + [\mathcal{O}_3^{ia},[G^{ib},\mathcal{O}_3^{kc}]] + [\mathcal{O}_3^{ia},[\mathcal{O}_3^{ib},G^{kc}]]) \nonumber \\
&  & \mbox{\hglue0.2truecm} = - \frac32 (N_c+N_f) \{G^{kc},T^8\} - 2(N_c+N_f) [J^2,[T^8,G^{kc}]] + \frac{N_f^2-4}{2N_f} d^{c8e} \mathcal{D}_3^{ke} + \frac12 (4N_f+3) d^{c8e} \mathcal{O}_3^{ke} \nonumber \\
&  & \mbox{\hglue0.6truecm} + \frac14 (11N_f+6) \{G^{kc},\{J^r,G^{r8}\}\} - \frac14 (5N_f+6) \{G^{k8},\{J^r,G^{rc}\}\} - \frac{N_f^2-4}{N_f} \{J^k,\{G^{r8},G^{rc}\}\} \nonumber \\
&  & \mbox{\hglue0.6truecm} - \frac{3N_cN_f(N_c+2N_f)-8N_f^2+32}{8N_f^2} \delta^{c8} \{J^2,J^k\} - \frac38 (N_c+N_f) d^{c8e} \mathcal{D}_4^{ke} + \frac{11}{4} (N_c+N_f) \{\mathcal{D}_2^{k8},\{J^r,G^{rc}\}\} \nonumber \\
&  & \mbox{\hglue0.6truecm} - \frac{13}{4} (N_c+N_f) \{J^2,\{G^{kc},T^8\}\} - \frac74 (N_c+N_f) \{J^2,[J^2,[T^8,G^{kc}]]\} \nonumber \\
&  & \mbox{\hglue0.6truecm} - \frac{3 [N_c N_f+6N_f^2+10N_f-32]}{64 N_f} \{J^2,[G^{kc},\{J^r,G^{r8}\}]\} + \frac{3[N_cN_f+6N_f^2+10N_f-32]}{64N_f} \{J^2,[G^{k8},\{J^r,G^{rc}\}]\} \nonumber \\
&  & \mbox{\hglue0.6truecm} - \frac{3[N_cN_f+6N_f^2+10N_f-32]}{64N_f} \{[J^2,G^{kc}],\{J^r,G^{r8}\}\} + \frac{3[N_cN_f+6N_f^2+10N_f-32]}{64N_f} \{[J^2,G^{k8}],\{J^r,G^{rc}\}\} \nonumber \\
&  & \mbox{\hglue0.6truecm} + \frac{3[N_cN_f+6N_f^2+10N_f-32]}{64N_f} \{J^k,[\{J^m,G^{mc}\},\{J^r,G^{r8}\}]\} + \frac{N_f^2+3N_f-8}{8N_f} d^{c8e} \mathcal{D}_5^{ke} + \frac14 (3N_f+8) d^{c8e} \mathcal{O}_5^{ke} \nonumber \\
&  & \mbox{\hglue0.6truecm} + 2 (N_f+3) \{J^2,\{G^{kc},\{J^r,G^{r8}\}\}\} - \frac14 (3N_f+8) \{J^2,\{G^{k8},\{J^r,G^{rc}\}\}\} - \frac{3}{16} \{J^2,\{J^k,\{T^c,T^8\}\}\} \nonumber \\
&  & \mbox{\hglue0.6truecm} - \frac{N_f^2+3N_f-16}{4N_f} \{J^2,\{J^k,\{G^{r8},G^{rc}\}\}\} - \frac{2N_f^2+5N_f+4}{4N_f} \{J^k,\{\{J^r,G^{rc}\},\{J^m,G^{m8}\}\}\} \nonumber \\
&  & \mbox{\hglue0.6truecm} + \frac{2N_f^2+3N_f-8}{4N_f^2} \delta^{c8}\{J^2,\{J^2,J^k\}\},
\end{eqnarray}

\begin{eqnarray}
&  & d^{ab8} ([\mathcal{D}_2^{ia},[\mathcal{D}_2^{ib},\mathcal{D}_3^{kc}]] + [\mathcal{D}_2^{ia},[\mathcal{D}_3^{ib},\mathcal{D}_2^{kc}]] + [\mathcal{D}_3^{ia},[\mathcal{D}_2^{ib},\mathcal{D}_2^{kc}]]) \nonumber \\
&  & \mbox{\hglue0.2truecm} = - \frac32 N_f d^{c8e} \mathcal{D}_3^{ke} + \frac{2(N_c+N_f)(N_f-2)}{N_f} \{\mathcal{D}_2^{kc},\{J^r,G^{r8}\}\} + \frac{(N_c+N_f)(N_f-4)}{N_f} \{\mathcal{D}_2^{k8},\{J^r,G^{rc}\}\} \nonumber \\
&  & \mbox{\hglue0.6truecm} + \frac14 \{J^2,[G^{kc},\{J^r,G^{r8}\}]\} - \frac14 \{J^2,[G^{k8},\{J^r,G^{rc}\}]\} + \frac14 \{[J^2,G^{kc}],\{J^r,G^{r8}\}\} - \frac14 \{[J^2,G^{k8}],\{J^r,G^{rc}\}\} \nonumber \\
&  & \mbox{\hglue0.6truecm} - \frac14 \{J^k,[\{J^m,G^{mc}\},\{J^r,G^{r8}\}]\} + \frac34 N_f d^{c8e} \mathcal{D}_5^{ke} + \frac{N_f-2}{N_f} \{J^2,\{J^k,\{T^c,T^8\}\}\} \nonumber \\
&  & \mbox{\hglue0.6truecm} + \{J^k,\{\{J^r,G^{rc}\},\{J^m,G^{m8}\}\}\},
\end{eqnarray}

\begin{eqnarray}
&  & d^{ab8} ([\mathcal{D}_2^{ia},[\mathcal{D}_2^{ib},\mathcal{O}_3^{kc}]] + [\mathcal{D}_2^{ia},[\mathcal{O}_3^{ib},\mathcal{D}_2^{kc}]] + [\mathcal{O}_3^{ia},[\mathcal{D}_2^{ib},\mathcal{D}_2^{kc}]]) \nonumber \\
&  & \mbox{\hglue0.2truecm} = - \frac32 N_f d^{c8e} \mathcal{O}_3^{ke} - \frac{(N_c+N_f)(N_f-2)}{N_f} \{\mathcal{D}_2^{kc},\{J^r,G^{r8}\}\} - \frac{(N_c+N_f)(N_f-4)}{2N_f} \{\mathcal{D}_2^{k8},\{J^r,G^{rc}\}\} \nonumber \\
&  & \mbox{\hglue0.6truecm} + \frac{(N_c+N_f)(N_f-4)}{2N_f} \{J^2,\{G^{kc},T^8\}\} + \frac{(N_c+N_f)(N_f-2)}{N_f} \{J^2,\{G^{k8},T^c\}\} + \frac14 (N_c+N_f) \{J^2,[J^2,[T^8,G^{kc}]]\} \nonumber \\
&  & \mbox{\hglue0.6truecm} - \frac{7}{16} \{J^2,[G^{kc},\{J^r,G^{r8}\}]\} + \frac{7}{16} \{J^2,[G^{k8},\{J^r,G^{rc}\}]\} - \frac{7}{16} \{[J^2,G^{kc}],\{J^r,G^{r8}\}\} \nonumber \\
&  & \mbox{\hglue0.6truecm} + \frac{7}{16} \{[J^2,G^{k8}],\{J^r,G^{rc}\}\} + \frac{7}{16} \{J^k,[\{J^m,G^{mc}\},\{J^r,G^{r8}\}]\} + \frac14 (N_f-2) d^{c8e} \mathcal{O}_5^{ke} \nonumber \\
&  & \mbox{\hglue0.6truecm} + \frac12 \{J^2,\{G^{kc},\{J^r,G^{r8}\}\}\} + \frac12 \{J^2,\{G^{k8},\{J^r,G^{rc}\}\}\} - \frac12 \{J^k,\{\{J^r,G^{rc}\},\{J^m,G^{m8}\}\}\},
\end{eqnarray}

\begin{eqnarray}
&  & d^{ab8} ([\mathcal{D}_2^{ia},[\mathcal{D}_3^{ib},\mathcal{D}_3^{kc}]] + [\mathcal{D}_3^{ia},[\mathcal{D}_2^{ib},\mathcal{D}_3^{kc}]] + [\mathcal{D}_3^{ia},[\mathcal{D}_3^{ib},\mathcal{D}_2^{kc}]]) \nonumber \\
&  & \mbox{\hglue0.2truecm} = - 3 (N_c+N_f) d^{c8e} \mathcal{D}_3^{ke} - 3 (N_f-2) d^{c8e} \mathcal{D}_4^{ke} + 2 (N_f+1) \{\mathcal{D}_2^{kc},\{J^r,G^{r8}\}\} - 6 \{\mathcal{D}_2^{k8},\{J^r,G^{rc}\}\} \nonumber \\
&  & \mbox{\hglue0.6truecm} + \frac{N_c(3N_f-4)+4 N_f^2-6N_f-4}{8N_f} \{J^2,[G^{kc},\{J^r,G^{r8}\}]\} - \frac{N_c(3N_f-4)+4 N_f^2-6N_f-4}{8N_f} \{J^2,[G^{k8},\{J^r,G^{rc}\}]\} \nonumber \\
&  & \mbox{\hglue0.6truecm} + \frac{N_c(3N_f-4)+4 N_f^2-6N_f-4}{8N_f} \{[J^2,G^{kc}],\{J^r,G^{r8}\}\} - \frac{N_c(3N_f-4)+4 N_f^2-6N_f-4}{8N_f} \{[J^2,G^{k8}],\{J^r,G^{rc}\}\} \nonumber \\
&  & \mbox{\hglue0.6truecm} - \frac{N_c(3N_f-4)+4 N_f^2-6N_f-4}{8N_f} \{J^k,[\{J^m,G^{mc}\},\{J^r,G^{r8}\}]\} + \frac32 (N_c+N_f) d^{c8e} \mathcal{D}_5^{ke} \nonumber \\
&  & \mbox{\hglue0.6truecm} + \frac12 (N_c+N_f) \{J^2,\{J^k,\{T^c,T^8\}\}\} + \frac{2(N_c+N_f)(N_f-2)}{N_f} \{J^k,\{\{J^r,G^{rc}\},\{J^m,G^{m8}\}\}\} \nonumber \\
&  & \mbox{\hglue0.6truecm} + \frac32 (N_f-2) d^{c8e} \mathcal{D}_6^{ke} - \frac{N_f+8}{N_f} \{J^2,\{\mathcal{D}_2^{kc},\{J^r,G^{r8}\}\}\} + \frac{7N_f-8}{N_f} \{J^2,\{\mathcal{D}_2^{k8},\{J^r,G^{rc}\}\}\},
\end{eqnarray}

\begin{eqnarray}
&  & d^{ab8} ([\mathcal{D}_2^{ia},[\mathcal{D}_3^{ib},\mathcal{O}_3^{kc}]] + [\mathcal{D}_3^{ia},[\mathcal{D}_2^{ib},\mathcal{O}_3^{kc}]] + [\mathcal{D}_3^{ia},[\mathcal{O}_3^{ib},\mathcal{D}_2^{kc}]] + [\mathcal{O}_3^{ia},[\mathcal{D}_3^{ib},\mathcal{D}_2^{kc}]] + [\mathcal{D}_2^{ia},[\mathcal{O}_3^{ib},\mathcal{D}_3^{kc}]] + [\mathcal{O}_3^{ia},[\mathcal{D}_2^{ib},\mathcal{D}_3^{kc}]]) \nonumber \\
&  & \mbox{\hglue0.2truecm} = - 6 (N_c+N_f) d^{c8e} \mathcal{O}_3^{ke} - 2 (N_f+1) \{\mathcal{D}_2^{kc},\{J^r,G^{r8}\}\} + 6 \{\mathcal{D}_2^{k8},\{J^r,G^{rc}\}\} - 6 \{J^2,\{G^{kc},T^8\}\} \nonumber \\
&  & \mbox{\hglue0.6truecm} + 2 (N_f+1) \{J^2,\{G^{k8},T^c\}\} + \frac{N_f^2+8}{N_f} \{J^2,[J^2,[T^8,G^{kc}]]\} \nonumber \\
&  & \mbox{\hglue0.6truecm} - \frac{N_c(29N_f-28)+42N_f^2-116N_f-16}{32N_f} \{J^2,[G^{kc},\{J^r,G^{r8}\}]\} \nonumber \\
&  & \mbox{\hglue0.6truecm} + \frac{N_c(29N_f-28)+42N_f^2-116N_f-16}{32N_f} \{J^2,[G^{k8},\{J^r,G^{rc}\}]\} \nonumber \\
&  & \mbox{\hglue0.6truecm} - \frac{N_c(29N_f-28)+42N_f^2-116N_f-16}{32N_f} \{[J^2,G^{kc}],\{J^r,G^{r8}\}\} \nonumber \\
&  & \mbox{\hglue0.6truecm} + \frac{N_c(29N_f-28)+42N_f^2-116N_f-16}{32N_f} \{[J^2,G^{k8}],\{J^r,G^{rc}\}\} \nonumber \\
&  & \mbox{\hglue0.6truecm} + \frac{N_c(29N_f-28)+42N_f^2-116N_f-16}{32N_f} \{J^k,[\{J^m,G^{mc}\},\{J^r,G^{r8}\}]\} \nonumber \\
&  & \mbox{\hglue0.6truecm} + \frac{(N_c+N_f)(N_f-4)}{N_f} d^{c8e} \mathcal{O}_5^{ke} + \frac{2(N_c+N_f)(N_f-2)}{N_f} \{J^2,\{G^{kc},\{J^r,G^{r8}\}\}\} \nonumber \\
&  & \mbox{\hglue0.6truecm} + \frac{2(N_c+N_f)(N_f-2)}{N_f} \{J^2,\{G^{k8},\{J^r,G^{rc}\}\}\} - \frac{2(N_c+N_f)(N_f-2)}{N_f} \{J^k,\{\{J^r,G^{rc}\},\{J^m,G^{m8}\}\}\} \nonumber \\
&  & \mbox{\hglue0.6truecm} - \frac{N_f^2+3N_f-8}{N_f} \{J^2,\{\mathcal{D}_2^{kc},\{J^r,G^{r8}\}\}\} - \frac{3N_f-8}{N_f} \{J^2,\{\mathcal{D}_2^{k8},\{J^r,G^{rc}\}\}\} + \frac{3N_f-8}{N_f} \{J^2,\{J^2,\{G^{kc},T^8\}\}\} \nonumber \\
&  & \mbox{\hglue0.6truecm} + \frac{N_f^2+3N_f-8}{N_f} \{J^2,\{J^2,\{G^{k8},T^c\}\}\} + \frac{N_f^2+4N_f-8}{2N_f} \{J^2,\{J^2,[J^2,[T^8,G^{kc}]]\}\} \nonumber \\
&  & \mbox{\hglue0.6truecm} - \frac14 \{J^2,\{J^2,[G^{kc},\{J^r,G^{r8}\}]\}\} + \frac14 \{J^2,\{J^2,[G^{k8},\{J^r,G^{rc}\}]\}\} - \frac14 \{J^2,\{[J^2,G^{kc}],\{J^r,G^{r8}\}\}\} \nonumber \\
&  & \mbox{\hglue0.6truecm} + \frac14 \{J^2,\{[J^2,G^{k8}],\{J^r,G^{rc}\}\}\} + \frac14 \{J^2,\{J^k,[\{J^m,G^{mc}\},\{J^r,G^{r8}\}]\}\},
\end{eqnarray}

\begin{eqnarray}
&  & d^{ab8} ([\mathcal{D}_2^{ia},[\mathcal{O}_3^{ib},\mathcal{O}_3^{kc}]] + [\mathcal{O}_3^{ia},[\mathcal{D}_2^{ib},\mathcal{O}_3^{kc}]] + [\mathcal{O}_3^{ia},[\mathcal{O}_3^{ib},\mathcal{D}_2^{kc}]]) \nonumber \\
&  & \mbox{\hglue0.2truecm} = 6 N_f d^{c8e} \mathcal{D}_2^{ke} - \frac{2(N_f+1)(N_c+N_f)}{N_f} d^{c8e} \mathcal{D}_3^{ke} - \frac34 (N_c+N_f) \{J^k,\{T^c,T^8\}\} \nonumber \\
&  & \mbox{\hglue0.6truecm} - \frac{2(N_c+N_f)(N_f-2)}{N_f} \{J^k,\{G^{r8},G^{rc}\}\} + \frac{2(N_c+N_f)(N_f-2)}{N_f^2} \delta^{c8}\{J^2,J^k\} + \frac14 (23N_f+24) d^{c8e} \mathcal{D}_4^{ke} \nonumber \\
&  & \mbox{\hglue0.6truecm} + \frac32 (N_f+4) \{\mathcal{D}_2^{kc},\{J^r,G^{r8}\}\} - 6 \{\mathcal{D}_2^{k8},\{J^r,G^{rc}\}\} \nonumber \\
&  & \mbox{\hglue0.6truecm} + \frac{N_c(49N_f+72)-12N_f^2-215N_f-24}{64N_f} \{J^2,[G^{kc},\{J^r,G^{r8}\}]\} \nonumber \\
&  & \mbox{\hglue0.6truecm} - \frac{N_c(49N_f+72)-12N_f^2-215N_f-24}{64N_f} \{J^2,[G^{k8},\{J^r,G^{rc}\}]\} \nonumber \\
&  & \mbox{\hglue0.6truecm} + \frac{N_c(49N_f+72)-12N_f^2-215N_f-24}{64N_f} \{[J^2,G^{kc}],\{J^r,G^{r8}\}\} \nonumber \\
&  & \mbox{\hglue0.6truecm} - \frac{N_c(49N_f+72)-12N_f^2-215N_f-24}{64N_f} \{[J^2,G^{k8}],\{J^r,G^{rc}\}\} \nonumber \\
&  & \mbox{\hglue0.6truecm} - \frac{N_c(49N_f+72)-12N_f^2-215N_f-24}{64N_f} \{J^k,[\{J^m,G^{mc}\},\{J^r,G^{r8}\}]\} - \frac{(N_c+N_f)(7N_f+8)}{8N_f} d^{c8e} \mathcal{D}_5^{ke} \nonumber \\
&  & \mbox{\hglue0.6truecm} - \frac58 (N_c+N_f) \{J^2,\{J^k,\{T^c,T^8\}\}\} - \frac{2(N_c+N_f)(N_f-2)}{N_f} \{J^2,\{J^k,\{G^{r8},G^{rc}\}\}\} \nonumber \\
&  & \mbox{\hglue0.6truecm} + \frac{(N_c+N_f)(N_f-2)}{2N_f} \{J^k,\{\{J^r,G^{rc}\},\{J^m,G^{m8}\}\}\} + \frac{(N_c+N_f)(N_f-2)}{N_f^2} \delta^{c8}\{J^2,\{J^2,J^k\}\} \nonumber \\
&  & \mbox{\hglue0.6truecm} + \frac14 (N_f+11) d^{c8e} \mathcal{D}_6^{ke} + \frac14 (2N_f+17) \{J^2,\{\mathcal{D}_2^{kc},\{J^r,G^{r8}\}\}\} - \frac{11}{4} \{J^2,\{\mathcal{D}_2^{k8},\{J^r,G^{rc}\}\}\} \nonumber \\
&  & \mbox{\hglue0.6truecm} - \frac{11}{32} \{J^2,\{J^2,[G^{kc},\{J^r,G^{r8}\}]\}\} + \frac{11}{32} \{J^2,\{J^2,[G^{k8},\{J^r,G^{rc}\}]\}\} - \frac{11}{32} \{J^2,\{[J^2,G^{kc}],\{J^r,G^{r8}\}\}\} \nonumber \\
&  & \mbox{\hglue0.6truecm} + \frac{11}{32} \{J^2,\{[J^2,G^{k8}],\{J^r,G^{rc}\}\}\} + \frac{11}{32} \{J^2,\{J^k,[\{J^m,G^{mc}\},\{J^r,G^{r8}\}]\}\},
\end{eqnarray}

\begin{eqnarray}
&  & d^{ab8} [\mathcal{D}_3^{ia},[\mathcal{D}_3^{ib},\mathcal{D}_3^{kc}]] \nonumber \\
&  & \mbox{\hglue0.2truecm} = -2 N_f d^{c8e} \mathcal{D}_3^{ke} + \frac{2N_c(N_c+2 N_f)}{N_f} \delta^{c8}\{J^2,J^k\} + 2 (N_c+N_f) d^{c8e} \mathcal{D}_4^{ke} - 4 (N_c+N_f) \{\mathcal{D}_2^{k8},\{J^r,G^{rc}\}\} \nonumber \\
&  & \mbox{\hglue0.6truecm} - \frac{N_c(4N_cN_f+24N_f^2+85N_f-104)-116N_f^2-432N_f+552}{16N_f} \{J^2,[G^{kc},\{J^r,G^{r8}\}]\} \nonumber \\
&  & \mbox{\hglue0.6truecm} + \frac{N_c(4N_cN_f+24N_f^2+85N_f-104)-116N_f^2-432N_f+552}{16N_f} \{J^2,[G^{k8},\{J^r,G^{rc}\}]\} \nonumber \\
&  & \mbox{\hglue0.6truecm} - \frac{N_c(4N_cN_f+24N_f^2+85N_f-104)-116N_f^2-432N_f+552}{16N_f} \{[J^2,G^{kc}],\{J^r,G^{r8}\}\} \nonumber \\
&  & \mbox{\hglue0.6truecm} + \frac{N_c(4N_cN_f+24N_f^2+85N_f-104)-116N_f^2-432N_f+552}{16N_f} \{[J^2,G^{k8}],\{J^r,G^{rc}\}\} \nonumber \\
&  & \mbox{\hglue0.6truecm} + \frac{N_c(4N_cN_f+24N_f^2+85N_f-104)-116N_f^2-432N_f+552}{16N_f} \{J^k,[\{J^m,G^{mc}\},\{J^r,G^{r8}\}]\} \nonumber \\
&  & \mbox{\hglue0.6truecm} - 2 d^{c8e} \mathcal{D}_5^{ke} + \{J^2,\{J^k,\{T^c,T^8\}\}\} -4 (N_f-1) \{J^2,\{J^k,\{G^{r8},G^{rc}\}\}\} \nonumber \\
&  & \mbox{\hglue0.6truecm} + 4 (N_f-1) \{J^k,\{\{J^r,G^{rc}\},\{J^m,G^{m8}\}\}\} - \frac{N_c(N_c+2N_f)+4}{N_f} \delta^{c8}\{J^2,\{J^2,J^k\}\} - (N_c+N_f) d^{c8e} \mathcal{D}_6^{ke} \nonumber \\
&  & \mbox{\hglue0.6truecm} + 4 (N_c+N_f) \{J^2,\{\mathcal{D}_2^{k8},\{J^r,G^{rc}\}\}\} + \frac{3N_cN_f-8}{8N_f} \{J^2,\{J^2,[G^{kc},\{J^r,G^{r8}\}]\}\} \nonumber \\
&  & \mbox{\hglue0.6truecm} - \frac{3N_cN_f-8}{8N_f} \{J^2,\{J^2,[G^{k8},\{J^r,G^{rc}\}]\}\} + \frac{3N_cN_f-8}{8N_f} \{J^2,\{[J^2,G^{kc}],\{J^r,G^{r8}\}\}\} \nonumber \\
&  & \mbox{\hglue0.6truecm} - \frac{3N_cN_f-8}{8N_f} \{J^2,\{[J^2,G^{k8}],\{J^r,G^{rc}\}\}\} - \frac{3N_cN_f-8}{8N_f} \{J^2,\{J^k,[\{J^m,G^{mc}\},\{J^r,G^{r8}\}]\}\} \nonumber \\
&  & \mbox{\hglue0.6truecm} + \frac12 (N_f+2) d^{c8e} \mathcal{D}_7^{ke} - \frac12 \{J^2,\{J^2,\{J^k,\{T^c,T^8\}\}\}\} + 2 (N_f-1) \{J^2,\{J^2,\{J^k,\{G^{r8},G^{rc}\}\}\}\} \nonumber \\
&  & \mbox{\hglue0.6truecm} - \frac{N_f^2-2N_f+8}{N_f} \{J^2,\{J^k,\{\{J^r,G^{rc}\},\{J^m,G^{m8}\}\}\}\} + \frac{2}{N_f} \delta^{c8} \{J^2,\{J^2,\{J^2,J^k\}\}\},
\end{eqnarray}

\begin{eqnarray}
&  & d^{ab8} ([\mathcal{D}_3^{ia},[\mathcal{D}_3^{ib},\mathcal{O}_3^{kc}]] + [\mathcal{O}_3^{ia},[\mathcal{D}_3^{ib},\mathcal{D}_3^{kc}]] + [\mathcal{D}_3^{ia},[\mathcal{O}_3^{ib},\mathcal{D}_3^{kc}]]) \nonumber \\
&  & \mbox{\hglue0.2truecm} = -6 N_f d^{c8e} \mathcal{O}_3^{ke} + 6 (N_c+N_f) \{\mathcal{D}_2^{k8},\{J^r,G^{rc}\}\} - 6 (N_c+N_f) \{J^2,\{G^{kc},T^8\}\} + 2 (N_c+N_f) \{J^2,[J^2,[T^8,G^{kc}]]\} \nonumber \\
&  & \mbox{\hglue0.6truecm} + \frac{26N_c^2N_f+48N_cN_f^2+949N_f-337N_cN_f+728N_c-194N_f^2-1972}{32 N_f} \{J^2,[G^{kc},\{J^r,G^{r8}\}]\} \nonumber \\
&  & \mbox{\hglue0.6truecm} - \frac{26N_c^2N_f+48N_cN_f^2+949N_f-337N_cN_f+728N_c-194N_f^2-1972}{32 N_f} \{J^2,[G^{k8},\{J^r,G^{rc}\}]\} \nonumber \\
&  & \mbox{\hglue0.6truecm} + \frac{26N_c^2N_f+48N_cN_f^2+949N_f-337N_cN_f+728N_c-194N_f^2-1972}{32 N_f} \{[J^2,G^{kc}],\{J^r,G^{r8}\}\} \nonumber \\
&  & \mbox{\hglue0.6truecm} - \frac{26N_c^2N_f+48N_cN_f^2+949N_f-337N_cN_f+728N_c-194N_f^2-1972}{32 N_f} \{[J^2,G^{k8}],\{J^r,G^{rc}\}\} \nonumber \\
&  & \mbox{\hglue0.6truecm} - \frac{26N_c^2N_f+48N_cN_f^2+949N_f-337N_cN_f+728N_c-194N_f^2-1972}{32 N_f} \{J^k,[\{J^m,G^{mc}\},\{J^r,G^{r8}\}]\} \nonumber \\
&  & \mbox{\hglue0.6truecm} + (N_f-14) d^{c8e} \mathcal{O}_5^{ke} + \frac{5N_f^2-10N_f+16}{N_f} \{J^2,\{G^{kc},\{J^r,G^{r8}\}\}\} + \frac{7N_f^2-2N_f-16}{N_f} \{J^2,\{G^{k8},\{J^r,G^{rc}\}\}\} \nonumber \\
&  & \mbox{\hglue0.6truecm} - 6 (N_f-1) \{J^k,\{\{J^r,G^{rc}\},\{J^m,G^{m8}\}\}\} - (N_c+N_f) \{J^2,\{\mathcal{D}_2^{k8},\{J^r,G^{rc}\}\}\} \nonumber \\
&  & \mbox{\hglue0.6truecm} + (N_c+N_f) \{J^2,\{J^2,\{G^{kc},T^8\}\}\} + 2 (N_c+N_f) \{J^2,\{J^2,[J^2,[T^8,G^{kc}]]\}\} \nonumber \\
&  & \mbox{\hglue0.6truecm} - \frac{6N_cN_f+2N_f^2-33N_f-44}{16N_f} \{J^2,\{J^2,[G^{kc},\{J^r,G^{r8}\}]\}\} \nonumber \\
&  & \mbox{\hglue0.6truecm} + \frac{6N_cN_f+2N_f^2-33N_f-44}{16N_f} \{J^2,\{J^2,[G^{k8},\{J^r,G^{rc}\}]\}\} \nonumber \\
&  & \mbox{\hglue0.6truecm} - \frac{6N_cN_f+2N_f^2-33N_f-44}{16N_f} \{J^2,\{[J^2,G^{kc}],\{J^r,G^{r8}\}\}\} \nonumber \\
&  & \mbox{\hglue0.6truecm} + \frac{6N_cN_f+2N_f^2-33N_f-44}{16N_f} \{J^2,\{[J^2,G^{k8}],\{J^r,G^{rc}\}\}\} \nonumber \\
&  & \mbox{\hglue0.6truecm} + \frac{6N_cN_f+2N_f^2-33N_f-44}{16N_f} \{J^2,\{J^k,[\{J^m,G^{mc}\},\{J^r,G^{r8}\}]\}\} \nonumber \\
&  & \mbox{\hglue0.6truecm} + \frac{(N_f+8)(N_f-4)}{2N_f} d^{c8e} \mathcal{O}_7^{ke} + \frac{6(N_f-4)}{N_f} \{J^2,\{J^2,\{G^{kc},\{J^r,G^{r8}\}\}\}\} \nonumber \\
&  & \mbox{\hglue0.6truecm} + \frac{N_f^2+6N_f-8}{N_f} \{J^2,\{J^2,\{G^{k8},\{J^r,G^{rc}\}\}\}\} - \frac{N_f^2+12N_f-32}{2N_f} \{J^2,\{J^k,\{\{J^r,G^{rc}\},\{J^m,G^{m8}\}\}\}\},
\end{eqnarray}

\begin{eqnarray}
&  & d^{ab8} ([\mathcal{D}_3^{ia},[\mathcal{O}_3^{ib},\mathcal{O}_3^{kc}]] + [\mathcal{O}_3^{ia},[\mathcal{D}_3^{ib},\mathcal{O}_3^{kc}]] + [\mathcal{O}_3^{ia},[\mathcal{O}_3^{ib},\mathcal{D}_3^{kc}]]) \nonumber \\
&  & \mbox{\hglue0.2truecm} = - \frac{12N_c(N_c+2N_f)}{N_f} \delta^{c8} J^k - 12(N_c+N_f) d^{c8e} \mathcal{D}_2^{ke} + 2(N_f-2) d^{c8e} \mathcal{D}_3^{ke} - 6 \{J^k,\{T^c,T^8\}\} \nonumber \\
&  & \mbox{\hglue0.6truecm} + 8 (N_f+1) \{J^k,\{G^{r8},G^{rc}\}\} - \frac{13N_c(N_c+2N_f)-16N_f+8}{N_f} \delta^{c8}\{J^2,J^k\} \nonumber \\
&  & \mbox{\hglue0.6truecm} - 13 (N_c+N_f) d^{c8e} \mathcal{D}_4^{ke} - 15 (N_c+N_f) \{\mathcal{D}_2^{k8},\{J^r,G^{rc}\}\} \nonumber \\
&  & \mbox{\hglue0.6truecm} + \frac{N_c(69N_cN_f-40N_f^2-695N_f-1032)+222N_f^2+1983N_f+4092}{64N_f} \{J^2,[G^{kc},\{J^r,G^{r8}\}]\} \nonumber \\
&  & \mbox{\hglue0.6truecm} - \frac{N_c(69N_cN_f-40N_f^2-695N_f-1032)+222N_f^2+1983N_f+4092}{64N_f} \{J^2,[G^{k8},\{J^r,G^{rc}\}]\} \nonumber \\
&  & \mbox{\hglue0.6truecm} + \frac{N_c(69N_cN_f-40N_f^2-695N_f-1032)+222N_f^2+1983N_f+4092}{64N_f} \{[J^2,G^{kc}],\{J^r,G^{r8}\}\} \nonumber \\
&  & \mbox{\hglue0.6truecm} - \frac{N_c(69N_cN_f-40N_f^2-695N_f-1032)+222N_f^2+1983N_f+4092}{64N_f} \{[J^2,G^{k8}],\{J^r,G^{rc}\}\} \nonumber \\
&  & \mbox{\hglue0.6truecm} - \frac{N_c(69N_cN_f-40N_f^2-695N_f-1032)+222N_f^2+1983N_f+4092}{64N_f} \{J^k,[\{J^m,G^{mc}\},\{J^r,G^{r8}\}]\} \nonumber \\
&  & \mbox{\hglue0.6truecm} + \frac{7N_f^2+2N_f-8}{2N_f} d^{c8e} \mathcal{D}_5^{ke} - \frac{13}{2} \{J^2,\{J^k,\{T^c,T^8\}\}\} + \frac{2(2N_f^2+7N_f+4)}{N_f} \{J^2,\{J^k,\{G^{r8},G^{rc}\}\}\} \nonumber \\
&  & \mbox{\hglue0.6truecm} + \frac12 (11N_f+8) \{J^k,\{\{J^r,G^{rc}\},\{J^m,G^{m8}\}\}\} - \frac{7N_cN_f(N_c +2N_f)-56N_f^2-8N_f+32}{4 N_f^2} \delta^{c8} \{J^2,\{J^2,J^k\}\} \nonumber \\
&  & \mbox{\hglue0.6truecm} - \frac74 (N_c+N_f) d^{c8e} \mathcal{D}_6^{ke} - 8 (N_c+N_f) \{J^2,\{\mathcal{D}_2^{k8},\{J^r,G^{rc}\}\}\} \nonumber \\
&  & \mbox{\hglue0.6truecm} + \frac{6N_cN_f+22N_f^2-5N_f-116}{32N_f} \{J^2,\{J^2,[G^{kc},\{J^r,G^{r8}\}]\}\} \nonumber \\
&  & \mbox{\hglue0.6truecm} - \frac{6N_cN_f+22N_f^2-5N_f-116}{32N_f} \{J^2,\{J^2,[G^{k8},\{J^r,G^{rc}\}]\}\} \nonumber \\
&  & \mbox{\hglue0.6truecm} + \frac{6N_cN_f+22N_f^2-5N_f-116}{32N_f} \{J^2,\{[J^2,G^{kc}],\{J^r,G^{r8}\}\}\} \nonumber \\
&  & \mbox{\hglue0.6truecm} - \frac{6N_cN_f+22N_f^2-5N_f-116}{32N_f} \{J^2,\{[J^2,G^{k8}],\{J^r,G^{rc}\}\}\} \nonumber \\
&  & \mbox{\hglue0.6truecm} - \frac{6N_cN_f+22N_f^2-5N_f-116}{32 N_f} \{J^2,\{J^k,[\{J^m,G^{mc}\},\{J^r,G^{r8}\}]\}\} \nonumber \\
&  & \mbox{\hglue0.6truecm} + \frac{(N_f+8)(N_f-1)}{4N_f} d^{c8e} \mathcal{D}_7^{ke} - \frac78 \{J^2,\{J^2,\{J^k,\{T^c,T^8\}\}\}\} - \frac{N_f^2-N_f-16}{2N_f} \{J^2,\{J^2,\{J^k,\{G^{r8},G^{rc}\}\}\}\} \nonumber \\
&  & \mbox{\hglue0.6truecm} + \frac{5N_f^2+26N_f-8}{4N_f} \{J^2,\{J^k,\{\{J^r,G^{rc}\},\{J^m,G^{m8}\}\}\}\} + \frac{2N_f^2+7N_f-8}{2N_f^2} \delta^{c8} \{J^2,\{J^2,\{J^2,J^k\}\}\},
\end{eqnarray}

\begin{eqnarray}
&  & d^{ab8} [\mathcal{O}_3^{ia},[\mathcal{O}_3^{ib},\mathcal{O}_3^{kc}]] \nonumber \\
&  & \mbox{\hglue0.2truecm} = \frac32 (N_c+N_f) \{\mathcal{D}_2^{k8},\{J^r,G^{rc}\}\} - \frac32 (N_c+N_f) \{J^2,\{G^{kc},T^8\}\} - 2 (N_c+N_f) \{J^2,[J^2,[T^8,G^{kc}]]\} \nonumber \\
&  & \mbox{\hglue0.6truecm} + \frac{N_cN_f(-91 N_c+1433 +16N_f)+560 N_c-4860 N_f-184 N_f^2-2340}{128N_f} \{J^2,[G^{kc},\{J^r,G^{r8}\}]\} \nonumber \\
&  & \mbox{\hglue0.6truecm} - \frac{N_cN_f(-91 N_c+1433 +16N_f)+560 N_c-4860 N_f-184 N_f^2-2340}{128N_f} \{J^2,[G^{k8},\{J^r,G^{rc}\}]\} \nonumber \\
&  & \mbox{\hglue0.6truecm} + \frac{N_cN_f(-91 N_c+1433 +16N_f)+560 N_c-4860 N_f-184 N_f^2-2340}{128N_f} \{[J^2,G^{kc}],\{J^r,G^{r8}\}\} \nonumber \\
&  & \mbox{\hglue0.6truecm} - \frac{N_cN_f(-91 N_c+1433 +16N_f)+560 N_c-4860 N_f-184 N_f^2-2340}{128N_f} \{[J^2,G^{k8}],\{J^r,G^{rc}\}\} \nonumber \\
&  & \mbox{\hglue0.6truecm} - \frac{N_cN_f(-91 N_c+1433 +16N_f)+560 N_c-4860 N_f-184 N_f^2-2340}{128N_f} \{J^k,[\{J^m,G^{mc}\},\{J^r,G^{r8}\}]\} \nonumber \\
&  & \mbox{\hglue0.6truecm} + \frac32 (N_f+1) d^{c8e} \mathcal{O}_5^{ke} + \frac14 (11N_f+6) \{J^2,\{G^{kc},\{J^r,G^{r8}\}\}\} - \frac14 (5N_f+6) \{J^2,\{G^{k8},\{J^r,G^{rc}\}\}\} \nonumber \\
&  & \mbox{\hglue0.6truecm} - \frac34 N_f \{J^k,\{\{J^r,G^{rc}\},\{J^m,G^{m8}\}\}\} + \frac94 (N_c+N_f) \{J^2,\{\mathcal{D}_2^{k8},\{J^r,G^{rc}\}\}\} \nonumber \\
&  & \mbox{\hglue0.6truecm} - \frac94 (N_c+N_f) \{J^2,\{J^2,\{G^{kc},T^8\}\}\} - \frac34 (N_c+N_f) \{J^2,\{J^2,[J^2,[T^8,G^{kc}]]\}\} \nonumber \\
&  & \mbox{\hglue0.6truecm} - \frac{3N_cN_f+20N_f^2+52 N_f-44}{64N_f} \{J^2,\{J^2,[G^{kc},\{J^r,G^{r8}\}]\}\} \nonumber \\
&  & \mbox{\hglue0.6truecm} + \frac{3N_cN_f+20N_f^2+52 N_f-44}{64N_f} \{J^2,\{J^2,[G^{k8},\{J^r,G^{rc}\}]\}\} \nonumber \\
&  & \mbox{\hglue0.6truecm} - \frac{3N_cN_f+20N_f^2+52 N_f-44}{64N_f} \{J^2,\{[J^2,G^{kc}],\{J^r,G^{r8}\}\}\} \nonumber \\
&  & \mbox{\hglue0.6truecm} + \frac{3N_cN_f+20N_f^2+52 N_f-44}{64N_f} \{J^2,\{[J^2,G^{k8}],\{J^r,G^{rc}\}\}\} \nonumber \\
&  & \mbox{\hglue0.6truecm} + \frac{3N_cN_f+20N_f^2+52 N_f-44}{64N_f} \{J^2,\{J^k,[\{J^m,G^{mc}\},\{J^r,G^{r8}\}]\}\} \nonumber \\
&  & \mbox{\hglue0.6truecm} + \frac14 (N_f+4) d^{c8e} \mathcal{O}_7^{ke} + (N_f+4) \{J^2,\{J^2,\{G^{kc},\{J^r,G^{r8}\}\}\}\} \nonumber \\
&  & \mbox{\hglue0.6truecm} - \frac14 (N_f+4) \{J^2,\{J^2,\{G^{k8},\{J^r,G^{rc}\}\}\}\} - \frac38 (N_f+4) \{J^2,\{J^k,\{\{J^r,G^{rc}\},\{J^m,G^{m8}\}\}\}\},
\end{eqnarray}

\section{\label{sec:r27}Reduction of flavor $\mathbf{27}$ operators}

Here, the complete reduction of the operator structure
\begin{equation*}
[A^{i8},[A^{i8},A^{kc}]]
\end{equation*}
at the physical value $N_c=3$ is provided. Individual structures read
\begin{equation}
[G^{i8},[G^{i8},G^{kc}]] = \frac14 f^{c8e} f^{8eg} G^{kg} + \frac12 d^{c8e} d^{8eg} G^{kg} + \frac{1}{N_f} \delta^{c8} G^{k8} + \frac{1}{2 N_f} d^{c88} J^k,
\end{equation}

\begin{eqnarray}
&  & [G^{i8},[G^{i8},\mathcal{D}_2^{kc}]] + [\mathcal{D}_2^{i8},[G^{i8},G^{kc}]] + [G^{i8},[\mathcal{D}_2^{i8},G^{kc}]] \nonumber \\
&  & \mbox{\hglue0.2truecm} = \frac94f^{c8e} f^{8eg} \mathcal{D}_2^{kg} + \frac12 d^{c8e} d^{8eg} \mathcal{D}_2^{kg} - \frac12 d^{ceg} d^{88e} \mathcal{D}_2^{kg} + \frac{3}{N_f} \delta^{c8} \mathcal{D}_2^{k8} + d^{c8e} \{G^{ke},T^8\} + \frac12 d^{88e} \{G^{ke},T^c\} \nonumber \\
&  & \mbox{\hglue0.6truecm} + i f^{c8e} [G^{k8},\{J^r,G^{re}\}],
\end{eqnarray}

\begin{eqnarray}
&  & [G^{i8},[G^{i8},\mathcal{D}_3^{kc}]] + [\mathcal{D}_3^{i8},[G^{i8},G^{kc}]] + [G^{i8},[\mathcal{D}_8^{ia},G^{kc}]] \nonumber \\
&  &\mbox{\hglue0.2truecm} = - \frac32 f^{c8e} f^{8eg} G^{kg} + \frac74 f^{c8e} f^{8eg} \mathcal{D}_3^{kg} + \frac32 d^{c8e} d^{8eg} \mathcal{D}_3^{kg} - d^{ceg} d^{88e} \mathcal{D}_3^{kg} + \frac{3}{N_f} \delta^{c8} \mathcal{D}_3^{k8} + d^{c8e} d^{8eg} \mathcal{O}_3^{kg} + \frac{1}{N_f} d^{c88} \{J^2,J^k\} \nonumber \\
&  & \mbox{\hglue0.6truecm} - 2 \{G^{kc},\{G^{r8},G^{r8}\}\} + 2 \{G^{k8},\{G^{r8},G^{rc}\}\} - 3 d^{c8e} \{J^k,\{G^{re},G^{r8}\}\} + d^{88e} \{J^k,\{G^{rc},G^{re}\}\} \nonumber \\
&  & \mbox{\hglue0.6truecm} + 4 d^{c8e} \{G^{ke},\{J^r,G^{r8}\}\} - d^{c8e} \{G^{k8},\{J^r,G^{re}\}\} + d^{88e} \{G^{ke},\{J^r,G^{rc}\}\} - \frac12 \epsilon^{kim} f^{c8e} \{T^e,\{J^i,G^{m8}\}\},
\end{eqnarray}

\begin{eqnarray}
&  & [G^{i8},[G^{i8},\mathcal{O}_3^{kc}]] + [\mathcal{O}_3^{i8},[G^{i8},G^{kc}]] + [G^{i8},[\mathcal{O}_3^{i8},G^{kc}]] \nonumber \\
&  &\mbox{\hglue0.2truecm} = \frac34 f^{c8e} f^{8eg} G^{kg} + \frac12 d^{c8e} d^{8eg} \mathcal{D}_3^{kg} + \frac{1}{N_f} \delta^{c8} \mathcal{D}_3^{k8} + \frac74 f^{c8e} f^{8eg} \mathcal{O}_3^{kg} + 2 d^{c8e} d^{8eg} \mathcal{O}_3^{kg} - d^{ceg} d^{88e} \mathcal{O}_3^{kg} + \frac{7}{N_f} \delta^{c8} \mathcal{O}_3^{k8} \nonumber \\
&  & \mbox{\hglue0.6truecm} + \frac{1}{N_f} d^{c88} \{J^2,J^k\} - \{G^{kc},\{G^{r8},G^{r8}\}\} - \{G^{k8},\{G^{r8},G^{rc}\}\} - \frac12 d^{c8e} \{J^k,\{G^{re},G^{r8}\}\} - \frac12 d^{88e} \{J^k,\{G^{rc},G^{re}\}\} \nonumber \\
&  & \mbox{\hglue0.6truecm} - d^{c8e} \{G^{ke},\{J^r,G^{r8}\}\} + \frac32 d^{c8e} \{G^{k8},\{J^r,G^{re}\}\} + d^{88e} \{G^{kc},\{J^r,G^{re}\}\} - \frac12 d^{88e} \{G^{ke},\{J^r,G^{rc}\}\} \nonumber \\
&  & \mbox{\hglue0.6truecm} + \frac34 \epsilon^{kim} f^{c8e} \{T^e,\{J^i,G^{m8}\}\},
\end{eqnarray}

\begin{eqnarray}
&  & [G^{i8},[\mathcal{D}_2^{i8},\mathcal{D}_2^{kc}]] + [\mathcal{D}_2^{i8},[G^{i8},\mathcal{D}_2^{kc}]] + [\mathcal{D}_2^{i8},[\mathcal{D}_2^{i8},G^{kc}]] \nonumber \\
&  & \mbox{\hglue0.2truecm} = - 2 f^{c8e} f^{8eg} G^{kg} + \frac34 f^{c8e} f^{8eg} \mathcal{D}_3^{kg} + \frac12 f^{c8e} f^{8eg} \mathcal{O}_3^{kg} + \frac12 \{G^{kc},\{T^8,T^8\}\} + \{G^{k8},\{T^c,T^8\}\} \nonumber \\
&  & \mbox{\hglue0.6truecm} - \frac12 \epsilon^{kim} f^{c8e} \{T^e,\{J^i,G^{m8}\}\},
\end{eqnarray}

\begin{eqnarray}
&  & [G^{i8},[\mathcal{D}_2^{i8},\mathcal{D}_3^{kc}]] + [\mathcal{D}_2^{i8},[G^{i8},\mathcal{D}_3^{kc}]] + [\mathcal{D}_2^{i8},[\mathcal{D}_3^{i8},G^{kc}]] + [\mathcal{D}_3^{i8},[\mathcal{D}_2^{i8},G^{kc}]] + [G^{i8},[\mathcal{D}_3^{i8},\mathcal{D}_2^{kc}]] + [\mathcal{D}_3^{i8},[G^{i8},\mathcal{D}_2^{kc}]] \nonumber \\
&  &\mbox{\hglue0.2truecm} = - 4 i f^{c8e} [G^{ke},\{J^r,G^{r8}\}] + 4 i f^{c8e} [G^{k8},\{J^r,G^{re}\}] + 2 d^{c8e} \{J^2,\{G^{ke},T^8\}\} + d^{88e} \{J^2,\{G^{ke},T^c\}\} \nonumber \\
&  & \mbox{\hglue0.6truecm} - 2 d^{c8e} \{\mathcal{D}_2^{k8},\{J^r,G^{re}\}\} - d^{88e} \{\mathcal{D}_2^{kc},\{J^r,G^{re}\}\} + 2 \{\{J^r,G^{rc}\},\{G^{k8},T^8\}\} + 2 \{\{J^r,G^{r8}\},\{G^{kc},T^8\}\} \nonumber \\
&  & \mbox{\hglue0.6truecm} + 2 \{\{J^r,G^{r8}\},\{G^{k8},T^c\}\} + 2 i f^{c8e} \{J^k,[\{J^m,G^{me}\},\{J^r,G^{r8}\}]\} - 2 i f^{c8e} \{\{J^r,G^{re}\},[J^2,G^{k8}]\} \nonumber \\
&  & \mbox{\hglue0.6truecm} + 2 i f^{c8e} \{J^2,[G^{ke},\{J^r,G^{r8}\}]\},
\end{eqnarray}

\begin{eqnarray}
&  & [G^{i8},[\mathcal{D}_2^{i8},\mathcal{O}_3^{kc}]] + [\mathcal{D}_2^{i8},[G^{i8},\mathcal{O}_3^{kc}]] + [\mathcal{D}_2^{i8},[\mathcal{O}_3^{i8},G^{kc}]] + [\mathcal{O}_3^{i8},[\mathcal{D}_2^{i8},G^{kc}]] + [G^{i8},[\mathcal{O}_3^{i8},\mathcal{D}_2^{kc}]] + [\mathcal{O}_3^{i8},[G^{i8},\mathcal{D}_2^{kc}]] \nonumber \\
&  &\mbox{\hglue0.2truecm} = 9 f^{c8e} f^{8eg} \mathcal{D}_2^{kg} + \frac{11}{2} f^{c8e} f^{8eg} \mathcal{D}_4^{kg} + d^{c8e} d^{8eg} \mathcal{D}_4^{kg} - d^{ceg} d^{88e} \mathcal{D}_4^{kg} + \frac{6}{N_f} \delta^{c8} \mathcal{D}_4^{k8} + d^{c8e} \{J^2,\{G^{ke},T^8\}\} \nonumber \\
&  & \mbox{\hglue0.6truecm} + \frac12 d^{88e} \{J^2,\{G^{ke},T^c\}\} - 2 \{\mathcal{D}_2^{kc},\{G^{r8},G^{r8}\}\} - 4 \{\mathcal{D}_2^{k8},\{G^{r8},G^{rc}\}\} + d^{c8e} \{\mathcal{D}_2^{k8},\{J^r,G^{re}\}\} \nonumber \\
&  & \mbox{\hglue0.6truecm} + \frac12 d^{88e} \{\mathcal{D}_2^{kc},\{J^r,G^{re}\}\} - 2 i f^{c8e} \{J^k,[\{J^m,G^{me}\},\{J^r,G^{r8}\}]\} + i f^{c8e} \{\{J^r,G^{re}\},[J^2,G^{k8}]\} \nonumber \\
&  & \mbox{\hglue0.6truecm} - i f^{c8e} \{\{J^r,G^{r8}\},[J^2,G^{ke}]\} - i f^{c8e} \{J^2,[G^{ke},\{J^r,G^{r8}\}]\} + 2 i f^{c8e} \{J^2,[G^{k8},\{J^r,G^{re}\}]\},
\end{eqnarray}

\begin{equation}
[\mathcal{D}_2^{i8},[\mathcal{D}_2^{i8},\mathcal{D}_2^{kc}]] = - f^{c8e} f^{8eg} \mathcal{D}_2^{kg} + \frac12 f^{c8e} f^{8eg} \mathcal{D}_4^{kg} + \frac12 \{\mathcal{D}_2^{kc},\{T^8,T^8\}\},
\end{equation}

\begin{eqnarray}
&  & [G^{i8},[\mathcal{D}_3^{i8},\mathcal{D}_3^{kc}]] + [\mathcal{D}_3^{i8},[G^{i8},\mathcal{D}_3^{kc}]] + [\mathcal{D}_3^{i8},[\mathcal{D}_3^{i8},G^{kc}]] \nonumber \\
&  &\mbox{\hglue0.2truecm} = - \frac92 f^{c8e} f^{8eg} G^{kg} - \frac14 i \epsilon^{kim} f^{c8e} f^{8eg} \{J^i,G^{mg}\} + \frac34 f^{c8e} f^{8eg} \mathcal{D}_3^{kg} + \frac{N_c}{8} i f^{c8e} d^{8eg} \mathcal{D}_3^{kg} + \frac{N_c}{8} i d^{c8e} f^{8eg} \mathcal{D}_3^{kg} \nonumber \\
&  & \mbox{\hglue0.6truecm} - \frac12 f^{c8e} f^{8eg} \mathcal{O}_3^{kg} + d^{c8e} d^{8eg} \mathcal{O}_3^{kg} - 3 d^{ceg} d^{88e} \mathcal{O}_3^{kg} - 6 \{G^{kc},\{G^{r8},G^{r8}\}\} + 6 \{G^{k8},\{G^{r8},G^{rc}\}\} \nonumber \\
&  & \mbox{\hglue0.6truecm} + 3 d^{c8e} \{J^k,\{G^{re},G^{r8}\}\} - 3 d^{88e} \{J^k,\{G^{rc},G^{re}\}\} - 2 d^{c8e} \{G^{ke},\{J^r,G^{r8}\}\} - d^{c8e} \{G^{k8},\{J^r,G^{re}\}\} \nonumber \\
&  & \mbox{\hglue0.6truecm} + 3 d^{88e} \{G^{kc},\{J^r,G^{re}\}\} + \frac{3(N_f-6)}{4} \epsilon^{kim} f^{c8e} \{T^e,\{J^i,G^{m8}\}\} - \frac14 i f^{c8e} d^{8eg} \mathcal{D}_4^{kg} - \frac{1}{2N_f} i \epsilon^{kim} \delta^{c8} \{J^2,\{J^i,G^{m8}\}\} \nonumber \\
&  & \mbox{\hglue0.6truecm} - \frac12 i f^{c8e} \{\mathcal{D}_2^{ke},\{J^r,G^{r8}\}\} + \frac32 i \epsilon^{kim} \{\{G^{r8},G^{rc}\},\{J^i,G^{m8}\}\} - i \epsilon^{kim} \{\{G^{r8},G^{r8}\},\{J^i,G^{mc}\}\} \nonumber \\
&  & \mbox{\hglue0.6truecm} + i \epsilon^{rim} \{G^{k8},\{J^r,\{G^{ic},G^{m8}\}\}\} - \frac14 i \epsilon^{rim} d^{c8e} \{J^k,\{J^r,\{G^{i8},G^{me}\}\}\} - \frac{3}{16} i \epsilon^{kim} f^{cae} f^{8eb} \{\{J^i,G^{m8}\},\{T^a,T^b\}\} \nonumber \\
&  & \mbox{\hglue0.6truecm} + \frac14 i f^{c8e} \{J^k,[\{J^m,G^{me}\},\{J^r,G^{r8}\}]\} + \frac14 i f^{c8e} \{\{J^r,G^{re}\},[J^2,G^{k8}]\} - \frac14 i f^{c8e} \{\{J^r,G^{r8}\},[J^2,G^{ke}]\} \nonumber \\
&  & \mbox{\hglue0.6truecm} - \frac14 i f^{c8e} \{J^2,[G^{ke},\{J^r,G^{r8}\}]\} + \frac14 i f^{c8e} \{J^2,[G^{k8},\{J^r,G^{re}\}]\} + \frac14 d^{c8e} \{J^2,[G^{ke},\{J^r,G^{r8}\}]\} \nonumber \\
&  & \mbox{\hglue0.6truecm} - \frac14 d^{c8e} \{J^2,[G^{k8},\{J^r,G^{re}\}]\} + \frac18 [G^{kc},\{\{J^m,G^{m8}\},\{J^r,G^{r8}\}\}] - \frac34 [G^{k8},\{\{J^m,G^{m8}\},\{J^r,G^{rc}\}\}] \nonumber \\
&  & \mbox{\hglue0.6truecm} + \frac34 \{\{J^m,G^{mc}\},[G^{k8},\{J^r,G^{r8}\}]\} + \frac14 i \epsilon^{kim} f^{cea} f^{e8b} \{\{J^i,G^{m8}\},\{G^{ra},G^{rb}\}\} + \frac32 f^{c8e} f^{8eg} \mathcal{D}_5^{kg} + d^{c8e} d^{8eg} \mathcal{O}_5^{kg} \nonumber \\
&  & \mbox{\hglue0.6truecm} + 2 \{J^2,\{G^{kc},\{G^{r8},G^{r8}\}\}\} - 2 \{J^2,\{G^{k8},\{G^{r8},G^{rc}\}\}\} - 3 d^{c8e} \{J^2,\{J^k,\{G^{re},G^{r8}\}\}\} \nonumber \\
&  & \mbox{\hglue0.6truecm} + 3 d^{88e} \{J^2,\{J^k,\{G^{rc},G^{re}\}\}\} + 8 d^{c8e} \{J^2,\{G^{ke},\{J^r,G^{r8}\}\}\} - d^{c8e} \{J^2,\{G^{k8},\{J^r,G^{re}\}\}\} \nonumber \\
&  & \mbox{\hglue0.6truecm} + 3 d^{88e} \{J^2,\{G^{ke},\{J^r,G^{rc}\}\}\} - \frac12 \epsilon^{kim} f^{c8e} \{J^2,\{T^e,\{J^i,G^{m8}\}\}\} + 6 \{G^{k8},\{\{J^m,G^{m8}\},\{J^r,G^{rc}\}\}\} \nonumber \\
&  & \mbox{\hglue0.6truecm} + 2 \{J^k,\{\{J^m,G^{mc}\},\{G^{r8},G^{r8}\}\}\} - 2 \{J^k,\{\{J^m,G^{m8}\},\{G^{r8},G^{rc}\}\}\} - 2 d^{c8e} \{\mathcal{D}_3^{ke},\{J^r,G^{r8}\}\} \nonumber \\
&  & \mbox{\hglue0.6truecm} - 3 d^{88e} \{\mathcal{D}_3^{kc},\{J^r,G^{re}\}\} - 2 \epsilon^{kim} f^{ab8} \{\{J^i,G^{m8}\},\{T^a,\{G^{rb},G^{rc}\}\}\} - 2 i \epsilon^{kil} [\{J^i,G^{l8}\},\{\{J^m,G^{m8}\},\{J^r,G^{rc}\}\}], \nonumber \\
\end{eqnarray}

\begin{eqnarray}
&  & [G^{i8},[\mathcal{D}_3^{i8},\mathcal{O}_3^{kc}]] + [\mathcal{D}_3^{i8},[G^{i8},\mathcal{O}_3^{kc}]] + [\mathcal{D}_3^{i8},[\mathcal{O}_3^{i8},G^{kc}]] + [\mathcal{O}_3^{i8},[\mathcal{D}_3^{i8},G^{kc}]] + [G^{i8},[\mathcal{O}_3^{i8},\mathcal{D}_3^{kc}]] + [\mathcal{O}_3^{i8},[G^{i8},\mathcal{D}_3^{kc}]] \nonumber \\
&  &\mbox{\hglue0.2truecm} = - \frac{39}{4} f^{c8e} f^{8eg} G^{kg} + \frac{5}{16} i \epsilon^{kim} f^{c8e} f^{8eg} \{J^i,G^{mg}\} + \frac12 f^{c8e} f^{8eg} \mathcal{D}_3^{kg} + 8 d^{c8e} d^{8eg} \mathcal{D}_3^{kg} - 4 d^{ceg} d^{88e} \mathcal{D}_3^{kg} \nonumber \\
&  & \mbox{\hglue0.6truecm} + \frac98 N_c i f^{c8e} d^{8eg} \mathcal{D}_3^{kg} + \frac98 N_c i d^{c8e} f^{8eg} \mathcal{D}_3^{kg} - \frac52 f^{c8e} f^{8eg} \mathcal{O}_3^{kg} - \frac{13}{2} d^{c8e} d^{8eg} \mathcal{O}_3^{kg} + \frac{13}{2} d^{ceg} d^{88e} \mathcal{O}_3^{kg} + \frac{8}{N_f} d^{c88} \{J^2,J^k\} \nonumber \\
&  & \mbox{\hglue0.6truecm} - 13 \{G^{kc},\{G^{r8},G^{r8}\}\} + 13 \{G^{k8},\{G^{r8},G^{rc}\}\} - \frac{19}{2} d^{c8e} \{J^k,\{G^{re},G^{r8}\}\} + \frac32 d^{88e} \{J^k,\{G^{rc},G^{re}\}\} \nonumber \\
&  & \mbox{\hglue0.6truecm} - 13 d^{c8e} \{G^{ke},\{J^r,G^{r8}\}\} + \frac{13}{2} d^{c8e} \{G^{k8},\{J^r,G^{re}\}\} - \frac{13}{2} d^{88e} \{G^{kc},\{J^r,G^{re}\}\} + 13 d^{88e} \{G^{ke},\{J^r,G^{rc}\}\} \nonumber \\
&  & \mbox{\hglue0.6truecm} - \frac14 (3N_f + 1) \epsilon^{kim} f^{c8e} \{T^e,\{J^i,G^{m8}\}\} - \frac94 i f^{c8e} d^{8eg} \mathcal{D}_4^{kg} - \frac{9}{2N_f} i \epsilon^{kim} \delta^{c8} \{J^2,\{J^i,G^{m8}\}\} \nonumber \\
&  & \mbox{\hglue0.6truecm} + \frac58 i f^{c8e} \{\mathcal{D}_2^{ke},\{J^r,G^{r8}\}\} + \frac{13}{4} i \epsilon^{kim} \{\{G^{r8},G^{rc}\},\{J^i,G^{m8}\}\} + \frac54 i \epsilon^{kim} \{\{G^{r8},G^{r8}\},\{J^i,G^{mc}\}\} \nonumber \\
&  & \mbox{\hglue0.6truecm} - \frac54 i \epsilon^{rim} \{G^{k8},\{J^r,\{G^{ic},G^{m8}\}\}\} - \frac94i \epsilon^{rim} d^{c8e} \{J^k,\{J^r,\{G^{i8},G^{me}\}\}\} - \frac{27}{16} i \epsilon^{kim} f^{cae} f^{8eb} \{\{J^i,G^{m8}\},\{T^a,T^b\}\} \nonumber \\
&  & \mbox{\hglue0.6truecm} - \frac{53}{8} i f^{c8e} \{J^k,[\{J^m,G^{me}\},\{J^r,G^{r8}\}]\} - \frac{53}{8} i f^{c8e} \{\{J^r,G^{re}\},[J^2,G^{k8}]\} + \frac{53}{8} i f^{c8e} \{\{J^r,G^{r8}\},[J^2,G^{ke}]\} \nonumber \\
&  & \mbox{\hglue0.6truecm} + \frac{53}{8} i f^{c8e} \{J^2,[G^{ke},\{J^r,G^{r8}\}]\} - \frac{53}{8} i f^{c8e} \{J^2,[G^{k8},\{J^r,G^{re}\}]\} + \frac94 d^{c8e} \{J^2,[G^{ke},\{J^r,G^{r8}\}]\} \nonumber \\
&  & \mbox{\hglue0.6truecm} - \frac94 d^{c8e} \{J^2,[G^{k8},\{J^r,G^{re}\}]\} - \frac{23}{16} [G^{kc},\{\{J^m,G^{m8}\},\{J^r,G^{r8}\}\}] - \frac{13}{8} [G^{k8},\{\{J^m,G^{m8}\},\{J^r,G^{rc}\}\}] \nonumber \\
&  & \mbox{\hglue0.6truecm} + \frac{13}{8} \{\{J^m,G^{mc}\},[G^{k8},\{J^r,G^{r8}\}]\} + \frac94 i \epsilon^{kim} f^{cea} f^{e8b} \{\{J^i,G^{m8}\},\{G^{ra},G^{rb}\}\} + 2 f^{c8e} f^{8eg} \mathcal{D}_5^{kg} + 3 d^{c8e} d^{8eg} \mathcal{D}_5^{kg} \nonumber \\
&  & \mbox{\hglue0.6truecm} - 2 d^{ceg} d^{88e} \mathcal{D}_5^{kg} + \frac{6}{N_f} \delta^{c8} \mathcal{D}_5^{k8} + d^{c8e} d^{8eg} \mathcal{O}_5^{kg} + \frac{2}{N_f} d^{c88} \{J^2,\{J^2,J^k\}\} - 6 \{J^2,\{G^{kc},\{G^{r8},G^{r8}\}\}\} \nonumber \\
&  & \mbox{\hglue0.6truecm} + 6 \{J^2,\{G^{k8},\{G^{r8},G^{rc}\}\}\} - 5 d^{c8e} \{J^2,\{J^k,\{G^{re},G^{r8}\}\}\} - d^{88e} \{J^2,\{J^k,\{G^{rc},G^{re}\}\}\} \nonumber \\
&  & \mbox{\hglue0.6truecm} - 7 d^{c8e} \{J^2,\{G^{ke},\{J^r,G^{r8}\}\}\} - d^{c8e} \{J^2,\{G^{k8},\{J^r,G^{re}\}\}\} - 10 d^{88e} \{J^2,\{G^{ke},\{J^r,G^{rc}\}\}\} \nonumber \\
&  & \mbox{\hglue0.6truecm} - \frac12 \epsilon^{kim} f^{c8e} \{J^2,\{T^e,\{J^i,G^{m8}\}\}\} + 13 \{G^{kc},\{\{J^m,G^{m8}\},\{J^r,G^{r8}\}\}\} - 13 \{G^{k8},\{\{J^m,G^{m8}\},\{J^r,G^{rc}\}\}\} \nonumber \\
&  & \mbox{\hglue0.6truecm} - 6 \{J^k,\{\{J^m,G^{mc}\},\{G^{r8},G^{r8}\}\}\} + \frac{13}{2} d^{c8e} \{\mathcal{D}_3^{ke},\{J^r,G^{r8}\}\} + \frac{15}{2} d^{88e} \{\mathcal{D}_3^{kc},\{J^r,G^{re}\}\} \nonumber \\
&  & \mbox{\hglue0.6truecm} + 2 \epsilon^{kim} f^{ab8} \{\{J^i,G^{m8}\},\{T^a,\{G^{rb},G^{rc}\}\}\} + 11 i \epsilon^{kil} [\{J^i,G^{l8}\},\{\{J^m,G^{m8}\},\{J^r,G^{rc}\}\}],
\end{eqnarray}

\begin{eqnarray}
&  & [G^{i8},[\mathcal{O}_3^{i8},\mathcal{O}_3^{kc}]] + [\mathcal{O}_3^{i8},[G^{i8},\mathcal{O}_3^{kc}]] + [\mathcal{O}_3^{i8},[\mathcal{O}_3^{i8},G^{kc}]] \nonumber \\
&  &\mbox{\hglue0.2truecm} = \frac{15}{4} f^{c8e} f^{8eg} G^{kg} - \frac{3}{32} i \epsilon^{kim} f^{c8e} f^{8eg} \{J^i,G^{mg}\} + \frac{25}{16} f^{c8e} f^{8eg} \mathcal{D}_3^{kg} + \frac12 d^{c8e} d^{8eg} \mathcal{D}_3^{kg} - \frac{45N_c}{64} i f^{c8e} d^{8eg} \mathcal{D}_3^{kg} \nonumber \\
&  & \mbox{\hglue0.6truecm} - \frac{45N_c}{64} i d^{c8e} f^{8eg} \mathcal{D}_3^{kg} + \frac{1}{N_f} \delta^{c8} \mathcal{D}_3^{k8} + \frac{35}{8} f^{c8e} f^{8eg} \mathcal{O}_3^{kg} + \frac{11}{2} d^{c8e} d^{8eg} \mathcal{O}_3^{kg} - \frac92 d^{ceg} d^{88e} \mathcal{O}_3^{kg} + \frac{12}{N_f} \delta^{c8} \mathcal{O}_3^{k8} \nonumber \\
&  & \mbox{\hglue0.6truecm} + \frac{1}{N_f} d^{c88} \{J^2,J^k\} + 3 \{G^{kc},\{G^{r8},G^{r8}\}\} - 5 \{G^{k8},\{G^{r8},G^{rc}\}\} - \frac52 d^{c8e} \{J^k,\{G^{re},G^{r8}\}\} \nonumber \\
&  & \mbox{\hglue0.6truecm} + \frac32 d^{88e} \{J^k,\{G^{rc},G^{re}\}\} + 2 d^{c8e} \{G^{ke},\{J^r,G^{r8}\}\} + \frac12 d^{c8e} \{G^{k8},\{J^r,G^{re}\}\} + \frac92 d^{88e} \{G^{kc},\{J^r,G^{re}\}\} \nonumber \\
&  & \mbox{\hglue0.6truecm} - 6 d^{88e} \{G^{ke},\{J^r,G^{rc}\}\} + \frac{3N_f + 32}{16} \epsilon^{kim} f^{c8e} \{T^e,\{J^i,G^{m8}\}\} + \frac{45}{32} i f^{c8e} d^{8eg} \mathcal{D}_4^{kg} \nonumber \\
&  & \mbox{\hglue0.6truecm} + \frac{45}{16N_f} i \epsilon^{kim} \delta^{c8} \{J^2,\{J^i,G^{m8}\}\} - \frac{3}{16} i f^{c8e} \{\mathcal{D}_2^{ke},\{J^r,G^{r8}\}\} - \frac{39}{16} i \epsilon^{kim} \{\{G^{r8},G^{rc}\},\{J^i,G^{m8}\}\} \nonumber \\
&  & \mbox{\hglue0.6truecm} - \frac38 i \epsilon^{kim} \{\{G^{r8},G^{r8}\},\{J^i,G^{mc}\}\} + \frac38 i \epsilon^{rim} \{G^{k8},\{J^r,\{G^{ic},G^{m8}\}\}\} + \frac{45}{32} i \epsilon^{rim} d^{c8e} \{J^k,\{J^r,\{G^{i8},G^{me}\}\}\} \nonumber \\
&  & \mbox{\hglue0.6truecm} + \frac{135}{128} i \epsilon^{kim} f^{cae} f^{8eb} \{\{J^i,G^{m8}\},\{T^a,T^b\}\} + \frac{79}{32} i f^{c8e} \{J^k,[\{J^m,G^{me}\},\{J^r,G^{r8}\}]\} \nonumber \\
&  & \mbox{\hglue0.6truecm} + \frac{79}{32} i f^{c8e} \{\{J^r,G^{re}\},[J^2,G^{k8}]\} - \frac{79}{32} i f^{c8e} \{\{J^r,G^{r8}\},[J^2,G^{ke}]\} - \frac{79}{32} i f^{c8e} \{J^2,[G^{ke},\{J^r,G^{r8}\}]\} \nonumber \\
&  & \mbox{\hglue0.6truecm} + \frac{79}{32} i f^{c8e} \{J^2,[G^{k8},\{J^r,G^{re}\}]\} - \frac{45}{32} d^{c8e} \{J^2,[G^{ke},\{J^r,G^{r8}\}]\} + \frac{45}{32} d^{c8e} \{J^2,[G^{k8},\{J^r,G^{re}\}]\} \nonumber \\
&  & \mbox{\hglue0.6truecm} + \frac{51}{64} [G^{kc},\{\{J^m,G^{m8}\},\{J^r,G^{r8}\}\}] + \frac{39}{32} [G^{k8},\{\{J^m,G^{m8}\},\{J^r,G^{rc}\}\}] - \frac{39}{32} \{\{J^m,G^{mc}\},[G^{k8},\{J^r,G^{r8}\}]\} \nonumber \\
&  & \mbox{\hglue0.6truecm} - \frac{45}{32} i \epsilon^{kim} f^{cea} f^{e8b} \{\{J^i,G^{m8}\},\{G^{ra},G^{rb}\}\} + \frac14 d^{c8e} d^{8eg} \mathcal{D}_5^{kg} + \frac{1}{2N_f} \delta^{c8} \mathcal{D}_5^{k8} + \frac{11}{4} f^{c8e} f^{8eg} \mathcal{O}_5^{kg} \nonumber \\
&  & \mbox{\hglue0.6truecm} + \frac{11}{4} d^{c8e} d^{8eg} \mathcal{O}_5^{kg} - 2 d^{ceg} d^{88e} \mathcal{O}_5^{kg} + \frac{11}{N_f} \delta^{c8} \mathcal{O}_5^{k8} + \frac{1}{2N_f} d^{c88} \{J^2,\{J^2,J^k\}\} - \frac72 \{J^2,\{G^{kc},\{G^{r8},G^{r8}\}\}\} \nonumber \\
&  & \mbox{\hglue0.6truecm} - \frac52 \{J^2,\{G^{k8},\{G^{r8},G^{rc}\}\}\} - \frac34 d^{c8e} \{J^2,\{J^k,\{G^{re},G^{r8}\}\}\} - \frac14 d^{88e} \{J^2,\{J^k,\{G^{rc},G^{re}\}\}\} \nonumber \\
&  & \mbox{\hglue0.6truecm} + \frac72 d^{c8e} \{J^2,\{G^{ke},\{J^r,G^{r8}\}\}\} + \frac{11}{4} d^{c8e} \{J^2,\{G^{k8},\{J^r,G^{re}\}\}\} + 2 d^{88e} \{J^2,\{G^{kc},\{J^r,G^{re}\}\}\} \nonumber \\
&  & \mbox{\hglue0.6truecm} + \frac{17}{4} d^{88e} \{J^2,\{G^{ke},\{J^r,G^{rc}\}\}\} + \frac{11}{8} \epsilon^{kim} f^{c8e} \{J^2,\{T^e,\{J^i,G^{m8}\}\}\} - \frac92 \{G^{kc},\{\{J^m,G^{m8}\},\{J^r,G^{r8}\}\}\} \nonumber \\
&  & \mbox{\hglue0.6truecm} + 5 \{G^{k8},\{\{J^m,G^{m8}\},\{J^r,G^{rc}\}\}\} + \frac32 \{J^k,\{\{J^m,G^{mc}\},\{G^{r8},G^{r8}\}\}\} + \frac12 \{J^k,\{\{J^m,G^{m8}\},\{G^{r8},G^{rc}\}\}\} \nonumber \\
&  & \mbox{\hglue0.6truecm} - \frac{11}{4} d^{c8e} \{\mathcal{D}_3^{ke},\{J^r,G^{r8}\}\} - 3 d^{88e} \{\mathcal{D}_3^{kc},\{J^r,G^{re}\}\} - \frac12 \epsilon^{kim} f^{ab8} \{\{J^i,G^{m8}\},\{T^a,\{G^{rb},G^{rc}\}\}\} \nonumber \\
&  & \mbox{\hglue0.6truecm} - 5 i \epsilon^{kil} [\{J^i,G^{l8}\},\{\{J^m,G^{m8}\},\{J^r,G^{rc}\}\}],
\end{eqnarray}

\begin{eqnarray}
&  & [\mathcal{D}_2^{i8},[\mathcal{D}_2^{i8},\mathcal{D}_3^{kc}]] + [\mathcal{D}_2^{i8},[\mathcal{D}_3^{i8},\mathcal{D}_2^{kc}]] + [\mathcal{D}_3^{i8},[\mathcal{D}_2^{i8},\mathcal{D}_2^{kc}]] \nonumber \\
&  &\mbox{\hglue0.2truecm} = - \frac52 f^{c8e} f^{8eg} \mathcal{D}_3^{kg} + \frac32 f^{c8e} f^{8eg} \mathcal{D}_5^{kg} + 2 \{\mathcal{D}_2^{kc},\{T^8,\{J^r,G^{r8}\}\}\} + \{\mathcal{D}_2^{k8},\{T^8,\{J^r,G^{rc}\}\}\},
\end{eqnarray}

\begin{eqnarray}
&  & [\mathcal{D}_2^{i8},[\mathcal{D}_2^{i8},\mathcal{O}_3^{kc}]] + [\mathcal{D}_2^{i8},[\mathcal{O}_3^{i8},\mathcal{D}_2^{kc}]] + [\mathcal{O}_3^{i8},[\mathcal{D}_2^{i8},\mathcal{D}_2^{kc}]] \nonumber \\
&  &\mbox{\hglue0.2truecm} = \frac14 f^{c8e} f^{8eg} \mathcal{D}_3^{kg} - 2 f^{c8e} f^{8eg} \mathcal{O}_3^{kg} + \frac12 f^{c8e} f^{8eg} \mathcal{O}_5^{kg} + \frac12 \{J^2,\{G^{kc},\{T^8,T^8\}\}\} + \{J^2,\{G^{k8},\{T^c,T^8\}\}\} \nonumber \\
&  & \mbox{\hglue0.6truecm} - \frac12 \epsilon^{kim} f^{c8e} \{J^2,\{T^e,\{J^i,G^{m8}\}\}\} - \{\mathcal{D}_2^{kc},\{T^8,\{J^r,G^{r8}\}\}\} - \frac12 \{\mathcal{D}_2^{k8},\{T^8,\{J^r,G^{rc}\}\}\},
\end{eqnarray}

\begin{eqnarray}
&  & [\mathcal{D}_2^{i8},[\mathcal{D}_3^{i8},\mathcal{D}_3^{kc}]] + [\mathcal{D}_3^{i8},[\mathcal{D}_2^{i8},\mathcal{D}_3^{kc}]] + [\mathcal{D}_3^{i8},[\mathcal{D}_3^{i8},\mathcal{D}_2^{kc}]] \nonumber \\
&  &\mbox{\hglue0.2truecm} = - 3 f^{c8e} f^{8eg} G^{kg} - \frac{1}{11} i \epsilon^{kim} f^{c8e} f^{8eg} \{J^i,G^{mg}\} - f^{c8e} f^{8eg} \mathcal{D}_3^{kg} + \frac{21N_c}{88} i f^{c8e} d^{8eg} \mathcal{D}_3^{kg} + \frac{21N_c}{88} i d^{c8e} f^{8eg} \mathcal{D}_3^{kg} \nonumber \\
&  & \mbox{\hglue0.6truecm} - 2 d^{c8e} d^{8eg} \mathcal{O}_3^{kg} + 2 d^{ceg} d^{88e} \mathcal{O}_3^{kg} - 4 \{G^{kc},\{G^{r8},G^{r8}\}\} + 4 \{G^{k8},\{G^{r8},G^{rc}\}\} + 2 d^{c8e} \{J^k,\{G^{re},G^{r8}\}\} \nonumber \\
&  & \mbox{\hglue0.6truecm} - 2 d^{88e} \{J^k,\{G^{rc},G^{re}\}\} - 4 d^{c8e} \{G^{ke},\{J^r,G^{r8}\}\} + 2 d^{c8e} \{G^{k8},\{J^r,G^{re}\}\} - 2 d^{88e} \{G^{kc},\{J^r,G^{re}\}\} \nonumber \\
&  & \mbox{\hglue0.6truecm} + 4 d^{88e} \{G^{ke},\{J^r,G^{rc}\}\} - \epsilon^{kim} f^{c8e} \{T^e,\{J^i,G^{m8}\}\} - \frac{21}{44} i f^{c8e} d^{8eg} \mathcal{D}_4^{kg} - \frac{21}{22N_f} i \epsilon^{kim} \delta^{c8} \{J^2,\{J^i,G^{m8}\}\} \nonumber \\
&  & \mbox{\hglue0.6truecm} - \frac{2}{11} i f^{c8e} \{\mathcal{D}_2^{ke},\{J^r,G^{r8}\}\} + \frac{29}{22} i \epsilon^{kim} \{\{G^{r8},G^{rc}\},\{J^i,G^{m8}\}\} - \frac{4}{11} i \epsilon^{kim} \{\{G^{r8},G^{r8}\},\{J^i,G^{mc}\}\} \nonumber \\
&  & \mbox{\hglue0.6truecm} + \frac{4}{11} i \epsilon^{rim} \{G^{k8},\{J^r,\{G^{ic},G^{m8}\}\}\} - \frac{21}{44} i \epsilon^{rim} d^{c8e} \{J^k,\{J^r,\{G^{i8},G^{me}\}\}\} \nonumber \\
&  & \mbox{\hglue0.6truecm} - \frac{63}{176} i \epsilon^{kim} f^{cae} f^{8eb} \{\{J^i,G^{m8}\},\{T^a,T^b\}\} - \frac{301}{44} i f^{c8e} \{J^k,[\{J^m,G^{me}\},\{J^r,G^{r8}\}]\} \nonumber \\
&  & \mbox{\hglue0.6truecm} - \frac{125}{44} i f^{c8e} \{\{J^r,G^{re}\},[J^2,G^{k8}]\} + \frac{125}{44} i f^{c8e} \{\{J^r,G^{r8}\},[J^2,G^{ke}]\} + \frac{125}{44} i f^{c8e} \{J^2,[G^{ke},\{J^r,G^{r8}\}]\} \nonumber \\
&  & \mbox{\hglue0.6truecm} - \frac{109}{44} i f^{c8e} \{J^2,[G^{k8},\{J^r,G^{re}\}]\} + \frac{21}{44} d^{c8e} \{J^2,[G^{ke},\{J^r,G^{r8}\}]\} - \frac{21}{44} d^{c8e} \{J^2,[G^{k8},\{J^r,G^{re}\}]\} \nonumber \\
&  & \mbox{\hglue0.6truecm} - \frac{13}{88} [G^{kc},\{\{J^m,G^{m8}\},\{J^r,G^{r8}\}\}] - \frac{29}{44} [G^{k8},\{\{J^m,G^{m8}\},\{J^r,G^{rc}\}\}] + \frac{29}{44} \{\{J^m,G^{mc}\},[G^{k8},\{J^r,G^{r8}\}]\} \nonumber \\
&  & \mbox{\hglue0.6truecm} + \frac{21}{44} i \epsilon^{kim} f^{cea} f^{e8b} \{\{J^i,G^{m8}\},\{G^{ra},G^{rb}\}\} - 4 d^{c8e} \{J^2,\{G^{ke},\{J^r,G^{r8}\}\}\} - 4 d^{88e} \{J^2,\{G^{ke},\{J^r,G^{rc}\}\}\} \nonumber \\
&  & \mbox{\hglue0.6truecm} + 4 \{G^{kc},\{\{J^m,G^{m8}\},\{J^r,G^{r8}\}\}\} - 4 \{G^{k8},\{\{J^m,G^{m8}\},\{J^r,G^{rc}\}\}\} + 2 d^{c8e} \{\mathcal{D}_3^{ke},\{J^r,G^{r8}\}\} \nonumber \\
&  & \mbox{\hglue0.6truecm} + 2 d^{88e} \{\mathcal{D}_3^{kc},\{J^r,G^{re}\}\} + 4 i \epsilon^{kil} [\{J^i,G^{l8}\},\{\{J^m,G^{m8}\},\{J^r,G^{rc}\}\}] - \frac{4}{11} d^{c8e} \{J^2,\{J^2,\{G^{ke},T^8\}\}\} \nonumber \\
&  & \mbox{\hglue0.6truecm} + \frac{4}{11} d^{c8e} \{J^2,\{\mathcal{D}_2^{k8},\{J^r,G^{re}\}\}\} + \frac{4}{11} \{J^2,\{\{J^r,G^{rc}\},\{G^{k8},T^8\}\}\} - \frac{4}{11} \{J^2,\{\{J^r,G^{r8}\},\{G^{kc},T^8\}\}\} \nonumber \\
&  & \mbox{\hglue0.6truecm} + 3 i f^{c8e} \{J^2,\{J^k,[\{J^m,G^{me}\},\{J^r,G^{r8}\}]\}\} + \frac{4}{11} i f^{c8e} \{J^2,\{\{J^r,G^{r8}\},[J^2,G^{ke}]\}\} \nonumber \\
&  & \mbox{\hglue0.6truecm} + 2 \{\mathcal{D}_2^{kc},\{\{J^m,G^{m8}\},\{J^r,G^{r8}\}\}\} + 4 \{\mathcal{D}_2^{k8},\{\{J^m,G^{mc}\},\{J^r,G^{r8}\}\}\} \nonumber \\
&  & \mbox{\hglue0.6truecm} - \frac{4}{11} i \epsilon^{kim} [\{T^8,\{J^r,G^{r8}\}\},\{J^2,\{J^i,G^{mc}\}\}],
\end{eqnarray}

\begin{eqnarray}
&  & [\mathcal{D}_2^{i8},[\mathcal{D}_3^{i8},\mathcal{O}_3^{kc}]] + [\mathcal{D}_3^{i8},[\mathcal{D}_2^{i8},\mathcal{O}_3^{kc}]] + [\mathcal{D}_3^{i8},[\mathcal{O}_3^{i8},\mathcal{D}_2^{kc}]] + [\mathcal{O}_3^{i8},[\mathcal{D}_3^{i8},\mathcal{D}_2^{kc}]] + [\mathcal{D}_2^{i8},[\mathcal{O}_3^{i8},\mathcal{D}_3^{kc}]] + [\mathcal{O}_3^{i8},[\mathcal{D}_2^{i8},\mathcal{D}_3^{kc}]] \nonumber \\
&  &\mbox{\hglue0.2truecm} = \frac{21}{2} f^{c8e} f^{8eg} G^{kg} + \frac{105}{176} i \epsilon^{kim} f^{c8e} f^{8eg} \{J^i,G^{mg}\} + \frac72 f^{c8e} f^{8eg} \mathcal{D}_3^{kg} - \frac{191N_c}{352} i f^{c8e} d^{8eg} \mathcal{D}_3^{kg} - \frac{191N_c}{352} i d^{c8e} f^{8eg} \mathcal{D}_3^{kg} \nonumber \\
&  & \mbox{\hglue0.6truecm} + 7 d^{c8e} d^{8eg} \mathcal{O}_3^{kg} - 7 d^{ceg} d^{88e} \mathcal{O}_3^{kg} + 14 \{G^{kc},\{G^{r8},G^{r8}\}\} - 14 \{G^{k8},\{G^{r8},G^{rc}\}\} - 7 d^{c8e} \{J^k,\{G^{re},G^{r8}\}\} \nonumber \\
&  & \mbox{\hglue0.6truecm} + 7 d^{88e} \{J^k,\{G^{rc},G^{re}\}\} + 14 d^{c8e} \{G^{ke},\{J^r,G^{r8}\}\} - 7 d^{c8e} \{G^{k8},\{J^r,G^{re}\}\} + 7 d^{88e} \{G^{kc},\{J^r,G^{re}\}\} \nonumber \\
&  & \mbox{\hglue0.6truecm} - 14 d^{88e} \{G^{ke},\{J^r,G^{rc}\}\} + \frac72 \epsilon^{kim} f^{c8e} \{T^e,\{J^i,G^{m8}\}\} + \frac{191}{176} i f^{c8e} d^{8eg} \mathcal{D}_4^{kg} + \frac{191}{88N_f} i \epsilon^{kim} \delta^{c8} \{J^2,\{J^i,G^{m8}\}\} \nonumber \\
&  & \mbox{\hglue0.6truecm} + \frac{105}{88} i f^{c8e} \{\mathcal{D}_2^{ke},\{J^r,G^{r8}\}\} - \frac{401}{88} i \epsilon^{kim} \{\{G^{r8},G^{rc}\},\{J^i,G^{m8}\}\} + \frac{105}{44} i \epsilon^{kim} \{\{G^{r8},G^{r8}\},\{J^i,G^{mc}\}\} \nonumber \\
&  & \mbox{\hglue0.6truecm} - \frac{105}{44} i \epsilon^{rim} \{G^{k8},\{J^r,\{G^{ic},G^{m8}\}\}\} + \frac{191}{176} i \epsilon^{rim} d^{c8e} \{J^k,\{J^r,\{G^{i8},G^{me}\}\}\} \nonumber \\
&  & \mbox{\hglue0.6truecm} + \frac{573}{704} i \epsilon^{kim} f^{cae} f^{8eb} \{\{J^i,G^{m8}\},\{T^a,T^b\}\} + \frac{2415}{176} i f^{c8e} \{J^k,[\{J^m,G^{me}\},\{J^r,G^{r8}\}]\} \nonumber \\
&  & \mbox{\hglue0.6truecm} + \frac{1711}{176} i f^{c8e} \{\{J^r,G^{re}\},[J^2,G^{k8}]\} - \frac{1711}{176} i f^{c8e} \{\{J^r,G^{r8}\},[J^2,G^{ke}]\} - \frac{2415}{176} i f^{c8e} \{J^2,[G^{ke},\{J^r,G^{r8}\}]\} \nonumber \\
&  & \mbox{\hglue0.6truecm} + \frac{2303}{176} i f^{c8e} \{J^2,[G^{k8},\{J^r,G^{re}\}]\} - \frac{191}{176} d^{c8e} \{J^2,[G^{ke},\{J^r,G^{r8}\}]\} + \frac{191}{176} d^{c8e} \{J^2,[G^{k8},\{J^r,G^{re}\}]\} \nonumber \\
&  & \mbox{\hglue0.6truecm} - \frac{19}{352} [G^{kc},\{\{J^m,G^{m8}\},\{J^r,G^{r8}\}\}] + \frac{401}{176} [G^{k8},\{\{J^m,G^{m8}\},\{J^r,G^{rc}\}\}] - \frac{401}{176}\{\{J^m,G^{mc}\},[G^{k8},\{J^r,G^{r8}\}]\} \nonumber \\
&  & \mbox{\hglue0.6truecm} - \frac{191}{176} i \epsilon^{kim} f^{cea} f^{e8b} \{\{J^i,G^{m8}\},\{G^{ra},G^{rb}\}\} + 14 d^{c8e} \{J^2,\{G^{ke},\{J^r,G^{r8}\}\}\} + 14 d^{88e} \{J^2,\{G^{ke},\{J^r,G^{rc}\}\}\} \nonumber \\
&  & \mbox{\hglue0.6truecm} - 14 \{G^{kc},\{\{J^m,G^{m8}\},\{J^r,G^{r8}\}\}\} + 14 \{G^{k8},\{\{J^m,G^{m8}\},\{J^r,G^{rc}\}\}\} - 7 d^{c8e} \{\mathcal{D}_3^{ke},\{J^r,G^{r8}\}\} \nonumber \\
&  & \mbox{\hglue0.6truecm} - 7 d^{88e} \{\mathcal{D}_3^{kc},\{J^r,G^{re}\}\} - 14i \epsilon^{kil} [\{J^i,G^{l8}\},\{\{J^m,G^{m8}\},\{J^r,G^{rc}\}\}] + \frac{29}{11} d^{c8e} \{J^2,\{J^2,\{G^{ke},T^8\}\}\} \nonumber \\
&  & \mbox{\hglue0.6truecm} + d^{88e} \{J^2,\{J^2,\{G^{ke},T^c\}\}\} - \frac{29}{11} d^{c8e} \{J^2,\{\mathcal{D}_2^{k8},\{J^r,G^{re}\}\}\} - d^{88e} \{J^2,\{\mathcal{D}_2^{kc},\{J^r,G^{re}\}\}\} \nonumber \\
&  & \mbox{\hglue0.6truecm} + \frac{15}{11} \{J^2,\{\{J^r,G^{rc}\},\{G^{k8},T^8\}\}\} + \frac{29}{11} \{J^2,\{\{J^r,G^{r8}\},\{G^{kc},T^8\}\}\} + 2 \{J^2,\{\{J^r,G^{r8}\},\{G^{k8},T^c\}\}\} \nonumber \\
&  & \mbox{\hglue0.6truecm} - i f^{c8e} \{J^2,\{J^k,[\{J^m,G^{me}\},\{J^r,G^{r8}\}]\}\} - 2 i f^{c8e} \{J^2,\{\{J^r,G^{re}\},[J^2,G^{k8}]\}\} \nonumber \\
&  & \mbox{\hglue0.6truecm} - \frac{7}{11} i f^{c8e} \{J^2,\{\{J^r,G^{r8}\},[J^2,G^{ke}]\}\} + 2 i f^{c8e} \{J^2,\{J^2,[G^{ke},\{J^r,G^{r8}\}]\}\} - 2 \{\mathcal{D}_2^{kc},\{\{J^m,G^{m8}\},\{J^r,G^{r8}\}\}\} \nonumber \\
&  & \mbox{\hglue0.6truecm} - 4 \{\mathcal{D}_2^{k8},\{\{J^m,G^{mc}\},\{J^r,G^{r8}\}\}\} + \frac{7}{11} i \epsilon^{kim} [\{T^8,\{J^r,G^{r8}\}\},\{J^2,\{J^i,G^{mc}\}\}],
\end{eqnarray}

\begin{eqnarray}
&  & [\mathcal{D}_2^{i8},[\mathcal{O}_3^{i8},\mathcal{O}_3^{kc}]] + [\mathcal{O}_3^{i8},[\mathcal{D}_2^{i8},\mathcal{O}_3^{kc}]] + [\mathcal{O}_3^{i8},[\mathcal{O}_3^{i8},\mathcal{D}_2^{kc}]] \nonumber \\
&  &\mbox{\hglue0.2truecm} = - \frac92 f^{c8e} f^{8eg} G^{kg} + 9 f^{c8e} f^{8eg} \mathcal{D}_2^{kg} - \frac{129}{176} i \epsilon^{kim} f^{c8e} f^{8eg} \{J^i,G^{mg}\} - \frac32 f^{c8e} f^{8eg} \mathcal{D}_3^{kg} + \frac{27N_c}{88} i f^{c8e} d^{8eg} \mathcal{D}_3^{kg} \nonumber \\
&  & \mbox{\hglue0.6truecm} + \frac{27N_c}{88} i d^{c8e} f^{8eg} \mathcal{D}_3^{kg} - 3 d^{c8e} d^{8eg} \mathcal{O}_3^{kg} + 3 d^{ceg} d^{88e} \mathcal{O}_3^{kg} - 6 \{G^{kc},\{G^{r8},G^{r8}\}\} + 6 \{G^{k8},\{G^{r8},G^{rc}\}\} \nonumber \\
&  & \mbox{\hglue0.6truecm} + 3 d^{c8e} \{J^k,\{G^{re},G^{r8}\}\} - 3 d^{88e} \{J^k,\{G^{rc},G^{re}\}\} - 6 d^{c8e} \{G^{ke},\{J^r,G^{r8}\}\} + 3 d^{c8e} \{G^{k8},\{J^r,G^{re}\}\} \nonumber \\
&  & \mbox{\hglue0.6truecm} - 3 d^{88e} \{G^{kc},\{J^r,G^{re}\}\} + 6 d^{88e} \{G^{ke},\{J^r,G^{rc}\}\} - \frac32 \epsilon^{kim} f^{c8e} \{T^e,\{J^i,G^{m8}\}\} + \frac{29}{2} f^{c8e} f^{8eg} \mathcal{D}_4^{kg} \nonumber \\
&  & \mbox{\hglue0.6truecm} + d^{c8e} d^{8eg} \mathcal{D}_4^{kg} - d^{ceg} d^{88e} \mathcal{D}_4^{kg} - \frac{27}{44} i f^{c8e} d^{8eg} \mathcal{D}_4^{kg} + \frac{6}{N_f} \delta^{c8} \mathcal{D}_4^{k8} - \frac{27}{22N_f} i \epsilon^{kim} \delta^{c8} \{J^2,\{J^i,G^{m8}\}\} \nonumber \\
&  & \mbox{\hglue0.6truecm} - 2 \{\mathcal{D}_2^{kc},\{G^{r8},G^{r8}\}\} - 4 \{\mathcal{D}_2^{k8},\{G^{r8},G^{rc}\}\} + 2 d^{c8e} \{\mathcal{D}_2^{k8},\{J^r,G^{re}\}\} + d^{88e} \{\mathcal{D}_2^{kc},\{J^r,G^{re}\}\} \nonumber \\
&  & \mbox{\hglue0.6truecm} - \frac{129}{88} i f^{c8e} \{\mathcal{D}_2^{ke},\{J^r,G^{r8}\}\} + \frac{183}{44} i \epsilon^{kim} \{\{G^{r8},G^{rc}\},\{J^i,G^{m8}\}\} - \frac{129}{44} i \epsilon^{kim} \{\{G^{r8},G^{r8}\},\{J^i,G^{mc}\}\} \nonumber \\
&  & \mbox{\hglue0.6truecm} + \frac{129}{44} i \epsilon^{rim} \{G^{k8},\{J^r,\{G^{ic},G^{m8}\}\}\} - \frac{27}{44} i \epsilon^{rim} d^{c8e} \{J^k,\{J^r,\{G^{i8},G^{me}\}\}\} \nonumber \\
&  & \mbox{\hglue0.6truecm} - \frac{81}{176} i \epsilon^{kim} f^{cae} f^{8eb} \{\{J^i,G^{m8}\},\{T^a,T^b\}\} - \frac{873}{88} i f^{c8e} \{J^k,[\{J^m,G^{me}\},\{J^r,G^{r8}\}]\} \nonumber \\
&  & \mbox{\hglue0.6truecm} - \frac{477}{88} i f^{c8e} \{\{J^r,G^{re}\},[J^2,G^{k8}]\} + \frac{477}{88} i f^{c8e} \{\{J^r,G^{r8}\},[J^2,G^{ke}]\} + \frac{477}{88} i f^{c8e} \{J^2,[G^{ke},\{J^r,G^{r8}\}]\} \nonumber \\
&  & \mbox{\hglue0.6truecm} - \frac{549}{88} i f^{c8e} \{J^2,[G^{k8},\{J^r,G^{re}\}]\} + \frac{27}{44} d^{c8e} \{J^2,[G^{ke},\{J^r,G^{r8}\}]\} - \frac{27}{44} d^{c8e} \{J^2,[G^{k8},\{J^r,G^{re}\}]\} \nonumber \\
&  & \mbox{\hglue0.6truecm} + \frac{75}{176} [G^{kc},\{\{J^m,G^{m8}\},\{J^r,G^{r8}\}\}] - \frac{183}{88} [G^{k8},\{\{J^m,G^{m8}\},\{J^r,G^{rc}\}\}] + \frac{183}{88} \{\{J^m,G^{mc}\},[G^{k8},\{J^r,G^{r8}\}]\} \nonumber \\
&  & \mbox{\hglue0.6truecm} + \frac{27}{44} i \epsilon^{kim} f^{cea} f^{e8b} \{\{J^i,G^{m8}\},\{G^{ra},G^{rb}\}\} - 6 d^{c8e} \{J^2,\{G^{ke},\{J^r,G^{r8}\}\}\} - 6 d^{88e} \{J^2,\{G^{ke},\{J^r,G^{rc}\}\}\} \nonumber \\
&  & \mbox{\hglue0.6truecm} + 6 \{G^{kc},\{\{J^m,G^{m8}\},\{J^r,G^{r8}\}\}\} - 6 \{G^{k8},\{\{J^m,G^{m8}\},\{J^r,G^{rc}\}\}\} + 3 d^{c8e} \{\mathcal{D}_3^{ke},\{J^r,G^{r8}\}\} \nonumber \\
&  & \mbox{\hglue0.6truecm} + 3 d^{88e} \{\mathcal{D}_3^{kc},\{J^r,G^{re}\}\} + 6i \epsilon^{kil} [\{J^i,G^{l8}\},\{\{J^m,G^{m8}\},\{J^r,G^{rc}\}\}] + \frac{13}{4} f^{c8e} f^{8eg} \mathcal{D}_6^{kg} + \frac12 d^{c8e} d^{8eg} \mathcal{D}_6^{kg} \nonumber \\
&  & \mbox{\hglue0.6truecm} - \frac12 d^{ceg} d^{88e} \mathcal{D}_6^{kg} + \frac{3}{N_f} \delta^{c8} \mathcal{D}_6^{k8} + \frac{9}{11} d^{c8e} \{J^2,\{J^2,\{G^{ke},T^8\}\}\} - 2 \{J^2,\{\mathcal{D}_2^{kc},\{G^{r8},G^{r8}\}\}\} \nonumber \\
&  & \mbox{\hglue0.6truecm} - 4 \{J^2,\{\mathcal{D}_2^{k8},\{G^{r8},G^{rc}\}\}\} + \frac{2}{11} d^{c8e} \{J^2,\{\mathcal{D}_2^{k8},\{J^r,G^{re}\}\}\} + \frac12 d^{88e} \{J^2,\{\mathcal{D}_2^{kc},\{J^r,G^{re}\}\}\} \nonumber \\
&  & \mbox{\hglue0.6truecm} - \frac{9}{11} \{J^2,\{\{J^r,G^{rc}\},\{G^{k8},T^8\}\}\} + \frac{9}{11} \{J^2,\{\{J^r,G^{r8}\},\{G^{kc},T^8\}\}\} - \frac54 i f^{c8e} \{J^2,\{J^k,[\{J^m,G^{me}\},\{J^r,G^{r8}\}]\}\} \nonumber \\
&  & \mbox{\hglue0.6truecm} + i f^{c8e} \{J^2,\{\{J^r,G^{re}\},[J^2,G^{k8}]\}\} - \frac{20}{11} i f^{c8e} \{J^2,\{\{J^r,G^{r8}\},[J^2,G^{ke}]\}\} - i f^{c8e} \{J^2,\{J^2,[G^{ke},\{J^r,G^{r8}\}]\}\} \nonumber \\
&  & \mbox{\hglue0.6truecm} + i f^{c8e} \{J^2,\{J^2,[G^{k8},\{J^r,G^{re}\}]\}\} + \frac12 \{\mathcal{D}_2^{kc},\{\{J^m,G^{m8}\},\{J^r,G^{r8}\}\}\} + \{\mathcal{D}_2^{k8},\{\{J^m,G^{mc}\},\{J^r,G^{r8}\}\}\} \nonumber \\
&  & \mbox{\hglue0.6truecm} + \frac{9}{11} i \epsilon^{kim} [\{T^8,\{J^r,G^{r8}\}\},\{J^2,\{J^i,G^{mc}\}\}]
\end{eqnarray}

\begin{eqnarray}
&  & [\mathcal{D}_3^{i8},[\mathcal{D}_3^{i8},\mathcal{D}_3^{kc}]] \nonumber \\
&  &\mbox{\hglue0.2truecm} = \frac{176N_c - 2433}{48} f^{c8e} f^{8eg} G^{kg} - \frac{6248 N_c + 3043}{6336} i \epsilon^{kim} f^{c8e} f^{8eg} \{J^i,G^{mg}\} + \frac{176N_c - 2793}{144} f^{c8e} f^{8eg} \mathcal{D}_3^{kg} \nonumber \\
&  & \mbox{\hglue0.6truecm} - \frac{5N_c(792N_c - 9151)}{6336} (i f^{c8e} d^{8eg} \mathcal{D}_3^{kg} + i d^{c8e} f^{8eg} \mathcal{D}_3^{kg}) + \frac{437}{144} f^{c8e} f^{8eg} \mathcal{O}_3^{kg} + \frac{176N_c - 2433}{72} d^{c8e} d^{8eg} \mathcal{O}_3^{kg} \nonumber \\
&  & \mbox{\hglue0.6truecm} - \frac{176N_c - 2433}{72} d^{ceg} d^{88e} \mathcal{O}_3^{kg} + \frac{176N_c - 2433}{36} \{G^{kc},\{G^{r8},G^{r8}\}\} - \frac{176N_c - 2433}{36} \{G^{k8},\{G^{r8},G^{rc}\}\} \nonumber \\
&  & \mbox{\hglue0.6truecm} - \frac{176N_c - 2433}{72} d^{c8e} \{J^k,\{G^{re},G^{r8}\}\} + \frac{176N_c - 2433}{72} d^{88e} \{J^k,\{G^{rc},G^{re}\}\} + \frac{176N_c - 2433}{36} d^{c8e} \{G^{ke},\{J^r,G^{r8}\}\} \nonumber \\
&  & \mbox{\hglue0.6truecm} - \frac{176N_c - 2433}{72} d^{c8e} \{G^{k8},\{J^r,G^{re}\}\} + \frac{176N_c - 2433}{72} d^{88e} \{G^{kc},\{J^r,G^{re}\}\} - \frac{176N_c - 2433}{36} d^{88e} \{G^{ke},\{J^r,G^{rc}\}\} \nonumber \\
&  & \mbox{\hglue0.6truecm} + \frac{704N_c + 879N_f - 13248}{576} \epsilon^{kim} f^{c8e} \{T^e,\{J^i,G^{m8}\}\} + \frac{5(792N_c - 9151)}{3168} i f^{c8e} d^{8eg} \mathcal{D}_4^{kg} \nonumber \\
&  & \mbox{\hglue0.6truecm} + \frac{1}{12} i \epsilon^{kim} f^{c8e} f^{8eg} \{J^2,\{J^i,G^{mg}\}\} + \frac{88N_c(N_c + 2N_f) + 5(792N_c - 9151)}{1584N_f} i \epsilon^{kim} \delta^{c8} \{J^2,\{J^i,G^{m8}\}\} \nonumber \\
&  & \mbox{\hglue0.6truecm} - \frac{6248N_c + 3043}{3168} i f^{c8e} \{\mathcal{D}_2^{ke},\{J^r,G^{r8}\}\} + \frac{1144N_c + 24399}{792} i \epsilon^{kim} \{\{G^{r8},G^{rc}\},\{J^i,G^{m8}\}\} \nonumber \\
&  & \mbox{\hglue0.6truecm} - \frac{6248N_c + 3043}{1584} i \epsilon^{kim} \{\{G^{r8},G^{r8}\},\{J^i,G^{mc}\}\} + \frac{6248N_c + 3043}{1584} i \epsilon^{rim} \{G^{k8},\{J^r,\{G^{ic},G^{m8}\}\}\} \nonumber \\
&  & \mbox{\hglue0.6truecm} + \frac{5(792N_c - 9151)}{3168} i \epsilon^{rim} d^{c8e} \{J^k,\{J^r,\{G^{i8},G^{me}\}\}\} + \frac{5(792N_c - 9151)}{4224} i \epsilon^{kim} f^{cae} f^{8eb} \{\{J^i,G^{m8}\},\{T^a,T^b\}\} \nonumber \\
&  & \mbox{\hglue0.6truecm} + \frac{1936N_c - 57473}{3168} i f^{c8e} \{J^k,[\{J^m,G^{me}\},\{J^r,G^{r8}\}]\} + \frac{1936N_c - 57473}{3168} i f^{c8e} \{\{J^r,G^{re}\},[J^2,G^{k8}]\} \nonumber \\
&  & \mbox{\hglue0.6truecm} - \frac{1936N_c - 57473}{3168} i f^{c8e} \{\{J^r,G^{r8}\},[J^2,G^{ke}]\} - \frac{1936N_c - 57473}{3168} i f^{c8e} \{J^2,[G^{ke},\{J^r,G^{r8}\}]\} \nonumber \\
&  & \mbox{\hglue0.6truecm} + \frac{1936N_c - 56161}{3168} i f^{c8e} \{J^2,[G^{k8},\{J^r,G^{re}\}]\} - \frac{5(792N_c - 9151)}{3168} d^{c8e} \{J^2,[G^{ke},\{J^r,G^{r8}\}]\} \nonumber \\
&  & \mbox{\hglue0.6truecm} + \frac{5(792N_c - 9151)}{3168} d^{c8e} \{J^2,[G^{k8},\{J^r,G^{re}\}]\} + \frac{1276N_c - 5339}{792} [G^{kc},\{\{J^m,G^{m8}\},\{J^r,G^{r8}\}\}] \nonumber \\
&  & \mbox{\hglue0.6truecm} - \frac{1144N_c + 24399}{1584} [G^{k8},\{\{J^m,G^{m8}\},\{J^r,G^{rc}\}\}] + \frac{1144N_c + 24399}{1584} \{\{J^m,G^{mc}\},[G^{k8},\{J^r,G^{r8}\}]\} \nonumber \\
&  & \mbox{\hglue0.6truecm} - \frac{5(792N_c - 9151)}{3168} i \epsilon^{kim} f^{cea} f^{e8b} \{\{J^i,G^{m8}\},\{G^{ra},G^{rb}\}\} - \frac16 f^{c8e} f^{8eg} \mathcal{D}_5^{kg} - \frac{1}{36}N_c(i f^{c8e} d^{8eg} \mathcal{D}_5^{kg} + i d^{c8e} f^{8eg} \mathcal{D}_5^{kg}) \nonumber \\
&  & \mbox{\hglue0.6truecm} + \frac23 d^{c8e} d^{8eg} \mathcal{O}_5^{kg} - \frac{14}{9} d^{ceg} d^{88e} \mathcal{O}_5^{kg} + \frac{341}{36} \{J^2,\{G^{kc},\{G^{r8},G^{r8}\}\}\} - \frac{341}{36} \{J^2,\{G^{k8},\{G^{r8},G^{rc}\}\}\} \nonumber \\
&  & \mbox{\hglue0.6truecm} + \frac{10}{3} d^{c8e} \{J^2,\{J^k,\{G^{re},G^{r8}\}\}\} - \frac{10}{3} d^{88e} \{J^2,\{J^k,\{G^{rc},G^{re}\}\}\} + \frac{352N_c - 5063}{72} d^{c8e} \{J^2,\{G^{ke},\{J^r,G^{r8}\}\}\} \nonumber \\
&  & \mbox{\hglue0.6truecm} - \frac23 d^{c8e} \{J^2,\{G^{k8},\{J^r,G^{re}\}\}\} + \frac{14}{9} d^{88e} \{J^2,\{G^{kc},\{J^r,G^{re}\}\}\} + \frac{352N_c - 5319}{72} d^{88e} \{J^2,\{G^{ke},\{J^r,G^{rc}\}\}\} \nonumber \\
&  & \mbox{\hglue0.6truecm} + \frac13 \epsilon^{kim} f^{c8e} \{J^2,\{T^e,\{J^i,G^{m8}\}\}\} - \frac{176N_c - 2433}{36} \{G^{kc},\{\{J^m,G^{m8}\},\{J^r,G^{r8}\}\}\} \nonumber \\
&  & \mbox{\hglue0.6truecm} + \frac{176N_c - 2433}{36} \{G^{k8},\{\{J^m,G^{m8}\},\{J^r,G^{rc}\}\}\} - \frac{581}{72} \{J^k,\{\{J^m,G^{mc}\},\{G^{r8},G^{r8}\}\}\} \nonumber \\
&  & \mbox{\hglue0.6truecm} + \frac{581}{72} \{J^k,\{\{J^m,G^{m8}\},\{G^{r8},G^{rc}\}\}\} - \frac{352N_c - 4871}{144} d^{c8e} \{\mathcal{D}_3^{ke},\{J^r,G^{r8}\}\} - \frac{352N_c - 5447}{144} d^{88e} \{\mathcal{D}_3^{kc},\{J^r,G^{re}\}\} \nonumber \\
&  & \mbox{\hglue0.6truecm} - \frac{293}{72} \epsilon^{kim} f^{ab8} \{\{J^i,G^{m8}\},\{T^a,\{G^{rb},G^{rc}\}\}\} + \frac{N_c + N_f}{36} i \epsilon^{kim} d^{c8e} \{J^2,\{T^e,\{J^i,G^{m8}\}\}\} \nonumber \\
&  & \mbox{\hglue0.6truecm} - \frac{11(32N_c - 469)}{72} i \epsilon^{kil} [\{J^i,G^{l8}\},\{\{J^m,G^{m8}\},\{J^r,G^{rc}\}\}] + \frac{1}{18} i f^{c8e} d^{8eg} \mathcal{D}_6^{kg} - \frac{41}{99} d^{c8e} \{J^2,\{J^2,\{G^{ke},T^8\}\}\} \nonumber \\
&  & \mbox{\hglue0.6truecm} + \frac{1}{9N_f} i \epsilon^{kim} \delta^{c8} \{J^2,\{J^2,\{J^i,G^{m8}\}\}\} + \frac{41}{99} d^{c8e} \{J^2,\{\mathcal{D}_2^{k8},\{J^r,G^{re}\}\}\} + \frac16 i f^{c8e} \{J^2,\{\mathcal{D}_2^{ke},\{J^r,G^{r8}\}\}\} \nonumber \\
&  & \mbox{\hglue0.6truecm} + \frac{41}{99} \{J^2,\{\{J^r,G^{rc}\},\{G^{k8},T^8\}\}\} - \frac{41}{99} \{J^2,\{\{J^r,G^{r8}\},\{G^{kc},T^8\}\}\}
 - \frac{1}{36} i \epsilon^{kim} \{J^2,\{\{T^c,T^8\},\{J^i,G^{m8}\}\}\} \nonumber
\end{eqnarray}
\begin{eqnarray}
&  & \mbox{\hglue0.6truecm} - \frac49 i \epsilon^{kim} \{J^2,\{\{G^{r8},G^{rc}\},\{J^i,G^{m8}\}\}\} + \frac13 i \epsilon^{kim} \{J^2,\{\{G^{r8},G^{r8}\},\{J^i,G^{mc}\}\}\} \nonumber \\
&  & \mbox{\hglue0.6truecm} - \frac13 i \epsilon^{rim} \{J^2,\{G^{k8},\{J^r,\{G^{ic},G^{m8}\}\}\}\} + \frac{1}{18} i \epsilon^{rim} d^{c8e} \{J^2,\{J^k,\{J^r,\{G^{i8},G^{me}\}\}\}\} \nonumber \\
&  & \mbox{\hglue0.6truecm} + \frac{1}{18} i \epsilon^{kim} f^{cae} f^{8eb} \{J^2,\{\{J^i,G^{m8}\},\{T^a,T^b\}\}\} - \frac56 i f^{c8e} \{J^2,\{J^k,[\{J^m,G^{me}\},\{J^r,G^{r8}\}]\}\} \nonumber \\
&  & \mbox{\hglue0.6truecm} - \frac56 i f^{c8e} \{J^2,\{\{J^r,G^{re}\},[J^2,G^{k8}]\}\} + \frac{247}{198} i f^{c8e} \{J^2,\{\{J^r,G^{r8}\},[J^2,G^{ke}]\}\} \nonumber \\
&  & \mbox{\hglue0.6truecm} + \frac56 i f^{c8e} \{J^2,\{J^2,[G^{ke},\{J^r,G^{r8}\}]\}\} - \frac56 i f^{c8e} \{J^2,\{J^2,[G^{k8},\{J^r,G^{re}\}]\}\} \nonumber \\
&  & \mbox{\hglue0.6truecm} - \frac{41}{99} i \epsilon^{kim} [\{T^8,\{J^r,G^{r8}\}\},\{J^2,\{J^i,G^{mc}\}\}] - \frac{1}{18} d^{c8e} \{J^2,\{J^2,[G^{ke},\{J^r,G^{r8}\}]\}\} \nonumber \\
&  & \mbox{\hglue0.6truecm} + \frac{1}{18} d^{c8e} \{J^2,\{J^2,[G^{k8},\{J^r,G^{re}\}]\}\} - \frac{1}{18}\{J^2,[G^{kc},\{\{J^m,G^{m8}\},\{J^r,G^{r8}\}\}]\} \nonumber \\
&  & \mbox{\hglue0.6truecm} + \frac29 \{J^2,[G^{k8},\{\{J^m,G^{m8}\},\{J^r,G^{rc}\}\}]\} - \frac29 \{J^2,\{\{J^m,G^{mc}\},[G^{k8},\{J^r,G^{r8}\}]\}\} \nonumber \\
&  & \mbox{\hglue0.6truecm} + f^{c8e} f^{8eg} \mathcal{D}_7^{kg} - 2 d^{c8e} \{J^2,\{J^2,\{J^k,\{G^{re},G^{r8}\}\}\}\} + 2 d^{88e} \{J^2,\{J^2,\{J^k,\{G^{rc},G^{re}\}\}\}\} \nonumber \\
&  & \mbox{\hglue0.6truecm} + \frac43 d^{c8e} \{J^2,\{J^2,\{G^{ke},\{J^r,G^{r8}\}\}\}\} + \frac43 d^{88e} \{J^2,\{J^2,\{G^{ke},\{J^r,G^{rc}\}\}\}\} \nonumber \\
&  & \mbox{\hglue0.6truecm} - \frac43 \{J^2,\{G^{kc},\{\{J^m,G^{m8}\},\{J^r,G^{r8}\}\}\}\} + \frac43 \{J^2,\{G^{k8},\{\{J^m,G^{m8}\},\{J^r,G^{rc}\}\}\}\} \nonumber \\
&  & \mbox{\hglue0.6truecm} + 2 \{J^2,\{J^k,\{\{J^m,G^{mc}\},\{G^{r8},G^{r8}\}\}\}\} - 2\{J^2,\{J^k,\{\{J^m,G^{m8}\},\{G^{r8},G^{rc}\}\}\}\} \nonumber \\
&  & \mbox{\hglue0.6truecm} + \frac13 d^{c8e} \{J^2,\{\mathcal{D}_3^{ke},\{J^r,G^{r8}\}\}\} - \frac53 d^{88e} \{J^2,\{\mathcal{D}_3^{kc},\{J^r,G^{re}\}\}\} \nonumber \\
&  & \mbox{\hglue0.6truecm} + \frac49 i \epsilon^{kil} \{J^2,[\{J^i,G^{l8}\},\{\{J^m,G^{m8}\},\{J^r,G^{rc}\}\}]\} + 2\{\mathcal{D}_3^{kc},\{\{J^m,G^{m8}\},\{J^r,G^{r8}\}\}\} \nonumber \\
&  & \mbox{\hglue0.6truecm} - \frac{16}{9} i \epsilon^{kil} \{J^2,\{J^i,\{J^r,[G^{l8},\{G^{r8},\{J^m,G^{mc}\}\}]\}\}\},
\end{eqnarray}

\begin{eqnarray}
&  & [\mathcal{D}_3^{i8},[\mathcal{O}_3^{i8},\mathcal{D}_3^{kc}]] + [\mathcal{D}_3^{i8},[\mathcal{D}_3^{i8},\mathcal{O}_3^{kc}]] + [\mathcal{O}_3^{i8},[\mathcal{D}_3^{i8},\mathcal{D}_3^{kc}]] \nonumber \\
&  &\mbox{\hglue0.2truecm} = \frac{416N_c-3489}{64} f^{c8e} f^{8eg} G^{kg} + \frac{59752N_c-241435}{8448} i \epsilon^{kim} f^{c8e} f^{8eg} \{J^i,G^{mg}\} + \frac{416N_c-3417}{192} f^{c8e} f^{8eg} \mathcal{D}_3^{kg} \nonumber \\
&  & \mbox{\hglue0.6truecm} - \frac{N_c(65736N_c-303023)}{8448} (i f^{c8e} d^{8eg} \mathcal{D}_3^{kg} + i d^{c8e} f^{8eg} \mathcal{D}_3^{kg}) - \frac{1699}{192} f^{c8e} f^{8eg} \mathcal{O}_3^{kg} + \frac{416N_c-3489}{96} d^{c8e} d^{8eg} \mathcal{O}_3^{kg} \nonumber \\
&  & \mbox{\hglue0.6truecm} - \frac{416N_c-3489}{96} d^{ceg} d^{88e} \mathcal{O}_3^{kg} + \frac{416N_c-3489}{48} \{G^{kc},\{G^{r8},G^{r8}\}\} - \frac{416N_c-3489}{48} \{G^{k8},\{G^{r8},G^{rc}\}\} \nonumber \\
&  & \mbox{\hglue0.6truecm} - \frac{416N_c-3489}{96} d^{c8e} \{J^k,\{G^{re},G^{r8}\}\} + \frac{416N_c-3489}{96} d^{88e} \{J^k,\{G^{rc},G^{re}\}\} \nonumber \\
&  & \mbox{\hglue0.6truecm} + \frac{416N_c-3489}{48} d^{c8e} \{G^{ke},\{J^r,G^{r8}\}\} - \frac{416N_c-3489}{96} d^{c8e} \{G^{k8},\{J^r,G^{re}\}\} \nonumber \\
&  & \mbox{\hglue0.6truecm} + \frac{416N_c-3489}{96} d^{88e} \{G^{kc},\{J^r,G^{re}\}\} - \frac{416N_c-3489}{48} d^{88e} \{G^{ke},\{J^r,G^{rc}\}\} \nonumber \\
&  & \mbox{\hglue0.6truecm} + \frac{1664N_c-345N_f-12576}{768} \epsilon^{kim} f^{c8e} \{T^e,\{J^i,G^{m8}\}\} + \frac{65736N_c-303023}{4224} i f^{c8e} d^{8eg} \mathcal{D}_4^{kg} \nonumber\\
&  & \mbox{\hglue0.6truecm} - \frac{15}{16} i \epsilon^{kim} f^{c8e} f^{8eg} \{J^2,\{J^i,G^{mg}\}\} + \frac{616N_c^2+88N_c(14N_f+747)-303023}{2112N_f} i \epsilon^{kim} \delta ^{c8} \{J^2,\{J^i,G^{m8}\}\} \nonumber \\
&  & \mbox{\hglue0.6truecm} + \frac{59752N_c-241435}{4224} i f^{c8e} \{\mathcal{D}_2^{ke},\{J^r,G^{r8}\}\} - \frac{62744N_c-272229}{1056} i \epsilon^{kim} \{\{G^{r8},G^{rc}\},\{J^i,G^{m8}\}\} \nonumber \\
&  & \mbox{\hglue0.6truecm} + \frac{59752N_c-241435}{2112} i \epsilon^{kim} \{\{G^{r8},G^{r8}\},\{J^i,G^{mc}\}\} + \frac{241435 - 59752 N_c}{2112} i \epsilon^{rim} \{G^{k8},\{J^r,\{G^{ic},G^{m8}\}\}\} \nonumber \\
&  & \mbox{\hglue0.6truecm} + \frac{65736N_c-303023}{4224} i \epsilon^{rim} d^{c8e} \{J^k,\{J^r,\{G^{i8},G^{me}\}\}\} + \frac{65736N_c-303023}{5632} i \epsilon^{kim} f^{cae} f^{8eb} \{\{J^i,G^{m8}\},\{T^a,T^b\}\} \nonumber \\
&  & \mbox{\hglue0.6truecm} + \frac{23056N_c-225641}{4224} i f^{c8e} \{J^k,[\{J^m,G^{me}\},\{J^r,G^{r8}\}]\} + \frac{23056N_c-225641}{4224} i f^{c8e} \{\{J^r,G^{re}\},[J^2,G^{k8}]\} \nonumber \\
&  & \mbox{\hglue0.6truecm} - \frac{23056N_c-225641}{4224} i f^{c8e} \{\{J^r,G^{r8}\},[J^2,G^{ke}]\} - \frac{23056N_c-225641}{4224} i f^{c8e} \{J^2,[G^{ke},\{J^r,G^{r8}\}]\} \nonumber \\
&  & \mbox{\hglue0.6truecm} + \frac{23056N_c-260233}{4224} i f^{c8e} \{J^2,[G^{k8},\{J^r,G^{re}\}]\} - \frac{65736N_c-303023}{4224} d^{c8e} \{J^2,[G^{ke},\{J^r,G^{r8}\}]\} \nonumber \\
&  & \mbox{\hglue0.6truecm} + \frac{65736N_c-303023}{4224} d^{c8e} \{J^2,[G^{k8},\{J^r,G^{re}\}]\} + \frac{1496N_c-15397}{2112}[G^{kc},\{\{J^m,G^{m8}\},\{J^r,G^{r8}\}\}] \nonumber \\
&  & \mbox{\hglue0.6truecm} + \frac{62744N_c-272229}{2112} [G^{k8},\{\{J^m,G^{m8}\},\{J^r,G^{rc}\}\}] - \frac{62744N_c-272229}{2112} \{\{J^m,G^{mc}\},[G^{k8},\{J^r,G^{r8}\}]\} \nonumber \\
&  & \mbox{\hglue0.6truecm} - \frac{65736N_c-303023}{4224} i \epsilon^{kim} f^{cea} f^{e8b} \{\{J^i,G^{m8}\},\{G^{ra},G^{rb}\}\} + \frac14 f^{c8e} f^{8eg} \mathcal{D}_5^{kg} - \frac{7N_c}{48} (i f^{c8e} d^{8eg} \mathcal{D}_5^{kg} + i d^{c8e} f^{8eg} \mathcal{D}_5^{kg}) \nonumber \\
&  & \mbox{\hglue0.6truecm} - \frac12 f^{c8e} f^{8eg} \mathcal{O}_5^{kg} - \frac32 d^{c8e} d^{8eg} \mathcal{O}_5^{kg} + \frac56 d^{ceg} d^{88e} \mathcal{O}_5^{kg} - \frac{643}{48} \{J^2,\{G^{kc},\{G^{r8},G^{r8}\}\}\} \nonumber \\
&  & \mbox{\hglue0.6truecm} + \frac{643}{48} \{J^2,\{G^{k8},\{G^{r8},G^{rc}\}\}\} - \frac12 d^{c8e} \{J^2,\{J^k,\{G^{re},G^{r8}\}\}\} + \frac12 d^{88e} \{J^2,\{J^k,\{G^{rc},G^{re}\}\}\} \nonumber \\
&  & \mbox{\hglue0.6truecm} + \frac{832N_c-7535}{96} d^{c8e} \{J^2,\{G^{ke},\{J^r,G^{r8}\}\}\} + \frac32 d^{c8e} \{J^2,\{G^{k8},\{J^r,G^{re}\}\}\} - \frac56 d^{88e} \{J^2,\{G^{kc},\{J^r,G^{re}\}\}\} \nonumber \\
&  & \mbox{\hglue0.6truecm} + \frac{832N_c-6255}{96} d^{88e} \{J^2,\{G^{ke},\{J^r,G^{rc}\}\}\} + \frac{3N_f-23}{4} \epsilon^{kim} f^{c8e} \{J^2,\{T^e,\{J^i,G^{m8}\}\}\} \nonumber \\
&  & \mbox{\hglue0.6truecm} - \frac{416N_c-3489}{48} \{G^{kc},\{\{J^m,G^{m8}\},\{J^r,G^{r8}\}\}\} + \frac{416N_c-3489}{48} \{G^{k8},\{\{J^m,G^{m8}\},\{J^r,G^{rc}\}\}\} \nonumber \\
&  & \mbox{\hglue0.6truecm} + \frac{691}{96} \{J^k,\{\{J^m,G^{mc}\},\{G^{r8},G^{r8}\}\}\} - \frac{691}{96} \{J^k,\{\{J^m,G^{m8}\},\{G^{r8},G^{rc}\}\}\} \nonumber \\
&  & \mbox{\hglue0.6truecm} - \frac{832N_c-7439}{192} d^{c8e} \{\mathcal{D}_3^{ke},\{J^r,G^{r8}\}\} - \frac{832N_c-6287}{192} d^{88e} \{\mathcal{D}_3^{kc},\{J^r,G^{re}\}\} \nonumber \\
&  & \mbox{\hglue0.6truecm} + \frac{115}{96} \epsilon^{kim} f^{ab8} \{\{J^i,G^{m8}\},\{T^a,\{G^{rb},G^{rc}\}\}\} + \frac{7(N_c+N_f)}{48} i \epsilon^{kim} d^{c8e} \{J^2,\{T^e,\{J^i,G^{m8}\}\}\} \nonumber \\
&  & \mbox{\hglue0.6truecm} - \frac{832N_c-6863}{96} i \epsilon^{kil} [\{J^i,G^{l8}\},\{\{J^m,G^{m8}\},\{J^r,G^{rc}\}\}] + \frac{7}{24} i f^{c8e} d^{8eg} \mathcal{D}_6^{kg} + \frac{1081}{132} d^{c8e} \{J^2,\{J^2,\{G^{ke},T^8\}\}\} \nonumber
\end{eqnarray}
\begin{eqnarray}
&  & \mbox{\hglue0.6truecm} + \frac{7}{12N_f} i \epsilon^{kim} \delta^{c8} \{J^2,\{J^2,\{J^i,G^{m8}\}\}\} - \frac{1081}{132} d^{c8e} \{J^2,\{\mathcal{D}_2^{k8},\{J^r,G^{re}\}\}\} - \frac{15}{8} i f^{c8e} \{J^2,\{\mathcal{D}_2^{ke},\{J^r,G^{r8}\}\}\} \nonumber \\
&  & \mbox{\hglue0.6truecm} - \frac{1081}{132} \{J^2,\{\{J^r,G^{rc}\},\{G^{k8},T^8\}\}\} + \frac{1081}{132} \{J^2,\{\{J^r,G^{r8}\},\{G^{kc},T^8\}\}\} - \frac{7}{48} i \epsilon^{kim} \{J^2,\{\{T^c,T^8\},\{J^i,G^{m8}\}\}\} \nonumber \\
&  & \mbox{\hglue0.6truecm} + \frac{19}{6} i \epsilon^{kim} \{J^2,\{\{G^{r8},G^{rc}\},\{J^i,G^{m8}\}\}\} - \frac{15}{4} i \epsilon^{kim} \{J^2,\{\{G^{r8},G^{r8}\},\{J^i,G^{mc}\}\}\} \nonumber \\
&  & \mbox{\hglue0.6truecm} + \frac{15}{4} i \epsilon^{rim} \{J^2,\{G^{k8},\{J^r,\{G^{ic},G^{m8}\}\}\}\} + \frac{7}{24} i \epsilon^{rim} d^{c8e} \{J^2,\{J^k,\{J^r,\{G^{i8},G^{me}\}\}\}\} \nonumber \\
&  & \mbox{\hglue0.6truecm} + \frac{7}{24} i \epsilon^{kim} f^{cae} f^{8eb} \{J^2,\{\{J^i,G^{m8}\},\{T^a,T^b\}\}\} - \frac34 i f^{c8e} \{J^2,\{J^k,[\{J^m,G^{me}\},\{J^r,G^{r8}\}]\}\} \nonumber \\
&  & \mbox{\hglue0.6truecm} - \frac34 i f^{c8e} \{J^2,\{\{J^r,G^{re}\},[J^2,G^{k8}]\}\} - \frac{491}{66} i f^{c8e} \{J^2,\{\{J^r,G^{r8}\},[J^2,G^{ke}]\}\} \nonumber \\
&  & \mbox{\hglue0.6truecm} + \frac34 i f^{c8e} \{J^2,\{J^2,[G^{ke},\{J^r,G^{r8}\}]\}\} - \frac34 i f^{c8e} \{J^2,\{J^2,[G^{k8},\{J^r,G^{re}\}]\}\} \nonumber \\
&  & \mbox{\hglue0.6truecm} + \frac{1081}{132} i \epsilon^{kim} [\{T^8,\{J^r,G^{r8}\}\},\{J^2,\{J^i,G^{mc}\}\}] - \frac{7}{24} d^{c8e} \{J^2,\{J^2,[G^{ke},\{J^r,G^{r8}\}]\}\} \nonumber \\
&  & \mbox{\hglue0.6truecm} + \frac{7}{24} d^{c8e} \{J^2,\{J^2,[G^{k8},\{J^r,G^{re}\}]\}\} + \frac{13}{12} \{J^2,[G^{kc},\{\{J^m,G^{m8}\},\{J^r,G^{r8}\}\}]\} \nonumber \\
&  & \mbox{\hglue0.6truecm} - \frac{19}{12} \{J^2,[G^{k8},\{\{J^m,G^{m8}\},\{J^r,G^{rc}\}\}]\} + \frac{19}{12}\{J^2,\{\{J^m,G^{mc}\},[G^{k8},\{J^r,G^{r8}\}]\}\} \nonumber \\
&  & \mbox{\hglue0.6truecm} + d^{c8e} d^{8eg} \mathcal{O}_7^{kg} + 2 \{J^2,\{J^2,\{G^{kc},\{G^{r8},G^{r8}\}\}\}\} - 2 \{J^2,\{J^2,\{G^{k8},\{G^{r8},G^{rc}\}\}\}\} \nonumber \\
&  & \mbox{\hglue0.6truecm} + 3 d^{c8e} \{J^2,\{J^2,\{G^{ke},\{J^r,G^{r8}\}\}\}\} - d^{c8e} \{J^2,\{J^2,\{G^{k8},\{J^r,G^{re}\}\}\}\} - 2 d^{88e} \{J^2,\{J^2,\{G^{ke},\{J^r,G^{rc}\}\}\}\} \nonumber \\
&  & \mbox{\hglue0.6truecm} - \frac12 \epsilon^{kim} f^{c8e} \{J^2,\{J^2,\{T^e,\{J^i,G^{m8}\}\}\}\} + 5 \{J^2,\{G^{kc},\{\{J^m,G^{m8}\},\{J^r,G^{r8}\}\}\}\} \nonumber \\
&  & \mbox{\hglue0.6truecm} + \{J^2,\{G^{k8},\{\{J^m,G^{m8}\},\{J^r,G^{rc}\}\}\}\} - \{J^2,\{J^k,\{\{J^m,G^{mc}\},\{G^{r8},G^{r8}\}\}\}\} \nonumber \\
&  & \mbox{\hglue0.6truecm} + \{J^2,\{J^k,\{\{J^m,G^{m8}\},\{G^{r8},G^{rc}\}\}\}\} - d^{c8e} \{J^2,\{\mathcal{D}_3^{ke},\{J^r,G^{r8}\}\}\} \nonumber \\
&  & \mbox{\hglue0.6truecm} + d^{88e} \{J^2,\{\mathcal{D}_3^{kc},\{J^r,G^{re}\}\}\} - 2 \epsilon^{kim} f^{ab8} \{J^2,\{\{J^i,G^{m8}\},\{T^a,\{G^{rb},G^{rc}\}\}\}\} \nonumber \\
&  & \mbox{\hglue0.6truecm} + \frac13 i \epsilon^{kil} \{J^2,[\{J^i,G^{l8}\},\{\{J^m,G^{m8}\},\{J^r,G^{rc}\}\}]\} - 3 \{\mathcal{D}_3^{kc},\{\{J^m,G^{m8}\},\{J^r,G^{r8}\}\}\} \nonumber \\
&  & \mbox{\hglue0.6truecm} + \frac83 i \epsilon^{kil} \{J^2,\{J^i,\{J^r,[G^{l8},\{G^{r8},\{J^m,G^{mc}\}\}]\}\}\},
\end{eqnarray}

\begin{eqnarray}
&  & [\mathcal{D}_3^{i8},[\mathcal{O}_3^{i8},\mathcal{O}_3^{kc}]] + [\mathcal{O}_3^{i8},[\mathcal{D}_3^{i8},\mathcal{O}_3^{kc}]] + [\mathcal{O}_3^{i8},[\mathcal{O}_3^{i8},\mathcal{D}_3^{kc}]] \nonumber \\
&  &\mbox{\hglue0.2truecm} = \frac{3(113N_c-137)}{8} f^{c8e} f^{8eg} G^{kg} - \frac{407N_c-13272}{704} i \epsilon^{kim} f^{c8e} f^{8eg} \{J^i,G^{mg}\} + \frac{113N_c-137}{8} f^{c8e} f^{8eg} \mathcal{D}_3^{kg} \nonumber \\
&  & \mbox{\hglue0.6truecm} + 8 d^{c8e} d^{8eg} \mathcal{D}_3^{kg} - 4 d^{ceg} d^{88e} \mathcal{D}_3^{kg} - \frac{N_c(4400N_c+39421)}{2816} (i f^{c8e} d^{8eg} \mathcal{D}_3^{kg} + i d^{c8e} f^{8eg} \mathcal{D}_3^{kg}) \nonumber \\
&  & \mbox{\hglue0.6truecm} - \frac{67}{16} f^{c8e} f^{8eg} \mathcal{O}_3^{kg} + \frac{113N_c-137}{4} d^{c8e} d^{8eg} \mathcal{O}_3^{kg} - \frac{113N_c-137}{4} d^{ceg} d^{88e} \mathcal{O}_3^{kg} + \frac{8}{N_f} d^{c88} \{J^2,J^k\} \nonumber \\
&  & \mbox{\hglue0.6truecm} + \frac{113N_c-137}{2} \{G^{kc},\{G^{r8},G^{r8}\}\} - \frac{113N_c-137}{2} \{G^{k8},\{G^{r8},G^{rc}\}\} - \frac{113N_c-73}{4} d^{c8e} \{J^k,\{G^{re},G^{r8}\}\} \nonumber \\
&  & \mbox{\hglue0.6truecm} + \frac{113N_c-105}{4} d^{88e} \{J^k,\{G^{rc},G^{re}\}\} + \frac{113N_c-137}{2} d^{c8e} \{G^{ke},\{J^r,G^{r8}\}\} - \frac{113N_c-137}{4} d^{c8e} \{G^{k8},\{J^r,G^{re}\}\} \nonumber \\
&  & \mbox{\hglue0.6truecm} + \frac{113N_c-137}{4} d^{88e} \{G^{kc},\{J^r,G^{re}\}\} - \frac{113N_c-137}{2} d^{88e} \{G^{ke},\{J^r,G^{rc}\}\} \nonumber \\
&  & \mbox{\hglue0.6truecm} + \frac{904N_c-57N_f-868}{64} \epsilon^{kim} f^{c8e} \{T^e,\{J^i,G^{m8}\}\} + \frac{4400N_c+39421}{1408} i f^{c8e} d^{8eg} \mathcal{D}_4^{kg} \nonumber \\
&  & \mbox{\hglue0.6truecm} + \frac54 i \epsilon^{kim} f^{c8e} f^{8eg} \{J^2,\{J^i,G^{mg}\}\} - \frac{1188N_c^2+88N_c(27N_f-50)-39421}{704 N_f} i \epsilon^{kim} \delta^{c8} \{J^2,\{J^i,G^{m8}\}\} \nonumber \\
&  & \mbox{\hglue0.6truecm} - \frac{407N_c-13272}{352} i f^{c8e} \{\mathcal{D}_2^{ke},\{J^r,G^{r8}\}\} - \frac{2772N_c+92509}{704} i \epsilon^{kim} \{\{G^{r8},G^{rc}\},\{J^i,G^{m8}\}\} \nonumber \\
&  & \mbox{\hglue0.6truecm} - \frac{407N_c-13272}{176} i \epsilon^{kim} \{\{G^{r8},G^{r8}\},\{J^i,G^{mc}\}\} + \frac{407N_c-13272}{176} i \epsilon^{rim} \{G^{k8},\{J^r,\{G^{ic},G^{m8}\}\}\} \nonumber \\
&  & \mbox{\hglue0.6truecm} + \frac{4400N_c+39421}{1408} i \epsilon^{rim} d^{c8e} \{J^k,\{J^r,\{G^{i8},G^{me}\}\}\} + \frac{3(4400N_c+39421)}{5632} i \epsilon^{kim} f^{cae} f^{8eb} \{\{J^i,G^{m8}\},\{T^a,T^b\}\} \nonumber \\
&  & \mbox{\hglue0.6truecm} + \frac{10879N_c-12735}{352}i f^{c8e} \{J^k,[\{J^m,G^{me}\},\{J^r,G^{r8}\}]\} + \frac{10879N_c-12735}{352}i f^{c8e} \{\{J^r,G^{re}\},[J^2,G^{k8}]\} \nonumber \\
&  & \mbox{\hglue0.6truecm} - \frac{10879N_c-12735}{352} i f^{c8e} \{\{J^r,G^{r8}\},[J^2,G^{ke}]\} - \frac{10879N_c-12735}{352} i f^{c8e} \{J^2,[G^{ke},\{J^r,G^{r8}\}]\} \nonumber \\
&  & \mbox{\hglue0.6truecm} + \frac{10879N_c-10415}{352} i f^{c8e} \{J^2,[G^{k8},\{J^r,G^{re}\}]\} - \frac{4400N_c+39421}{1408} d^{c8e} \{J^2,[G^{ke},\{J^r,G^{r8}\}]\} \nonumber \\
&  & \mbox{\hglue0.6truecm} + \frac{4400N_c+39421}{1408} d^{c8e} \{J^2,[G^{k8},\{J^r,G^{re}\}]\} + \frac{6028N_c-13667}{2816} [G^{kc},\{\{J^m,G^{m8}\},\{J^r,G^{r8}\}\}] \nonumber \\
&  & \mbox{\hglue0.6truecm} + \frac{2772N_c+92509}{1408} [G^{k8},\{\{J^m,G^{m8}\},\{J^r,G^{rc}\}\}] - \frac{2772N_c+92509}{1408} \{\{J^m,G^{mc}\},[G^{k8},\{J^r,G^{r8}\}]\} \nonumber \\
&  & \mbox{\hglue0.6truecm} - \frac{4400N_c+39421}{1408} i \epsilon^{kim} f^{cea} f^{e8b} \{\{J^i,G^{m8}\},\{G^{ra},G^{rb}\}\} + \frac{41}{8} f^{c8e} f^{8eg} \mathcal{D}_5^{kg} + 11 d^{c8e} d^{8eg} \mathcal{D}_5^{kg} - 6 d^{ceg} d^{88e} \mathcal{D}_5^{kg} \nonumber \\
&  & \mbox{\hglue0.6truecm} + \frac{27N_c}{32} (i f^{c8e} d^{8eg} \mathcal{D}_5^{kg} + i d^{c8e} f^{8eg} \mathcal{D}_5^{kg}) + \frac{6}{N_f}\delta^{c8} \mathcal{D}_5^{k8} - f^{c8e} f^{8eg} \mathcal{O}_5^{kg} - 2 d^{c8e} d^{8eg} \mathcal{O}_5^{kg} + \frac{11}{2} d^{ceg} d^{88e} \mathcal{O}_5^{kg} \nonumber \\
&  & \mbox{\hglue0.6truecm} + \frac{10}{N_f} d^{c88} \{J^2,\{J^2,J^k\}\} - \frac{35}{4} \{J^2,\{G^{kc},\{G^{r8},G^{r8}\}\}\} + \frac{35}{4}\{J^2,\{G^{k8},\{G^{r8},G^{rc}\}\}\} \nonumber \\
&  & \mbox{\hglue0.6truecm} - 16 d^{c8e} \{J^2,\{J^k,\{G^{re},G^{r8}\}\}\} + 2 d^{88e} \{J^2,\{J^k,\{G^{rc},G^{re}\}\}\} + \frac{452N_c-561}{8} d^{c8e} \{J^2,\{G^{ke},\{J^r,G^{r8}\}\}\} \nonumber \\
&  & \mbox{\hglue0.6truecm} + 2 d^{c8e} \{J^2,\{G^{k8},\{J^r,G^{re}\}\}\} - \frac{11}{2} d^{88e} \{J^2,\{G^{kc},\{J^r,G^{re}\}\}\} + \frac{452N_c-469}{8} d^{88e} \{J^2,\{G^{ke},\{J^r,G^{rc}\}\}\} \nonumber \\
&  & \mbox{\hglue0.6truecm} - \frac{3N_f-8}{4} \epsilon^{kim} f^{c8e} \{J^2,\{T^e,\{J^i,G^{m8}\}\}\} - \frac{113N_c-137}{2} \{G^{kc},\{\{J^m,G^{m8}\},\{J^r,G^{r8}\}\}\} \nonumber \\
&  & \mbox{\hglue0.6truecm} + \frac{113N_c-137}{2} \{G^{k8},\{\{J^m,G^{m8}\},\{J^r,G^{rc}\}\}\} - \frac{77}{8} \{J^k,\{\{J^m,G^{mc}\},\{G^{r8},G^{r8}\}\}\} \nonumber \\
&  & \mbox{\hglue0.6truecm} + \frac{29}{8} \{J^k,\{\{J^m,G^{m8}\},\{G^{r8},G^{rc}\}\}\} - \frac{452N_c-545}{16} d^{c8e} \{\mathcal{D}_3^{ke},\{J^r,G^{r8}\}\} - \frac{452N_c-593}{16} d^{88e} \{\mathcal{D}_3^{kc},\{J^r,G^{re}\}\} \nonumber \\
&  & \mbox{\hglue0.6truecm} + \frac{19}{8} \epsilon^{kim} f^{ab8} \{\{J^i,G^{m8}\},\{T^a,\{G^{rb},G^{rc}\}\}\} - \frac{27(N_c+N_f)}{32} i \epsilon^{kim} d^{c8e} \{J^2,\{T^e,\{J^i,G^{m8}\}\}\} \nonumber \\
&  & \mbox{\hglue0.6truecm} - \frac{452N_c-529}{8} i \epsilon^{kil} [\{J^i,G^{l8}\},\{\{J^m,G^{m8}\},\{J^r,G^{rc}\}\}] - \frac{27}{16} i f^{c8e} d^{8eg} \mathcal{D}_6^{kg} \nonumber
\end{eqnarray}
\begin{eqnarray}
&  & \mbox{\hglue0.6truecm} - \frac{145}{22} d^{c8e} \{J^2,\{J^2,\{G^{ke},T^8\}\}\} - \frac{27}{8N_f} i \epsilon^{kim} \delta^{c8} \{J^2,\{J^2,\{J^i,G^{m8}\}\}\} + \frac{145}{22} d^{c8e} \{J^2,\{\mathcal{D}_2^{k8},\{J^r,G^{re}\}\}\} \nonumber \\
&  & \mbox{\hglue0.6truecm} + \frac52 i f^{c8e} \{J^2,\{\mathcal{D}_2^{ke},\{J^r,G^{r8}\}\}\} + \frac{145}{22} \{J^2,\{\{J^r,G^{rc}\},\{G^{k8},T^8\}\}\} - \frac{145}{22} \{J^2,\{\{J^r,G^{r8}\},\{G^{kc},T^8\}\}\} \nonumber \\
&  & \mbox{\hglue0.6truecm} + \frac{27}{32} i \epsilon^{kim} \{J^2,\{\{T^c,T^8\},\{J^i,G^{m8}\}\}\} - \frac{13}{8} i \epsilon^{kim} \{J^2,\{\{G^{r8},G^{rc}\},\{J^i,G^{m8}\}\}\} \nonumber \\
&  & \mbox{\hglue0.6truecm} + 5 i \epsilon^{kim} \{J^2,\{\{G^{r8},G^{r8}\},\{J^i,G^{mc}\}\}\} - 5 i \epsilon^{rim} \{J^2,\{G^{k8},\{J^r,\{G^{ic},G^{m8}\}\}\}\} \nonumber \\
&  & \mbox{\hglue0.6truecm} - \frac{27}{16} i \epsilon^{rim} d^{c8e} \{J^2,\{J^k,\{J^r,\{G^{i8},G^{me}\}\}\}\} - \frac{27}{16} i \epsilon^{kim} f^{cae} f^{8eb} \{J^2,\{\{J^i,G^{m8}\},\{T^a,T^b\}\}\} \nonumber \\
&  & \mbox{\hglue0.6truecm} - \frac52 i f^{c8e} \{J^2,\{J^k,[\{J^m,G^{me}\},\{J^r,G^{r8}\}]\}\} - \frac52 i f^{c8e} \{J^2,\{\{J^r,G^{re}\},[J^2,G^{k8}]\}\} \nonumber \\
&  & \mbox{\hglue0.6truecm} + \frac{100}{11} i f^{c8e} \{J^2,\{\{J^r,G^{r8}\},[J^2,G^{ke}]\}\} + \frac52 i f^{c8e} \{J^2,\{J^2,[G^{ke},\{J^r,G^{r8}\}]\}\} - \frac52i f^{c8e} \{J^2,\{J^2,[G^{k8},\{J^r,G^{re}\}]\}\} \nonumber \\
&  & \mbox{\hglue0.6truecm} - \frac{145}{22} i \epsilon^{kim} [\{T^8,\{J^r,G^{r8}\}\},\{J^2,\{J^i,G^{mc}\}\}] + \frac{27}{16} d^{c8e} \{J^2,\{J^2,[G^{ke},\{J^r,G^{r8}\}]\}\} \nonumber \\
&  & \mbox{\hglue0.6truecm} - \frac{27}{16} d^{c8e} \{J^2,\{J^2,[G^{k8},\{J^r,G^{re}\}]\}\} - \frac{67}{32} \{J^2,[G^{kc},\{\{J^m,G^{m8}\},\{J^r,G^{r8}\}\}]\} \nonumber \\
&  & \mbox{\hglue0.6truecm} + \frac{13}{16} \{J^2,[G^{k8},\{\{J^m,G^{m8}\},\{J^r,G^{rc}\}\}]\} - \frac{13}{16} \{J^2,\{\{J^m,G^{mc}\},[G^{k8},\{J^r,G^{r8}\}]\}\} + f^{c8e} f^{8eg} \mathcal{D}_7^{kg} \nonumber \\
&  & \mbox{\hglue0.6truecm} + \frac32 d^{c8e} d^{8eg} \mathcal{D}_7^{kg} - d^{ceg} d^{88e} \mathcal{D}_7^{kg} + \frac{3}{N_f} \delta^{c8} \mathcal{D}_7^{k8} + \frac{1}{N_f}d^{c88} \{J^2,\{J^2,\{J^2,J^k\}\}\} - 4 \{J^2,\{J^2,\{G^{kc},\{G^{r8},G^{r8}\}\}\}\} \nonumber \\
&  & \mbox{\hglue0.6truecm} + 4 \{J^2,\{J^2,\{G^{k8},\{G^{r8},G^{rc}\}\}\}\} - \frac72 d^{c8e} \{J^2,\{J^2,\{J^k,\{G^{re},G^{r8}\}\}\}\} - \frac12 d^{88e} \{J^2,\{J^2,\{J^k,\{G^{rc},G^{re}\}\}\}\} \nonumber \\
&  & \mbox{\hglue0.6truecm} - 2 d^{c8e} \{J^2,\{J^2,\{G^{ke},\{J^r,G^{r8}\}\}\}\} - 2 d^{88e} \{J^2,\{J^2,\{G^{ke},\{J^r,G^{rc}\}\}\}\} + 4 \{J^2,\{G^{kc},\{\{J^m,G^{m8}\},\{J^r,G^{r8}\}\}\}\} \nonumber \\
&  & \mbox{\hglue0.6truecm} - 4 \{J^2,\{G^{k8},\{\{J^m,G^{m8}\},\{J^r,G^{rc}\}\}\}\} - \frac92 \{J^2,\{J^k,\{\{J^m,G^{mc}\},\{G^{r8},G^{r8}\}\}\}\} \nonumber \\
&  & \mbox{\hglue0.6truecm} - \frac32 \{J^2,\{J^k,\{\{J^m,G^{m8}\},\{G^{r8},G^{rc}\}\}\}\} + \frac{11}{4} d^{c8e} \{J^2,\{\mathcal{D}_3^{ke},\{J^r,G^{r8}\}\}\} + \frac94 d^{88e} \{J^2,\{\mathcal{D}_3^{kc},\{J^r,G^{re}\}\}\} \nonumber \\
&  & \mbox{\hglue0.6truecm} + 2 \epsilon^{kim} f^{ab8} \{J^2,\{\{J^i,G^{m8}\},\{T^a,\{G^{rb},G^{rc}\}\}\}\} - 5 i \epsilon^{kil} \{J^2,[\{J^i,G^{l8}\},\{\{J^m,G^{m8}\},\{J^r,G^{rc}\}\}]\} \nonumber \\
&  & \mbox{\hglue0.6truecm} + \frac32 \{\mathcal{D}_3^{kc},\{\{J^m,G^{m8}\},\{J^r,G^{r8}\}\}\} + 7 i \epsilon^{kil} \{J^2,\{J^i,\{J^r,[G^{l8},\{G^{r8},\{J^m,G^{mc}\}\}]\}\}\},
\end{eqnarray}

\begin{eqnarray}
&  & [\mathcal{O}_3^{i8},[\mathcal{O}_3^{i8},\mathcal{O}_3^{kc}]] \nonumber \\
&  &\mbox{\hglue0.2truecm} = - \frac{17872N_c-37221}{768} f^{c8e} f^{8eg} G^{kg} - \frac{196856N_c-92903}{101376} i \epsilon^{kim} f^{c8e} f^{8eg} \{J^i,G^{mg}\} - \frac{17872N_c-38589}{2304} f^{c8e} f^{8eg} \mathcal{D}_3^{kg} \nonumber \\
&  & \mbox{\hglue0.6truecm} + \frac{N_c(323928N_c-612391)}{101376} (i f^{c8e} d^{8eg} \mathcal{D}_3^{kg} + i d^{c8e} f^{8eg} \mathcal{D}_3^{kg}) + \frac{3431}{2304} f^{c8e} f^{8eg} \mathcal{O}_3^{kg} - \frac{17872N_c-37221}{1152} d^{c8e} d^{8eg} \mathcal{O}_3^{kg} \nonumber \\
&  & \mbox{\hglue0.6truecm} + \frac{17872N_c-37221}{1152} d^{ceg} d^{88e} \mathcal{O}_3^{kg} - \frac{17872N_c-37221}{576} \{G^{kc},\{G^{r8},G^{r8}\}\} + \frac{17872N_c-37221}{576} \{G^{k8},\{G^{r8},G^{rc}\}\} \nonumber \\
&  & \mbox{\hglue0.6truecm} + \frac{17872N_c-37221}{1152} d^{c8e} \{J^k,\{G^{re},G^{r8}\}\} - \frac{17872N_c-37221}{1152} d^{88e} \{J^k,\{G^{rc},G^{re}\}\} \nonumber \\
&  & \mbox{\hglue0.6truecm} - \frac{17872N_c-37221}{576} d^{c8e} \{G^{ke},\{J^r,G^{r8}\}\} + \frac{17872N_c-37221}{1152} d^{c8e} \{G^{k8},\{J^r,G^{re}\}\} \nonumber \\
&  & \mbox{\hglue0.6truecm} - \frac{17872N_c-37221}{1152} d^{88e} \{G^{kc},\{J^r,G^{re}\}\} + \frac{17872N_c-37221}{576} d^{88e} \{G^{ke},\{J^r,G^{rc}\}\} \nonumber \\
&  & \mbox{\hglue0.6truecm} - \frac{71488N_c-2085N_f-140544}{9216} \epsilon^{kim} f^{c8e} \{T^e,\{J^i,G^{m8}\}\} - \frac{323928N_c-612391}{50688} i f^{c8e} d^{8eg} \mathcal{D}_4^{kg} \nonumber \\
&  & \mbox{\hglue0.6truecm} - \frac{77}{192} i \epsilon^{kim} f^{c8e} f^{8eg} \{J^2,\{J^i,G^{mg}\}\} + \frac{24904N_c^2+88N_c(566N_f-3681)+612391}{25344N_f} i \epsilon^{kim} \delta^{c8} \{J^2,\{J^i,G^{m8}\}\} \nonumber \\
&  & \mbox{\hglue0.6truecm} - \frac{196856N_c-92903}{50688} i f^{c8e} \{\mathcal{D}_2^{ke},\{J^r,G^{r8}\}\} + \frac{260392N_c-352647}{12672} i \epsilon^{kim} \{\{G^{r8},G^{rc}\},\{J^i,G^{m8}\}\} \nonumber \\
&  & \mbox{\hglue0.6truecm} - \frac{196856N_c-92903}{25344} i \epsilon^{kim} \{\{G^{r8},G^{r8}\},\{J^i,G^{mc}\}\} + \frac{196856N_c-92903}{25344} i \epsilon^{rim} \{G^{k8},\{J^r,\{G^{ic},G^{m8}\}\}\} \nonumber \\
&  & \mbox{\hglue0.6truecm} - \frac{323928N_c-612391}{50688} i \epsilon^{rim} d^{c8e} \{J^k,\{J^r,\{G^{i8},G^{me}\}\}\} - \frac{323928N_c-612391}{67584} i \epsilon^{kim} f^{cae} f^{8eb} \{\{J^i,G^{m8}\},\{T^a,T^b\}\} \nonumber \\
&  & \mbox{\hglue0.6truecm} - \frac{696344N_c-1094845}{50688} i f^{c8e} \{J^k,[\{J^m,G^{me}\},\{J^r,G^{r8}\}]\} - \frac{696344N_c-1094845}{50688} i f^{c8e} \{\{J^r,G^{re}\},[J^2,G^{k8}]\} \nonumber \\
&  & \mbox{\hglue0.6truecm} + \frac{696344N_c-1094845}{50688} i f^{c8e} \{\{J^r,G^{r8}\},[J^2,G^{ke}]\} + \frac{696344N_c-1094845}{50688} i f^{c8e} \{J^2,[G^{ke},\{J^r,G^{r8}\}]\} \nonumber \\
&  & \mbox{\hglue0.6truecm} - \frac{696344N_c-1035869}{50688} i f^{c8e} \{J^2,[G^{k8},\{J^r,G^{re}\}]\} + \frac{323928N_c-612391}{50688} d^{c8e} \{J^2,[G^{ke},\{J^r,G^{r8}\}]\} \nonumber \\
&  & \mbox{\hglue0.6truecm} - \frac{323928N_c-612391}{50688} d^{c8e} \{J^2,[G^{k8},\{J^r,G^{re}\}]\} - \frac{3971N_c-16234}{3168} [G^{kc},\{\{J^m,G^{m8}\},\{J^r,G^{r8}\}\}] \nonumber \\
&  & \mbox{\hglue0.6truecm} - \frac{260392N_c-352647}{25344} [G^{k8},\{\{J^m,G^{m8}\},\{J^r,G^{rc}\}\}] + \frac{260392N_c-352647}{25344} \{\{J^m,G^{mc}\},[G^{k8},\{J^r,G^{r8}\}]\} \nonumber \\
&  & \mbox{\hglue0.6truecm} + \frac{323928N_c-612391}{50688} i \epsilon^{kim} f^{cea} f^{e8b} \{\{J^i,G^{m8}\},\{G^{ra},G^{rb}\}\} + \frac{19}{48} f^{c8e} f^{8eg} \mathcal{D}_5^{kg} \nonumber \\
&  & \mbox{\hglue0.6truecm} - \frac{283N_c}{576} (i f^{c8e} d^{8eg} \mathcal{D}_5^{kg} + i d^{c8e} f^{8eg} \mathcal{D}_5^{kg}) + \frac{29}{8} f^{c8e} f^{8eg} \mathcal{O}_5^{kg} + \frac{91}{24} d^{c8e} d^{8eg} \mathcal{O}_5^{kg} - \frac{343}{72} d^{ceg} d^{88e} \mathcal{O}_5^{kg} + \frac{12}{N_f} \delta^{c8} \mathcal{O}_5^{k8} \nonumber \\
&  & \mbox{\hglue0.6truecm} + \frac{455}{576} \{J^2,\{G^{kc},\{G^{r8},G^{r8}\}\}\} - \frac{1607}{576} \{J^2,\{G^{k8},\{G^{r8},G^{rc}\}\}\} - \frac{19}{24} d^{c8e} \{J^2,\{J^k,\{G^{re},G^{r8}\}\}\} \nonumber \\
&  & \mbox{\hglue0.6truecm} + \frac{19}{24} d^{88e} \{J^2,\{J^k,\{G^{rc},G^{re}\}\}\} - \frac{35744N_c-72115}{1152} d^{c8e} \{J^2,\{G^{ke},\{J^r,G^{r8}\}\}\} + \frac{53}{24} d^{c8e} \{J^2,\{G^{k8},\{J^r,G^{re}\}\}\} \nonumber \\
&  & \mbox{\hglue0.6truecm} + \frac{343}{72} d^{88e} \{J^2,\{G^{kc},\{J^r,G^{re}\}\}\} - \frac{35744N_c-68499}{1152} d^{88e} \{J^2,\{G^{ke},\{J^r,G^{rc}\}\}\} \nonumber \\
&  & \mbox{\hglue0.6truecm} + \frac{9N_f+55}{48} \epsilon^{kim} f^{c8e} \{J^2,\{T^e,\{J^i,G^{m8}\}\}\} + \frac{17872N_c-37221}{576} \{G^{kc},\{\{J^m,G^{m8}\},\{J^r,G^{r8}\}\}\} \nonumber \\
&  & \mbox{\hglue0.6truecm} - \frac{17872N_c-37221}{576} \{G^{k8},\{\{J^m,G^{m8}\},\{J^r,G^{rc}\}\}\} + \frac{457}{1152}\{J^k,\{\{J^m,G^{mc}\},\{G^{r8},G^{r8}\}\}\} \nonumber \\
&  & \mbox{\hglue0.6truecm} + \frac{695}{1152}\{J^k,\{\{J^m,G^{m8}\},\{G^{r8},G^{rc}\}\}\} + \frac{35744N_c-73747}{2304} d^{c8e} \{\mathcal{D}_3^{ke},\{J^r,G^{r8}\}\} \nonumber \\
&  & \mbox{\hglue0.6truecm} + \frac{35744N_c-74899}{2304} d^{88e} \{\mathcal{D}_3^{kc},\{J^r,G^{re}\}\} - \frac{695}{1152} \epsilon^{kim} f^{ab8} \{\{J^i,G^{m8}\},\{T^a,\{G^{rb},G^{rc}\}\}\} \nonumber \\
&  & \mbox{\hglue0.6truecm} + \frac{283(N_c+N_f)}{576} i \epsilon^{kim} d^{c8e} \{J^2,\{T^e,\{J^i,G^{m8}\}\}\} + \frac{35744N_c-73747}{1152} i \epsilon^{kil} [\{J^i,G^{l8}\},\{\{J^m,G^{m8}\},\{J^r,G^{rc}\}\}] \nonumber
\end{eqnarray}
\begin{eqnarray}
&  & \mbox{\hglue0.6truecm} + \frac{283}{288} i f^{c8e} d^{8eg} \mathcal{D}_6^{kg} + \frac{1843}{1584} d^{c8e} \{J^2,\{J^2,\{G^{ke},T^8\}\}\} + \frac{283}{144 N_f} i \epsilon^{kim} \delta^{c8} \{J^2,\{J^2,\{J^i,G^{m8}\}\}\} \nonumber \\
&  & \mbox{\hglue0.6truecm} - \frac{1843}{1584} d^{c8e} \{J^2,\{\mathcal{D}_2^{k8},\{J^r,G^{re}\}\}\} - \frac{77}{96} i f^{c8e} \{J^2,\{\mathcal{D}_2^{ke},\{J^r,G^{r8}\}\}\} - \frac{1843}{1584}\{J^2,\{\{J^r,G^{rc}\},\{G^{k8},T^8\}\}\} \nonumber \\
&  & \mbox{\hglue0.6truecm} + \frac{1843}{1584}\{J^2,\{\{J^r,G^{r8}\},\{G^{kc},T^8\}\}\} - \frac{283}{576} i \epsilon^{kim} \{J^2,\{\{T^c,T^8\},\{J^i,G^{m8}\}\}\} \nonumber \\
&  & \mbox{\hglue0.6truecm} - \frac{13}{36} i \epsilon^{kim} \{J^2,\{\{G^{r8},G^{rc}\},\{J^i,G^{m8}\}\}\} - \frac{77}{48} i \epsilon^{kim} \{J^2,\{\{G^{r8},G^{r8}\},\{J^i,G^{mc}\}\}\} \nonumber \\
&  & \mbox{\hglue0.6truecm} + \frac{77}{48} i \epsilon^{rim} \{J^2,\{G^{k8},\{J^r,\{G^{ic},G^{m8}\}\}\}\} + \frac{283}{288} i \epsilon^{rim} d^{c8e} \{J^2,\{J^k,\{J^r,\{G^{i8},G^{me}\}\}\}\} \nonumber \\
&  & \mbox{\hglue0.6truecm} + \frac{283}{288} i \epsilon^{kim} f^{cae} f^{8eb} \{J^2,\{\{J^i,G^{m8}\},\{T^a,T^b\}\}\} + \frac{73}{96} i f^{c8e} \{J^2,\{J^k,[\{J^m,G^{me}\},\{J^r,G^{r8}\}]\}\} \nonumber \\
&  & \mbox{\hglue0.6truecm} + \frac{73}{96} i f^{c8e} \{J^2,\{\{J^r,G^{re}\},[J^2,G^{k8}]\}\} - \frac{6095}{3168} i f^{c8e} \{J^2,\{\{J^r,G^{r8}\},[J^2,G^{ke}]\}\} \nonumber \\
&  & \mbox{\hglue0.6truecm} - \frac{73}{96} i f^{c8e} \{J^2,\{J^2,[G^{ke},\{J^r,G^{r8}\}]\}\} + \frac{73}{96} i f^{c8e} \{J^2,\{J^2,[G^{k8},\{J^r,G^{re}\}]\}\} \nonumber \\
&  & \mbox{\hglue0.6truecm} + \frac{1843}{1584} i \epsilon^{kim} [\{T^8,\{J^r,G^{r8}\}\},\{J^2,\{J^i,G^{mc}\}\}] - \frac{283}{288} d^{c8e} \{J^2,\{J^2,[G^{ke},\{J^r,G^{r8}\}]\}\} \nonumber \\
&  & \mbox{\hglue0.6truecm} + \frac{283}{288} d^{c8e} \{J^2,\{J^2,[G^{k8},\{J^r,G^{re}\}]\}\} + \frac{257}{288} \{J^2,[G^{kc},\{\{J^m,G^{m8}\},\{J^r,G^{r8}\}\}]\} \nonumber \\
&  & \mbox{\hglue0.6truecm} + \frac{13}{72} \{J^2,[G^{k8},\{\{J^m,G^{m8}\},\{J^r,G^{rc}\}\}]\} - \frac{13}{72} \{J^2,\{\{J^m,G^{mc}\},[G^{k8},\{J^r,G^{r8}\}]\}\} \nonumber \\
&  & \mbox{\hglue0.6truecm} + \frac54 f^{c8e} f^{8eg} \mathcal{O}_7^{kg} + \frac54 d^{c8e} d^{8eg} \mathcal{O}_7^{kg} - d^{ceg} d^{88e} \mathcal{O}_7^{kg} + \frac{5}{N_f} \delta^{c8} \mathcal{O}_7^{k8} - \frac52 \{J^2,\{J^2,\{G^{kc},\{G^{r8},G^{r8}\}\}\}\} \nonumber \\
&  & \mbox{\hglue0.6truecm} - \frac32 \{J^2,\{J^2,\{G^{k8},\{G^{r8},G^{rc}\}\}\}\} + \frac{13}{12} d^{c8e} \{J^2,\{J^2,\{G^{ke},\{J^r,G^{r8}\}\}\}\} + \frac54 d^{c8e} \{J^2,\{J^2,\{G^{k8},\{J^r,G^{re}\}\}\}\} \nonumber \\
&  & \mbox{\hglue0.6truecm} + d^{88e} \{J^2,\{J^2,\{G^{kc},\{J^r,G^{re}\}\}\}\} + \frac43 d^{88e} \{J^2,\{J^2,\{G^{ke},\{J^r,G^{rc}\}\}\}\} + \frac58 \epsilon^{kim} f^{c8e} \{J^2,\{J^2,\{T^e,\{J^i,G^{m8}\}\}\}\} \nonumber \\
&  & \mbox{\hglue0.6truecm} - \frac{13}{12} \{J^2,\{G^{kc},\{\{J^m,G^{m8}\},\{J^r,G^{r8}\}\}\}\} + \frac{19}{12} \{J^2,\{G^{k8},\{\{J^m,G^{m8}\},\{J^r,G^{rc}\}\}\}\} \nonumber \\
&  & \mbox{\hglue0.6truecm} + \frac54 \{J^2,\{J^k,\{\{J^m,G^{mc}\},\{G^{r8},G^{r8}\}\}\}\} + \frac34 \{J^2,\{J^k,\{\{J^m,G^{m8}\},\{G^{r8},G^{rc}\}\}\}\} \nonumber \\
&  & \mbox{\hglue0.6truecm} - \frac76 d^{c8e} \{J^2,\{\mathcal{D}_3^{ke},\{J^r,G^{r8}\}\}\} - \frac76 d^{88e} \{J^2,\{\mathcal{D}_3^{kc},\{J^r,G^{re}\}\}\} - \frac12 \epsilon^{kim} f^{ab8} \{J^2,\{\{J^i,G^{m8}\},\{T^a,\{G^{rb},G^{rc}\}\}\}\} \nonumber \\
&  & \mbox{\hglue0.6truecm} + \frac{85}{36} i \epsilon^{kil} \{J^2,[\{J^i,G^{l8}\},\{\{J^m,G^{m8}\},\{J^r,G^{rc}\}\}]\} - \frac14 \{\mathcal{D}_3^{kc},\{\{J^m,G^{m8}\},\{J^r,G^{r8}\}\}\} \nonumber \\
&  & \mbox{\hglue0.6truecm} - \frac{71}{18} i \epsilon^{kil} \{J^2,\{J^i,\{J^r,[G^{l8},\{G^{r8},\{J^m,G^{mc}\}\}]\}\}\}.
\end{eqnarray}

\section{\label{sec:bases}Operator bases for flavor $\mathbf{8}$ and flavor $\mathbf{27}$ representations}

Loop graphs \ref{fig:l1}(a,b,c) contribute to the renormalization of the baryon axial vector current in the way indicated by Eq.~(\ref{eq:dasplit}). The pertinent flavor $\mathbf{8}$ and flavor $\mathbf{27}$ operator bases can be organized into $n$-body operators for $n=1,\ldots,7$. Constructing these operator bases is a rather involved task, because all participating operators must be linearly independent. Constructing complete operator bases is far beyond the scope of the present paper.

The operator basis for the flavor $\mathbf{8}$ representation can conveniently be chosen as
\begin{eqnarray}
\begin{array}{lll}
O_{1}^{kc} = d^{c8e} G^{ke}, &
O_{2}^{kc} = \delta^{c8} J^k, \nonumber \\
O_{3}^{kc} = d^{c8e} \mathcal{D}_2^{ke}, &
O_{4}^{kc} = \{G^{kc},T^8\}, \nonumber \\
O_{5}^{kc} = \{G^{k8},T^c\}, &
O_{6}^{kc} = [J^2,[T^8,G^{kc}]], \nonumber \\
O_{7}^{kc} = d^{c8e} \mathcal{D}_3^{ke}, &
O_{8}^{kc} = d^{c8e} \mathcal{O}_3^{ke}, \nonumber \\
O_{9}^{kc} = \{G^{kc},\{J^r,G^{r8}\}\}, &
O_{10}^{kc} = \{G^{k8},\{J^r,G^{rc}\}\}, \nonumber \\ 
O_{11}^{kc} = \{J^k,\{T^c,T^8\}\}, &
O_{12}^{kc} = \{J^k,\{G^{rc},G^{r8}\}\}, \nonumber \\
O_{13}^{kc} = \delta^{c8} \{J^2,J^k\}, &
O_{14}^{kc} = d^{c8e} \mathcal{D}_4^{ke}, \nonumber \\
O_{15}^{kc} = \{\mathcal{D}_2^{kc},\{J^r,G^{r8}\}\}, &
O_{16}^{kc} = \{\mathcal{D}_2^{k8},\{J^r,G^{rc}\}\}, \nonumber \\ 
O_{17}^{kc} = \{J^2,\{G^{kc},T^8\}\}, &
O_{18}^{kc} = \{J^2,\{G^{k8},T^c\}\}, \nonumber \\
O_{19}^{kc} = \{J^2,[J^2,[T^8,G^{kc}]]\}, &
O_{20}^{kc} = \{J^2,[G^{kc},\{J^r,G^{r8}\}]\}, \nonumber \\
O_{21}^{kc} = \{J^2,[G^{k8},\{J^r,G^{rc}\}]\}, &
O_{22}^{kc} = \{[J^2,G^{kc}],\{J^r,G^{r8}\}], \nonumber \\
O_{23}^{kc} = \{[J^2,G^{k8}],\{J^r,G^{rc}\}], &
O_{24}^{kc} = \{J^k,[\{J^m,G^{mc}\},\{J^r,G^{r8}\}]\}, \nonumber \\
O_{25}^{kc} = d^{c8e} \mathcal{D}_5^{ke}, &
O_{26}^{kc} = d^{c8e} \mathcal{O}_5^{ke}, \nonumber \\
O_{27}^{kc} = \{J^2,\{G^{kc},\{J^r,G^{r8}\}\}\}, &
O_{28}^{kc} = \{J^2,\{G^{k8},\{J^r,G^{rc}\}\}\}, \nonumber \\
O_{29}^{kc} = \{J^2,\{J^k,\{T^c,T^8\}\}\}, &
O_{30}^{kc} = \{J^2,\{J^k,\{G^{rc},G^{r8}\}\}\}, \nonumber \\
O_{31}^{kc} = \{J^k,\{\{J^m,G^{mc}\},\{J^r,G^{r8}\}\}\}, &
O_{32}^{kc} = \delta^{c8} \{J^2,\{J^2,J^k\}\}, \nonumber \\
O_{33}^{kc} = d^{c8e} \mathcal{D}_6^{ke}, &
O_{34}^{kc} = \{J^2,\{\mathcal{D}_2^{kc},\{J^r,G^{r8}\}\}\}, \nonumber \\
O_{35}^{kc} = \{J^2,\{\mathcal{D}_2^{k8},\{J^r,G^{rc}\}\}\}, &
O_{36}^{kc} = \{J^2,\{J^2,\{G^{kc},T^8\}\}\}, \nonumber \\
O_{37}^{kc} = \{J^2,\{J^2,\{G^{k8},T^c\}\}\}, &
O_{38}^{kc} = \{J^2,\{J^2,[J^2,[T^8,G^{kc}]]\}\}, \nonumber \\
O_{39}^{kc} = \{J^2,\{J^2,[G^{kc},\{J^r,G^{r8}\}]\}\}, &
O_{40}^{kc} = \{J^2,\{J^2,[G^{k8},\{J^r,G^{rc}\}]\}\}, \nonumber \\ 
O_{41}^{kc} = \{J^2,\{[J^2,G^{kc}],\{J^r,G^{r8}\}\}\}, &
O_{42}^{kc} = \{J^2,\{[J^2,G^{k8}],\{J^r,G^{rc}\}\}\}, \nonumber \\
O_{43}^{kc} = \{J^2,\{J^k,[\{J^m,G^{mc}\},\{J^r,G^{r8}\}]\}\}, &
O_{44}^{kc} = d^{c8e} \mathcal{D}_7^{ke}, \nonumber \\
O_{45}^{kc} = d^{c8e} \mathcal{O}_7^{ke}, &
O_{46}^{kc} = \{J^2,\{J^2,\{G^{kc},\{J^r,G^{r8}\}\}\}\}, \nonumber \\
O_{47}^{kc} = \{J^2,\{J^2,\{G^{k8},\{J^r,G^{rc}\}\}\}\}, &
O_{48}^{kc} = \{J^2,\{J^2,\{J^k,\{T^c,T^8\}\}\}\}, \nonumber \\
O_{49}^{kc} = \{J^2,\{J^2,\{J^k,\{G^{rc},G^{r8}\}\}\}\}, &
O_{50}^{kc} = \{J^2,\{J^k,\{\{J^m,G^{mc}\},\{J^r,G^{r8}\}\}\}\}, \nonumber \\
O_{51}^{kc} = \delta^{c8} \{J^2,\{J^2,\{J^2,J^k\}\}\}. &
\end{array}
\end{eqnarray}

For the flavor $\mathbf{27}$ representation, the construction of a linearly independent operator basis becomes more difficult because of 
the considerable number of operator identities among them. Some of these identities can be outlined as
\begin{equation}
i f^{c8e} d^{8eg} G^{kg} + i d^{c8e} f^{8eg} G^{kg} = \frac{i}{N_c} f^{c8e} d^{8eg} \mathcal{D}_2^{kg} + \frac{i}{N_c} d^{c8e} f^{8eg} \mathcal{D}_2^{kg},
\end{equation}
\begin{equation}
\{\mathcal{D}_2^{k8},\{T^c,T^8\}\} - \{\mathcal{D}_2^{kc},\{T^8,T^8\}\} + f^{c8e} f^{8eg} \mathcal{D}_2^{kg} = 0,
\end{equation}
\begin{equation}
i d^{c8e} f^{8eg} \mathcal{O}_3^{kg} + i f^{c8e} d^{8eg} \mathcal{O}_3^{kg} = 0,
\end{equation}
\begin{equation}
\frac{i}{2} f^{c8e} d^{8eg} \mathcal{D}_3^{kg} + \frac{i}{2} d^{c8e} f^{8eg} \mathcal{D}_3^{kg} = \frac{1}{N_c} i f^{c8e} d^{8eg} \mathcal{D}_4^{kg} + \frac{1}{N_c} i d^{c8e} f^{8eg} \mathcal{D}_4^{kg},
\end{equation}
\begin{equation}
i d^{c8e} f^{8eg} \mathcal{O}_5^{kg} + i f^{c8e} d^{8eg} \mathcal{O}_5^{kg} = 0,
\end{equation}
\begin{equation}
\frac{i}{2} f^{c8e} d^{8eg} \mathcal{D}_5^{kg} + \frac{i}{2} d^{c8e} f^{8eg} \mathcal{D}_5^{kg} = \frac{1}{N_c} i f^{c8e} d^{8eg} \mathcal{D}_6^{kg} + \frac{1}{N_c} i d^{c8e} f^{8eg} \mathcal{D}_6^{kg},
\end{equation}
\begin{equation}
[G^{k8},\{\{J^r,G^{r8}\},\{J^m,G^{mc}\}\}] - \{\{J^r,G^{rc}\},[G^{k8},\{J^m,G^{m8}\}]\} = \{\{J^r,G^{r8}\},[G^{k8},\{J^m,G^{mc}\}]\}
\end{equation}
\begin{equation}
d^{c8e} \{\mathcal{D}_3^{ke},\{J^r, G^{r8}\}\} - d^{c8e} \{\mathcal{D}_3^{k8},\{J^r, G^{re}\}\} = 0,
\end{equation}
\begin{equation}
d^{88e} \{\mathcal{D}_3^{kc},\{J^r,G^{re}\}\} - d^{88e} \{\mathcal{D}_3^{ke},\{J^r,G^{rc}\}\} = 0.
\end{equation}
The above list, however, is not exhaustive at all, but helps identify some linearly dependent operators. After a careful analysis, the operator basis for the flavor $\mathbf{27}$ representation is conveniently chosen as
\begin{eqnarray}
\begin{array}{lll}
T_{1}^{kc} = f^{c8e} f^{8eg} G^{kg}, &
T_{2}^{kc} = d^{c8e} d^{8eg} G^{kg}, \nonumber \\
T_{3}^{kc} = \delta^{c8} G^{k8}, &
T_{4}^{kc} = d^{c88} J^k, \nonumber \\
T_{5}^{kc} = f^{c8e} f^{8eg} \mathcal{D}_2^{kg}, &
T_{6}^{kc} = d^{c8e} d^{8eg} \mathcal{D}_2^{kg}, \nonumber \\
T_{7}^{kc} = d^{ceg} d^{88e} \mathcal{D}_2^{kg}, &
T_{8}^{kc} = \delta^{c8} \mathcal{D}_2^{k8}, \nonumber \\
T_{9}^{kc} = d^{c8e} \{G^{ke},T^8\}, &
T_{10}^{kc} = d^{88e} \{G^{ke},T^c\}, \nonumber \\
T_{11}^{kc} = i \epsilon^{kim} f^{c8e} f^{8eg} \{J^i,G^{mg}\}, &
T_{12}^{kc} = i f^{c8e} [G^{ke},\{J^r,G^{r8}\}], \nonumber \\
T_{13}^{kc} = i f^{c8e} [G^{k8},\{J^r,G^{re}\}], &
T_{14}^{kc} = f^{c8e} f^{8eg} \mathcal{D}_3^{kg}, \nonumber \\
T_{15}^{kc} = d^{c8e} d^{8eg} \mathcal{D}_3^{kg}, &
T_{16}^{kc} = d^{ceg} d^{88e} \mathcal{D}_3^{kg}, \nonumber \\
T_{17}^{kc} = i f^{c8e} d^{8eg} \mathcal{D}_3^{kg}, &
T_{18}^{kc} = i d^{c8e} f^{8eg} \mathcal{D}_3^{kg}, \nonumber \\
T_{19}^{kc} = \delta^{c8} \mathcal{D}_3^{k8}, &
T_{20}^{kc} = f^{c8e} f^{8eg} \mathcal{O}_3^{kg}, \nonumber \\
T_{21}^{kc} = d^{c8e} d^{8eg} \mathcal{O}_3^{kg}, &
T_{22}^{kc} = d^{ceg} d^{88e} \mathcal{O}_3^{kg}, \nonumber \\
T_{23}^{kc} = \delta^{c8} \mathcal{O}_3^{k8}, &
T_{24}^{kc} = d^{c88} \{J^2,J^k\}, \nonumber \\
T_{25}^{kc} = \{G^{kc},\{T^8,T^8\}\}, &
T_{26}^{kc} = \{G^{k8},\{T^c,T^8\}\}, \nonumber \\
T_{27}^{kc} = \{G^{kc},\{G^{r8},G^{r8}\}\}, &
T_{28}^{kc} = \{G^{k8},\{G^{rc},G^{r8}\}\}, \nonumber \\
T_{29}^{kc} = d^{c8e} \{J^k,\{G^{re},G^{r8}\}\}, &
T_{30}^{kc} = d^{88e} \{J^k,\{G^{rc},G^{re}\}\}, \nonumber \\
T_{31}^{kc} = d^{c8e} \{G^{ke},\{J^r,G^{r8}\}\}, &
T_{32}^{kc} = d^{c8e} \{G^{k8},\{J^r,G^{re}\}\}, \nonumber \\
T_{33}^{kc} = d^{88e} \{G^{kc},\{J^r,G^{re}\}\}, &
T_{34}^{kc} = d^{88e} \{G^{ke},\{J^r,G^{rc}\}\}, \nonumber \\
T_{35}^{kc} = \epsilon^{kim} f^{c8e} \{T^e,\{J^i,G^{m8}\}\}, &
T_{36}^{kc} = f^{c8e} f^{8eg} \mathcal{D}_4^{kg}, \nonumber \\
T_{37}^{kc} = d^{c8e} d^{8eg} \mathcal{D}_4^{kg}, &
T_{38}^{kc} = d^{ceg} d^{88e} \mathcal{D}_4^{kg}, \nonumber \\
T_{39}^{kc} = i f^{c8e} d^{8eg} \mathcal{D}_4^{kg}, &
T_{40}^{kc} = \delta^{c8} \mathcal{D}_4^{k8}, \nonumber \\
T_{41}^{kc} = \{J^2,T_{9}^{kc}\}, &
T_{42}^{kc} = \{J^2,T_{10}^{kc}\}, \nonumber \\
T_{43}^{kc} = \{J^2,T_{11}^{kc}\}, &
T_{44}^{kc} = i \epsilon^{kim} \delta^{c8} \{J^2,\{J^i,G^{m8}\}\}, \nonumber \\
T_{45}^{kc} = \{\mathcal{D}_2^{kc},\{T^8,T^8\}\}, &
T_{46}^{kc} = \{\mathcal{D}_2^{kc},\{G^{r8},G^{r8}\}\}, \nonumber \\
T_{47}^{kc} = \{\mathcal{D}_2^{k8},\{G^{rc},G^{r8}\}\}, &
T_{48}^{kc} = d^{c8e} \{\mathcal{D}_2^{k8},\{J^r,G^{re}\}\}, \nonumber \\
T_{49}^{kc} = d^{88e} \{\mathcal{D}_2^{kc},\{J^r,G^{re}\}\}, &
T_{50}^{kc} = i f^{c8e} \{\mathcal{D}_2^{ke},\{J^r,G^{r8}\}\}, \nonumber \\
T_{51}^{kc} = \{\{J^r,G^{rc}\},\{G^{k8},T^8\}\}, &
T_{52}^{kc} = \{\{J^r,G^{r8}\},\{G^{kc},T^8\}\}, \nonumber \\
T_{53}^{kc} = \{\{J^r,G^{r8}\},\{G^{k8},T^c\}\}, &
T_{54}^{kc} = i \epsilon^{kim} \{\{J^i,G^{m8}\},\{G^{r8},G^{rc}\}\}, \nonumber \\
T_{55}^{kc} = i \epsilon^{kim} \{\{J^i,G^{mc}\},\{G^{r8},G^{r8}\}\}, &
T_{56}^{kc} = i \epsilon^{rim} \{G^{k8},\{J^r,\{G^{ic},G^{m8}\}\}\}, \nonumber \\
T_{57}^{kc} = i \epsilon^{rim} d^{c8e} \{J^k,\{J^r,\{G^{i8},G^{me}\}\}\}, &
T_{58}^{kc} = i \epsilon^{kim} f^{cae} f^{8eb} \{\{J^i,G^{m8}\},\{T^a,T^b\}\}, \nonumber \\
T_{59}^{kc} = i f^{c8e} \{J^k,[\{J^i,G^{ie}\},\{J^r,G^{r8}\}]\}, &
T_{60}^{kc} = i f^{c8e} \{\{J^r,G^{re}\},[J^2,G^{k8}]\}, \nonumber \\
T_{61}^{kc} = i f^{c8e} \{\{J^r,G^{r8}\},[J^2,G^{ke}]\}, &
T_{62}^{kc} = \{J^2,T_{12}^{kc}\}, \nonumber \\
T_{63}^{kc} = \{J^2,T_{13}^{kc}\}, &
T_{64}^{kc} = d^{c8e} \{J^2,[G^{ke},\{J^r,G^{r8}\}]\}, \nonumber \\
T_{65}^{kc} = d^{c8e} \{J^2,[G^{k8},\{J^r,G^{re}\}]\}, &
T_{66}^{kc} = [G^{kc},\{\{J^m,G^{m8}\},\{J^r,G^{r8}\}\}], \nonumber \\
T_{67}^{kc} = [G^{k8},\{\{J^m,G^{m8}\},\{J^r,G^{rc}\}\}], &
T_{68}^{kc} = \{\{J^m,G^{mc}\},[G^{k8},\{J^r,G^{r8}\}]\}, \nonumber \\
T_{69}^{kc} = i \epsilon^{kim} f^{cea} f^{e8b} \{\{J^i,G^{m8}\},\{G^{ra},G^{rb}\}\}, &
T_{70}^{kc} = f^{c8e} f^{8eg} \mathcal{D}_5^{kg}, \nonumber \\
T_{71}^{kc} = d^{c8e} d^{8eg} \mathcal{D}_5^{kg}, &
T_{72}^{kc} = d^{ceg} d^{88e} \mathcal{D}_5^{kg}, \nonumber \\
T_{73}^{kc} = i f^{c8e} d^{8eg} \mathcal{D}_5^{kg}, &
T_{74}^{kc} = i d^{c8e} f^{8eg} \mathcal{D}_5^{kg}, \nonumber \\
T_{75}^{kc} = \delta^{c8} \mathcal{D}_5^{k8}, &
T_{76}^{kc} = f^{c8e} f^{8eg} \mathcal{O}_5^{kg}, \nonumber \\
T_{77}^{kc} = d^{c8e} d^{8eg} \mathcal{O}_5^{kg}, &
T_{78}^{kc} = d^{ceg} d^{88e} \mathcal{O}_5^{kg}, \nonumber \\
T_{79}^{kc} = \delta^{c8} \mathcal{O}_5^{k8}, &
T_{80}^{kc} = d^{c88} \{J^2,\{J^2,J^k\}\}, \nonumber \\
T_{81}^{kc} = \{J^2,T_{25}^{kc}\}, &
T_{82}^{kc} = \{J^2,T_{26}^{kc}\}, \nonumber \\
T_{83}^{kc} = \{J^2,T_{27}^{kc}\}, &
T_{84}^{kc} = \{J^2,T_{28}^{kc}\}, \nonumber \\
T_{85}^{kc} = \{J^2,T_{29}^{kc}\}, &
T_{86}^{kc} = \{J^2,T_{30}^{kc}\}, \nonumber \\
T_{87}^{kc} = \{J^2,T_{31}^{kc}\}, &
T_{88}^{kc} = \{J^2,T_{32}^{kc}\}, \nonumber \\
T_{89}^{kc} = \{J^2,T_{33}^{kc}\}, &
T_{90}^{kc} = \{J^2,T_{34}^{kc}\}, \nonumber \\
T_{91}^{kc} = \{J^2,T_{35}^{kc}\}, &
T_{92}^{kc} = \{G^{kc},\{\{J^m,G^{m8}\},\{J^r,G^{r8}\}\}\}, \nonumber \\
T_{93}^{kc} = \{G^{k8},\{\{J^m,G^{m8}\},\{J^r,G^{rc}\}\}\}, &
T_{94}^{kc} = \{J^k,\{\{J^m,G^{mc}\},\{G^{r8},G^{r8}\}\}\}, \nonumber \\
T_{95}^{kc} = \{J^k,\{\{J^m,G^{m8}\},\{G^{r8},G^{rc}\}\}\}, &
T_{96}^{kc} = \{\mathcal{D}_2^{kc},\{T^8,\{J^r,G^{r8}\}\}\}, \nonumber \\
T_{97}^{kc} = \{\mathcal{D}_2^{k8},\{T^8,\{J^r,G^{rc}\}\}\}, &
T_{98}^{kc} = d^{c8e} \{\mathcal{D}_3^{ke},\{J^r,G^{r8}\}\}, \nonumber \\
T_{99}^{kc} = d^{88e} \{\mathcal{D}_3^{kc},\{J^r,G^{re}\}\}, &
T_{100}^{kc} = \epsilon^{kim} f^{ab8} \{\{J^i,G^{m8}\},\{T^a,\{G^{rb},G^{rc}\}\}\}, \nonumber \\
\end{array}
\end{eqnarray}
\begin{eqnarray}
\begin{array}{lll}
T_{101}^{kc} = i \epsilon^{kim} d^{c8e} \{J^2,\{T^e,\{J^i,G^{m8}\}\}\}, &
T_{102}^{kc} = i \epsilon^{kil} [\{J^i,G^{l8}\},\{\{J^m,G^{m8}\},\{J^r,G^{rc}\}\}], \nonumber \\
T_{103}^{kc} = f^{c8e} f^{8eg} \mathcal{D}_6^{kg}, &
T_{104}^{kc} = d^{c8e} d^{8eg} \mathcal{D}_6^{kg}, \nonumber \\
T_{105}^{kc} = d^{ceg} d^{88e} \mathcal{D}_6^{kg}, &
T_{106}^{kc} = i f^{c8e} d^{8eg} \mathcal{D}_6^{kg}, \nonumber \\
T_{107}^{kc} = \delta^{c8} \mathcal{D}_6^{k8}, &
T_{108}^{kc} = \{J^2,\{J^2,T_{9}^{kc}\}\}, \nonumber \\
T_{109}^{kc} = \{J^2,\{J^2,T_{10}^{kc}\}\}, &
T_{110}^{kc} = \{J^2,T_{44}^{kc}\}, \nonumber \\
T_{111}^{kc} = \{J^2,T_{46}^{kc}\}, &
T_{112}^{kc} = \{J^2,T_{47}^{kc}\}, \nonumber \\
T_{113}^{kc} = \{J^2,T_{48}^{kc}\}, &
T_{114}^{kc} = \{J^2,T_{49}^{kc}\}, \nonumber \\
T_{115}^{kc} = \{J^2,T_{50}^{kc}\}, &
T_{116}^{kc} = \{J^2,T_{51}^{kc}\}, \nonumber \\
T_{117}^{kc} = \{J^2,T_{52}^{kc}\}, &
T_{118}^{kc} = \{J^2,T_{53}^{kc}\}, \nonumber \\
T_{119}^{kc} = i \epsilon^{kim} \{J^2,\{\{T^c,T^8\},\{J^i,G^{m8}\}\}\}, &
T_{120}^{kc} = i \epsilon^{kim} \{J^2,\{\{G^{r8},G^{rc}\},\{J^i,G^{m8}\}\}\}, \nonumber \\
T_{121}^{kc} = i \epsilon^{kim} \{J^2,\{\{G^{r8},G^{r8}\},\{J^i,G^{mc}\}\}\}, &
T_{122}^{kc} = i \epsilon^{rim} \{J^2,T_{56}^{kc}\}, \nonumber \\
T_{123}^{kc} = \{J^2,T_{57}^{kc}\}, &
T_{124}^{kc} = \{J^2,T_{58}^{kc}\}, \nonumber \\
T_{125}^{kc} = \{J^2,T_{59}^{kc}\}, &
T_{126}^{kc} = \{J^2,T_{60}^{kc}\}, \nonumber \\
T_{127}^{kc} = \{J^2,T_{61}^{kc}\}, &
T_{128}^{kc} = \{J^2,\{J^2,T_{12}^{kc}\}\}, \nonumber \\
T_{129}^{kc} = \{J^2,\{J^2,T_{13}^{kc}\}\}, &
T_{130}^{kc} = \{\mathcal{D}_2^{kc},\{\{J^m,G^{m8}\},\{J^r,G^{r8}\}\}\}, \nonumber \\
T_{131}^{kc} = \{\mathcal{D}_2^{k8},\{\{J^m,G^{mc}\},\{J^r,G^{r8}\}\}\}, &
T_{132}^{kc} = i \epsilon^{kim} [\{T^8,\{J^r,G^{r8}\}\},\{J^2,\{J^i,G^{mc}\}\}], \nonumber \\
T_{133}^{kc} = \{J^2,T_{64}^{kc}\}, &
T_{134}^{kc} = \{J^2,T_{65}^{kc}\}, \nonumber \\
T_{135}^{kc} = \{J^2,T_{66}^{kc}\}, &
T_{136}^{kc} = \{J^2,T_{67}^{kc}\}, \nonumber \\
T_{137}^{kc} = \{J^2,T_{68}^{kc}\}, &
T_{138}^{kc} = f^{c8e} f^{8eg} \mathcal{D}_7^{kg}, \nonumber \\
T_{139}^{kc} = d^{c8e} d^{8eg} \mathcal{D}_7^{kg}, &
T_{140}^{kc} = d^{ceg} d^{88e} \mathcal{D}_7^{kg}, \nonumber \\
T_{141}^{kc} = \delta^{c8} \mathcal{D}_7^{k8}, &
T_{142}^{kc} = f^{c8e} f^{8eg} \mathcal{O}_7^{kg}, \nonumber \\
T_{143}^{kc} = d^{c8e} d^{8eg} \mathcal{O}_7^{kg}, &
T_{144}^{kc} = d^{ceg} d^{88e} \mathcal{O}_7^{kg}, \nonumber \\
T_{145}^{kc} = \delta^{c8} \mathcal{O}_7^{k8}, &
T_{146}^{kc} = \{J^2,\{J^2,\{J^2,T_{4}^{kc}\}\}\}, \nonumber \\
T_{147}^{kc} = \{J^2,\{J^2,T_{27}^{kc}\}\}, &
T_{148}^{kc} = \{J^2,\{J^2,T_{28}^{kc}\}\}, \nonumber \\
T_{149}^{kc} = \{J^2,\{J^2,T_{29}^{kc}\}\}, &
T_{150}^{kc} = \{J^2,\{J^2,T_{30}^{kc}\}\}, \nonumber \\
T_{151}^{kc} = \{J^2,\{J^2,T_{31}^{kc}\}\}, &
T_{152}^{kc} = \{J^2,\{J^2,T_{32}^{kc}\}\}, \nonumber \\
T_{153}^{kc} = \{J^2,\{J^2,T_{33}^{kc}\}\}, &
T_{154}^{kc} = \{J^2,\{J^2,T_{34}^{kc}\}\}, \nonumber \\
T_{155}^{kc} = \{J^2,\{J^2,T_{35}^{kc}\}\}, &
T_{156}^{kc} = \{J^2,T_{92}^{kc}\}, \nonumber \\
T_{157}^{kc} = \{J^2,T_{93}^{kc}\}, &
T_{158}^{kc} = \{J^2,T_{94}^{kc}\}, \nonumber \\
T_{159}^{kc} = \{J^2,T_{95}^{kc}\}, &
T_{160}^{kc} = \{J^2,T_{98}^{kc}\}, \nonumber \\
T_{161}^{kc} = \{J^2,T_{99}^{kc}\}, &
T_{162}^{kc} = \{J^2,T_{100}^{kc}\}, \nonumber \\
T_{163}^{kc} = \{J^2,T_{102}^{kc}\}, &
T_{164}^{kc} = \{\mathcal{D}_3^{kc},\{\{J^m,G^{m8}\},\{J^r,G^{r8}\}\}\}, \nonumber \\
T_{165}^{kc} = i \epsilon^{kil} \{J^2,\{J^i,\{J^m,[G^{l8},\{G^{m8},\{J^r,G^{rc}\}\}]\}\}\}. &
\end{array}
\end{eqnarray}

Nontrivial matrix elements of singlet, $\mathbf{8}$, and $\mathbf{27}$ operators between $SU(6)$ baryon states are listed in Tables \ref{t:mtx1}, \ref{t:mtx8}, and \ref{t:mtx27}, respectively. Subsets of $m$-body operators $\{X_m\}$ yield trivial matrix elements in two possible ways: Either $\langle X_i\rangle = 0$ or $\langle X_j \rangle = \langle\{J^2,X_{j-2}\}\rangle$. Trivial matrix elements are not listed in these tables.

\begin{table*}
\caption{\label{t:mtx1}Matrix elements of flavor singlet operators.}
\begin{ruledtabular}
\begin{tabular}{rrrrrrrr}
& $np$ & $\Xi^\pm \Lambda$ & $\Lambda p $ & $\Sigma^-n$ & $\Xi^-\Lambda$ & $\Xi^-\Sigma^0$ & $\Xi^0\Sigma^+$ \\
\hline
$\langle S_{1}^{kc}\rangle$ & $\frac56$ & $\frac{1}{\sqrt{6}}$ & $-\frac12 \sqrt{\frac32}$ & $\frac16$ & $\frac{1}{2 \sqrt{6}}$ & $\frac{5}{6 \sqrt{2}}$ & $\frac56$ \\
$\langle S_{2}^{kc}\rangle$ & $\frac12$ & $0$ & $-\frac12 \sqrt{\frac32}$ & $-\frac12$ & $\frac12 \sqrt{\frac32}$ & $\frac{1}{2 \sqrt{2}}$ & $\frac12$ \\
$\langle S_{3}^{kc}\rangle$ & $\frac52$ & $\sqrt{\frac32}$ & $-\frac32 \sqrt{\frac32}$ & $\frac12$ & $\frac12 \sqrt{\frac32}$ & $\frac{5}{2 \sqrt{2}}$ & $\frac52$ \\
\end{tabular}
\end{ruledtabular}
\end{table*}

\begin{table*}
\caption{\label{t:mtx8}Matrix elements of flavor $\mathbf{8}$ operators.}
\begin{ruledtabular}
\begin{tabular}{rrrrrrrr}
& $np$ & $\Xi^\pm \Lambda$ & $\Lambda p $ & $\Sigma^-n$ & $\Xi^-\Lambda$ & $\Xi^-\Sigma^0$ & $\Xi^0\Sigma^+$ \\
\hline
$\langle O_{1}^{kc}\rangle$ & $\frac{5}{6 \sqrt{3}}$ & $\frac{1}{3 \sqrt{2}}$ & $\frac{1}{4 \sqrt{2}}$ & $-\frac{1}{12 \sqrt{3}}$ & $-\frac{1}{12 \sqrt{2}}$ & $-\frac{5}{12 \sqrt{6}}$ & $-\frac{5}{12 \sqrt{3}}$ \\
$\langle O_{3}^{kc}\rangle$ & $\frac{1}{2 \sqrt{3}}$ & $0$ & $\frac{1}{4 \sqrt{2}}$ & $\frac{1}{4 \sqrt{3}}$ & $-\frac{1}{4 \sqrt{2}}$ & $-\frac{1}{4 \sqrt{6}}$ & $-\frac{1}{4 \sqrt{3}}$ \\
$\langle O_{4}^{kc}\rangle$ & $\frac{5}{2 \sqrt{3}}$ & $0$ & $-\frac{3}{4 \sqrt{2}}$ & $\frac{1}{4 \sqrt{3}}$ & $-\frac{1}{4 \sqrt{2}}$ & $-\frac{5}{4 \sqrt{6}}$ & $-\frac{5}{4 \sqrt{3}}$ \\
$\langle O_{5}^{kc}\rangle$ & $\frac{1}{2 \sqrt{3}}$ & $0$ & $\frac{1}{4 \sqrt{2}}$ & $-\frac{\sqrt{3}}{4}$ & $-\frac{5}{4 \sqrt{2}}$ & $-\frac{1}{4 \sqrt{6}}$ & $-\frac{1}{4 \sqrt{3}}$ \\
$\langle O_{7}^{kc}\rangle$ & $\frac{5}{2 \sqrt{3}}$ & $\frac{1}{\sqrt{2}}$ & $\frac{3}{4 \sqrt{2}}$ & $-\frac{1}{4 \sqrt{3}}$ & $-\frac{1}{4 \sqrt{2}}$ & $-\frac{5}{4 \sqrt{6}}$ & $-\frac{5}{4 \sqrt{3}}$ \\
$\langle O_{9}^{kc}\rangle$ & $\frac{5}{4 \sqrt{3}}$ & $0$ & $\frac{3}{8 \sqrt{2}}$ & $\frac{\sqrt{3}}{8}$ & $-\frac{5}{8 \sqrt{2}}$ & $-\frac{5}{8 \sqrt{6}}$ & $-\frac{5}{8 \sqrt{3}}$ \\
$\langle O_{10}^{kc}\rangle$ & $\frac{5}{4 \sqrt{3}}$ & $0$ & $\frac{3}{8 \sqrt{2}}$ & $\frac{\sqrt{3}}{8}$ & $-\frac{5}{8 \sqrt{2}}$ & $-\frac{5}{8 \sqrt{6}}$ & $-\frac{5}{8 \sqrt{3}}$ \\
$\langle O_{11}^{kc}\rangle$ & $\sqrt{3}$ & $0$ & $-\frac{3}{2 \sqrt{2}}$ & $-\frac{\sqrt{3}}{2}$ & $-\frac{3}{2 \sqrt{2}}$ & $-\frac12 \sqrt{\frac32}$ & $-\frac{\sqrt{3}}{2}$ \\
$\langle O_{12}^{kc}\rangle$ & $\frac{5}{4 \sqrt{3}}$ & $-\frac{1}{\sqrt{2}}$ & $\frac{3}{8 \sqrt{2}}$ & $\frac{11}{8 \sqrt{3}}$ & $-\frac{13}{8 \sqrt{2}}$ & $-\frac{5}{8 \sqrt{6}}$ & $-\frac{5}{8 \sqrt{3}}$ \\
$\langle O_{15}^{kc}\rangle$ & $\frac{\sqrt{3}}{4}$ & $0$ & $\frac{3}{8 \sqrt{2}}$ & $-\frac{3 \sqrt{3}}{8}$ & $-\frac{15}{8 \sqrt{2}}$ & $-\frac18 \sqrt{\frac32}$ & $-\frac{\sqrt{3}}{8}$ \\
$\langle O_{16}^{kc}\rangle$ & $\frac{5 \sqrt{3}}{4}$ & $0$ & $-\frac{9}{8 \sqrt{2}}$ & $\frac{\sqrt{3}}{8}$ & $-\frac{3}{8 \sqrt{2}}$ & $-\frac58 \sqrt{\frac32}$ & $-\frac{5 \sqrt{3}}{8}$ \\
$\langle O_{20}^{kc}\rangle$ & $0$ & $\frac{3}{2 \sqrt{2}}$ & $\frac{27}{16 \sqrt{2}}$ & $\frac{\sqrt{3}}{16}$ & $-\frac{3}{16 \sqrt{2}}$ & $-\frac{25}{16} \sqrt{\frac32}$ & $-\frac{25 \sqrt{3}}{16}$ \\
$\langle O_{21}^{kc}\rangle$ & $0$ & $-\frac{3}{2 \sqrt{2}}$ & $-\frac{27}{16 \sqrt{2}}$ & $-\frac{\sqrt{3}}{16}$ & $\frac{3}{16 \sqrt{2}}$ & $\frac{25}{16} \sqrt{\frac32}$ & $\frac{25 \sqrt{3}}{16}$ \\
$\langle O_{24}^{kc}\rangle$ & $0$ & $\frac{3}{\sqrt{2}}$ & $\frac{27}{8 \sqrt{2}}$ & $\frac{\sqrt{3}}{8}$ & $-\frac{3}{8 \sqrt{2}}$ & $-\frac{25}{8} \sqrt{\frac32}$ & $-\frac{25 \sqrt{3}}{8}$ \\
$\langle O_{31}^{kc}\rangle$ & $\frac{5 \sqrt{3}}{4}$ & $0$ & $\frac{9}{8 \sqrt{2}}$ & $\frac{3 \sqrt{3}}{8}$ & $-\frac{15}{8 \sqrt{2}}$ & $-\frac58 \sqrt{\frac32}$ & $-\frac{5 \sqrt{3}}{8}$ \\
\end{tabular}
\end{ruledtabular}
\end{table*}

\begin{table*}
\caption{\label{t:mtx27}Matrix elements of flavor $\mathbf{27}$ operators.}
\begin{ruledtabular}
\begin{tabular}{rrrrrrrr}
& $np$ & $\Xi^\pm \Lambda$ & $\Lambda p $ & $\Sigma^-n$ & $\Xi^-\Lambda$ & $\Xi^-\Sigma^0$ & $\Xi^0\Sigma^+$ \\
\hline
$\langle T_{1}^{kc}\rangle$ & $0$ & $0$ & $-\frac38 \sqrt{\frac32}$ & $\frac18$ & $\frac18 \sqrt{\frac32}$ & $\frac{5}{8 \sqrt{2}}$ & $\frac58$ \\
$\langle T_{2}^{kc}\rangle$ & $\frac{5}{18}$ & $\frac{1}{3 \sqrt{6}}$ & $-\frac{1}{8 \sqrt{6}}$ & $\frac{1}{72}$ & $\frac{1}{24 \sqrt{6}}$ & $\frac{5}{72 \sqrt{2}}$ & $\frac{5}{72}$ \\
$\langle T_{5}^{kc}\rangle$ & $0$ & $0$ & $-\frac38 \sqrt{\frac32}$ & $-\frac38$ & $\frac38 \sqrt{\frac32}$ & $\frac{3}{8 \sqrt{2}}$ & $\frac38$ \\
$\langle T_{6}^{kc}\rangle$ & $\frac16$ & $0$ & $-\frac{1}{8 \sqrt{6}}$ & $-\frac{1}{24}$ & $\frac{1}{8 \sqrt{6}}$ & $\frac{1}{24 \sqrt{2}}$ & $\frac{1}{24}$ \\
$\langle T_{7}^{kc}\rangle$ & $-\frac16$ & $0$ & $-\frac{1}{4 \sqrt{6}}$ & $-\frac{1}{12}$ & $\frac{1}{4 \sqrt{6}}$ & $\frac{1}{12 \sqrt{2}}$ & $\frac{1}{12}$ \\
$\langle T_{9}^{kc}\rangle$ & $\frac56$ & $0$ & $\frac18 \sqrt{\frac32}$ & $-\frac{1}{24}$ & $\frac{1}{8 \sqrt{6}}$ & $\frac{5}{24 \sqrt{2}}$ & $\frac{5}{24}$ \\
$\langle T_{10}^{kc}\rangle$ & $-\frac16$ & $0$ & $-\frac{1}{4 \sqrt{6}}$ & $\frac14$ & $\frac{5}{4 \sqrt{6}}$ & $\frac{1}{12 \sqrt{2}}$ & $\frac{1}{12}$ \\
$\langle T_{12}^{kc}\rangle$ & $0$ & $0$ & $-\frac{9}{16} \sqrt{\frac32}$ & $-\frac{1}{16}$ & $\frac{1}{16} \sqrt{\frac32}$ & $\frac{25}{16 \sqrt{2}}$ & $\frac{25}{16}$ \\
$\langle T_{13}^{kc}\rangle$ & $0$ & $0$ & $\frac{9}{16} \sqrt{\frac32}$ & $\frac{1}{16}$ & $-\frac{1}{16} \sqrt{\frac32}$ & $-\frac{25}{16 \sqrt{2}}$ & $-\frac{25}{16}$ \\
$\langle T_{14}^{kc}\rangle$ & $0$ & $0$ & $-\frac98 \sqrt{\frac32}$ & $\frac38$ & $\frac38 \sqrt{\frac32}$ & $\frac{15}{8 \sqrt{2}}$ & $\frac{15}{8}$ \\
$\langle T_{15}^{kc}\rangle$ & $\frac56$ & $\frac{1}{\sqrt{6}}$ & $-\frac18 \sqrt{\frac32}$ & $\frac{1}{24}$ & $\frac{1}{8 \sqrt{6}}$ & $\frac{5}{24 \sqrt{2}}$ & $\frac{5}{24}$ \\
$\langle T_{16}^{kc}\rangle$ & $-\frac56$ & $-\frac{1}{\sqrt{6}}$ & $-\frac14 \sqrt{\frac32}$ & $\frac{1}{12}$ & $\frac{1}{4 \sqrt{6}}$ & $\frac{5}{12 \sqrt{2}}$ & $\frac{5}{12}$ \\
$\langle T_{17}^{kc}\rangle$ & $0$ & $0$ & $-\frac38 \sqrt{\frac32}$ & $\frac18$ & $\frac18 \sqrt{\frac32}$ & $\frac{5}{8 \sqrt{2}}$ & $\frac58$ \\
$\langle T_{18}^{kc}\rangle$ & $0$ & $0$ & $\frac38 \sqrt{\frac32}$ & $-\frac18$ & $-\frac18 \sqrt{\frac32}$ & $-\frac{5}{8 \sqrt{2}}$ & $-\frac58$ \\
$\langle T_{25}^{kc}\rangle$ & $\frac52$ & $0$ & $-\frac34 \sqrt{\frac32}$ & $\frac14$ & $\frac14 \sqrt{\frac32}$ & $\frac{5}{4 \sqrt{2}}$ & $\frac54$ \\
$\langle T_{26}^{kc}\rangle$ & $\frac12$ & $0$ & $\frac18 \sqrt{\frac32}$ & $-\frac38$ & $\frac58 \sqrt{\frac32}$ & $\frac{1}{8 \sqrt{2}}$ & $\frac18$ \\
$\langle T_{27}^{kc}\rangle$ & $\frac{5}{24}$ & $\sqrt{\frac23}$ & $-\frac{5}{16} \sqrt{\frac32}$ & $\frac{13}{48}$ & $\frac{7}{16} \sqrt{\frac32}$ & $\frac{145}{48 \sqrt{2}}$ & $\frac{145}{48}$ \\
$\langle T_{28}^{kc}\rangle$ & $\frac{5}{24}$ & $0$ & $-\frac{1}{32} \sqrt{\frac32}$ & $\frac{11}{32}$ & $\frac{65}{32 \sqrt{6}}$ & $\frac{5}{96 \sqrt{2}}$ & $\frac{5}{96}$ \\
$\langle T_{29}^{kc}\rangle$ & $\frac{5}{12}$ & $-\frac{1}{\sqrt{6}}$ & $-\frac{1}{16} \sqrt{\frac32}$ & $-\frac{11}{48}$ & $\frac{13}{16 \sqrt{6}}$ & $\frac{5}{48 \sqrt{2}}$ & $\frac{5}{48}$ \\
$\langle T_{30}^{kc}\rangle$ & $-\frac{5}{12}$ & $\frac{1}{\sqrt{6}}$ & $-\frac18 \sqrt{\frac32}$ & $-\frac{11}{24}$ & $\frac{13}{8 \sqrt{6}}$ & $\frac{5}{24 \sqrt{2}}$ & $\frac{5}{24}$ \\
$\langle T_{31}^{kc}\rangle$ & $\frac{5}{12}$ & $0$ & $-\frac{1}{16} \sqrt{\frac32}$ & $-\frac{1}{16}$ & $\frac{5}{16 \sqrt{6}}$ & $\frac{5}{48 \sqrt{2}}$ & $\frac{5}{48}$ \\
$\langle T_{32}^{kc}\rangle$ & $\frac{5}{12}$ & $0$ & $-\frac{1}{16} \sqrt{\frac32}$ & $-\frac{1}{16}$ & $\frac{5}{16 \sqrt{6}}$ & $\frac{5}{48 \sqrt{2}}$ & $\frac{5}{48}$ \\
$\langle T_{33}^{kc}\rangle$ & $-\frac{5}{12}$ & $0$ & $-\frac18 \sqrt{\frac32}$ & $-\frac18$ & $\frac{5}{8 \sqrt{6}}$ & $\frac{5}{24 \sqrt{2}}$ & $\frac{5}{24}$ \\
$\langle T_{34}^{kc}\rangle$ & $-\frac{5}{12}$ & $0$ & $-\frac18 \sqrt{\frac32}$ & $-\frac18$ & $\frac{5}{8 \sqrt{6}}$ & $\frac{5}{24 \sqrt{2}}$ & $\frac{5}{24}$ \\
$\langle T_{45}^{kc}\rangle$ & $\frac32$ & $0$ & $-\frac34 \sqrt{\frac32}$ & $-\frac34$ & $\frac34 \sqrt{\frac32}$ & $\frac{3}{4 \sqrt{2}}$ & $\frac34$ \\
$\langle T_{46}^{kc}\rangle$ & $\frac18$ & $0$ & $-\frac{5}{16} \sqrt{\frac32}$ & $-\frac{13}{16}$ & $\frac{21}{16} \sqrt{\frac32}$ & $\frac{29}{16 \sqrt{2}}$ & $\frac{29}{16}$ \\
$\langle T_{47}^{kc}\rangle$ & $\frac58$ & $0$ & $\frac{3}{32} \sqrt{\frac32}$ & $\frac{11}{32}$ & $\frac{13}{32} \sqrt{\frac32}$ & $\frac{5}{32 \sqrt{2}}$ & $\frac{5}{32}$ \\
$\langle T_{48}^{kc}\rangle$ & $\frac54$ & $0$ & $\frac{3}{16} \sqrt{\frac32}$ & $-\frac{1}{16}$ & $\frac{1}{16} \sqrt{\frac32}$ & $\frac{5}{16 \sqrt{2}}$ & $\frac{5}{16}$ \\
$\langle T_{49}^{kc}\rangle$ & $-\frac14$ & $0$ & $-\frac18 \sqrt{\frac32}$ & $\frac38$ & $\frac58 \sqrt{\frac32}$ & $\frac{1}{8 \sqrt{2}}$ & $\frac18$ \\
$\langle T_{50}^{kc}\rangle$ & $0$ & $0$ & $-\frac{3}{16} \sqrt{\frac32}$ & $\frac{9}{16}$ & $\frac{15}{16} \sqrt{\frac32}$ & $\frac{3}{16 \sqrt{2}}$ & $\frac{3}{16}$ \\
$\langle T_{51}^{kc}\rangle$ & $\frac54$ & $0$ & $-\frac38 \sqrt{\frac32}$ & $\frac18$ & $\frac38 \sqrt{\frac32}$ & $\frac{15}{8 \sqrt{2}}$ & $\frac{15}{8}$ \\
$\langle T_{52}^{kc}\rangle$ & $\frac54$ & $0$ & $\frac{3}{16} \sqrt{\frac32}$ & $\frac{3}{16}$ & $\frac{5}{16} \sqrt{\frac32}$ & $\frac{5}{16 \sqrt{2}}$ & $\frac{5}{16}$ \\
$\langle T_{53}^{kc}\rangle$ & $\frac14$ & $0$ & $-\frac{1}{16} \sqrt{\frac32}$ & $-\frac{9}{16}$ & $\frac{25}{16} \sqrt{\frac32}$ & $\frac{1}{16 \sqrt{2}}$ & $\frac{1}{16}$ \\
$\langle T_{56}^{kc}\rangle$ & $0$ & $0$ & $\frac{3}{16} \sqrt{\frac32}$ & $\frac{3}{16}$ & $\frac{5}{16} \sqrt{\frac32}$ & $-\frac{11}{16 \sqrt{2}}$ & $-\frac{11}{16}$ \\
$\langle T_{57}^{kc}\rangle$ & $0$ & $\sqrt{\frac32}$ & $-\frac38 \sqrt{\frac32}$ & $\frac18$ & $-\frac18 \sqrt{\frac32}$ & $\frac{11}{8 \sqrt{2}}$ & $\frac{11}{8}$ \\
$\langle T_{59}^{kc}\rangle$ & $0$ & $0$ & $-\frac{27}{16} \sqrt{\frac32}$ & $-\frac{3}{16}$ & $\frac{3}{16} \sqrt{\frac32}$ & $\frac{75}{16 \sqrt{2}}$ & $\frac{75}{16}$ \\
$\langle T_{64}^{kc}\rangle$ & $0$ & $\frac12 \sqrt{\frac32}$ & $-\frac{9}{32} \sqrt{\frac32}$ & $-\frac{1}{32}$ & $\frac{1}{32} \sqrt{\frac32}$ & $\frac{25}{32 \sqrt{2}}$ & $\frac{25}{32}$ \\
$\langle T_{65}^{kc}\rangle$ & $0$ & $-\frac12 \sqrt{\frac32}$ & $\frac{9}{32} \sqrt{\frac32}$ & $\frac{1}{32}$ & $-\frac{1}{32} \sqrt{\frac32}$ & $-\frac{25}{32 \sqrt{2}}$ & $-\frac{25}{32}$ \\
\end{tabular}
\end{ruledtabular}
\end{table*}

\begin{table*}
\caption{Matrix elements of flavor $\mathbf{27}$ operators. Continuation of above table.}
\begin{ruledtabular}
\begin{tabular}{rrrrrrrr}
& $np$ & $\Xi^\pm \Lambda$ & $\Lambda p $ & $\Sigma^-n$ & $\Xi^-\Lambda$ & $\Xi^-\Sigma^0$ & $\Xi^0\Sigma^+$ \\
\hline
$\langle T_{66}^{kc}\rangle$ & $0$ & $0$ & $-\frac{9}{16} \sqrt{\frac32}$ & $\frac{3}{16}$ & $\frac{5}{16} \sqrt{\frac32}$ & $\frac{25}{16 \sqrt{2}}$ & $\frac{25}{16}$ \\
$\langle T_{67}^{kc}\rangle$ & $0$ & $0$ & $\frac{9}{32} \sqrt{\frac32}$ & $-\frac{3}{32}$ & $-\frac{5}{32} \sqrt{\frac32}$ & $-\frac{25}{32 \sqrt{2}}$ & $-\frac{25}{32}$ \\
$\langle T_{92}^{kc}\rangle$ & $\frac58$ & $\sqrt{\frac32}$ & $-\frac{15}{16} \sqrt{\frac32}$ & $\frac{5}{16}$ & $\frac{13}{16} \sqrt{\frac32}$ & $\frac{65}{16 \sqrt{2}}$ & $\frac{65}{16}$ \\
$\langle T_{93}^{kc}\rangle$ & $\frac58$ & $0$ & $-\frac{3}{32} \sqrt{\frac32}$ & $\frac{9}{32}$ & $\frac{25}{32} \sqrt{\frac32}$ & $\frac{5}{32 \sqrt{2}}$ & $\frac{5}{32}$ \\
$\langle T_{94}^{kc}\rangle$ & $\frac58$ & $\sqrt{6}$ & $-\frac{15}{16} \sqrt{\frac32}$ & $\frac{13}{16}$ & $\frac{21}{16} \sqrt{\frac32}$ & $\frac{145}{16 \sqrt{2}}$ & $\frac{145}{16}$ \\
$\langle T_{95}^{kc}\rangle$ & $\frac58$ & $0$ & $-\frac{3}{32} \sqrt{\frac32}$ & $\frac{33}{32}$ & $\frac{65}{32} \sqrt{\frac32}$ & $\frac{5}{32 \sqrt{2}}$ & $\frac{5}{32}$ \\
$\langle T_{96}^{kc}\rangle$ & $\frac34$ & $0$ & $-\frac38 \sqrt{\frac32}$ & $-\frac38$ & $\frac98 \sqrt{\frac32}$ & $\frac{9}{8 \sqrt{2}}$ & $\frac98$ \\
$\langle T_{97}^{kc}\rangle$ & $\frac{15}{4}$ & $0$ & $-\frac{9}{16} \sqrt{\frac32}$ & $\frac{3}{16}$ & $\frac{3}{16} \sqrt{\frac32}$ & $\frac{15}{16 \sqrt{2}}$ & $\frac{15}{16}$ \\
$\langle T_{98}^{kc}\rangle$ & $\frac54$ & $0$ & $-\frac{3}{16} \sqrt{\frac32}$ & $-\frac{3}{16}$ & $\frac{5}{16} \sqrt{\frac32}$ & $\frac{5}{16 \sqrt{2}}$ & $\frac{5}{16}$ \\
$\langle T_{99}^{kc}\rangle$ & $-\frac54$ & $0$ & $-\frac38 \sqrt{\frac32}$ & $-\frac38$ & $\frac58 \sqrt{\frac32}$ & $\frac{5}{8 \sqrt{2}}$ & $\frac58$ \\
$\langle T_{130}^{kc}\rangle$ & $\frac38$ & $0$ & $-\frac{15}{16} \sqrt{\frac32}$ & $-\frac{15}{16}$ & $\frac{39}{16} \sqrt{\frac32}$ & $\frac{39}{16 \sqrt{2}}$ & $\frac{39}{16}$ \\
$\langle T_{131}^{kc}\rangle$ & $\frac{15}{8}$ & $0$ & $\frac{9}{32} \sqrt{\frac32}$ & $\frac{9}{32}$ & $\frac{15}{32} \sqrt{\frac32}$ & $\frac{15}{32 \sqrt{2}}$ & $\frac{15}{32}$ \\
\end{tabular}
\end{ruledtabular}
\end{table*}

\end{document}